\documentclass[article,twocolumn]{emulateapj}

\usepackage{graphicx,amssymb,amsmath,amsfonts}
\usepackage{color,subfigure}
\usepackage{natbib}
\usepackage[section]{placeins}
\usepackage{cases}
\usepackage[mathscr]{euscript}
\usepackage{mathrsfs}

\usepackage{gensymb}

\newcommand{\appropto}{\mathrel{\vcenter{
  \offinterlineskip\halign{\hfil$##$\cr
    \propto\cr\noalign{\kern2pt}\sim\cr\noalign{\kern-2pt}}}}}

\def\app#1#2{%
  \mathrel{%
    \setbox0=\hbox{$#1\sim$}%
    \setbox2=\hbox{%
      \rlap{\hbox{$#1\propto$}}%
      \lower1.1\ht0\box0%
    }%
    \raise0.25\ht2\box2%
  }%
}

\renewcommand{\th}{$^{\rm th}$}

\ifdefined\kms
\renewcommand{\kms}{\,\rm km\ s^{-1}}
\else
\newcommand{\kms}{\,\rm km\ s^{-1}}
\fi
\newcommand{\erg}{\,\rm erg}

\newcommand{\Hz}{\,\rm{Hz}}

\let\AAold\AA
\renewcommand{\AA}{\text{\AAold}}
\newcommand{\mic}{\,\mbox{$\mu$m}}
\newcommand{\cm}{\,{\rm cm}}

\newcommand{\kpc}{\,{\rm kpc}}
\newcommand{\Mpc}{\,{\rm Mpc}}

\newcommand{\ryd}{\,{\rm Ryd}}
\newcommand{\s}{\,{\rm s}}

\newcommand{\yr}{\,{\rm yr}}
\newcommand{\Myr}{\,{\rm Myr}}
\newcommand{\Gyr}{\,{\rm Gyr}}

\newcommand{\K}{\,{\rm K}}

\newcommand{\msun}{\,{\rm M_{\odot}}}
\newcommand{\zsun}{\,{\rm Z_{\odot}}}

\renewcommand{\mp}{m_{\rm p}}

\newcommand{\dex}{\,{\rm dex}}

\DeclareRobustCommand{\ion}[2]{%
\relax\ifmmode
\ifx\testbx\f@series
{\mathbf{#1\,\mathsc{#2}}}\else
{\mathrm{#1\,\mathsc{#2}}}\fi
\else\textup{#1\,{\mdseries\textsc{#2}}}%
\fi}

\newcommand{\hi}{\text{H~{\sc i}}}
\newcommand{\hii}{\text{H~{\sc ii}}}

\newcommand{\niip}{\text{N~{\sc ii}}}
\newcommand{\niiip}{\text{N~{\sc iii}}}
\newcommand{\nvp}{\text{N~{\sc v}}}

\newcommand{\siiip}{\text{S~{\sc iii}}}

\newcommand{\Siiip}{\text{Si~{\sc ii}}}
\newcommand{\Siiiip}{\text{Si~{\sc iii}}}

\newcommand{\oiiip}{\text{O~{\sc iii}}}

\newcommand{\oivp}{\text{O~{\sc iv}}}
\newcommand{\ovp}{\text{O~{\sc v}}}
\newcommand{\ovip}{\text{O~{\sc vi}}}

\newcommand{\oviip}{\text{O~{\sc vii}}}
\newcommand{\oviiip}{\text{O~{\sc viii}}}

\newcommand{\neviiip}{\text{Ne~{\sc viii}}}

\newcommand{\mgip}{\text{Mg~{\sc i}}}
\newcommand{\mgiip}{\text{Mg~{\sc ii}}}

\newcommand{\ciiip}{\text{C~{\sc iii}}}
\newcommand{\ciip}{\text{C~{\sc ii}}}
\newcommand{\civp}{\text{C~{\sc iv}}}

\newcommand{\Lya}{\text{Ly$\alpha$}}
\newcommand{\Ha}{\text{H$\alpha$}}

\ifdefined\aap
\else
\newcommand{\aap}{A\&A}
\newcommand{\araa}{ARA\&A}
\newcommand{\apjl}{ApJ}
\newcommand{\apjs}{ApJS}
\newcommand{\apj}{ApJ}
\newcommand{\aj}{AJ}
\newcommand{\mnras}{MNRAS}

\fi

\newcommand{\cloudy}{{\sc cloudy}}

\newcommand{\HST}{{\it HST}}

\newcommand{\Mdot}{{\dot M}}

\newcommand{\ebv}{E(B-V)}

\newcommand{\nH}{n_{\rm H}}
\newcommand{\NH}{N_{\rm H}}

\newcommand{\Mhalo}{M_{\rm halo}}
\newcommand{\NHI}{N_{\rm HI}}
\newcommand{\rvir}{R_{\rm vir}}

\newcommand{\novi}{N_{\rm \ovip}}

\newcommand{\Rimp}{R_{\perp}}
\newcommand{\Lstar}{L^*}

\newcommand{\nhi}{n_{\rm \hi}}

\newcommand{\Mstar}{M_{\rm star}}

\newcommand{\el}{{\rm X}}
\renewcommand{\ion}{\el^{i+}}

\newcommand{\supovi}{^{(\ovip)}}

\newcommand{\Movi}{M_{\ovip}}
\newcommand{\Mgasovi}{M_{\rm gas}\supovi}

\newcommand{\lovi}{l_{\ovip}}
\newcommand{\bovi}{b_{\ovip}}
\newcommand{\fovi}{f_{\ovip}}

\newcommand{\Novi}{N_{\ovip}}

\renewcommand{\novi}{n_{\ovip}}

\newcommand{\sLstar}{$\sim$$\Lstar$}

\newcommand{\fhi}{f_{\hi}}
\newcommand{\Omegab}{\Omega_{\rm b}}
\newcommand{\OmegaM}{\Omega_{\rm M}}
\newcommand{\Tvir}{T_{\rm vir}}

\renewcommand{\d}{{\rm d}}

\newcommand{\Rovi}{R_{\ovip}}

\renewcommand{\Mdot}{{\dot M}}

\renewcommand{\vr}{v_{\rm r}}

\renewcommand{\ebv}{E_{\rm B-V}}

\newcommand{\scriptM}{\mathscr{M}}

\newcommand{\opv}{{\rm O}^{5+}}
\newcommand{\nopv}{n_{{\rm O}^{5+}}}

\begin{document}

\title{
Does circumgalactic \ovip\ trace low-pressure gas beyond the accretion shock?\\clues from \hi\ and low-ion absorption, line kinematics, and dust extinction
} 
\author{
Jonathan~Stern\altaffilmark{1}\altaffilmark{3}\footnotemark[*]\footnotemark[\textdagger],
Claude-Andr{\'e}~Faucher-Gigu{\`e}re\altaffilmark{1},
Joseph~F.~Hennawi\altaffilmark{2}\altaffilmark{3},
Zachary~Hafen\altaffilmark{1},
Sean~D.~Johnson\altaffilmark{4}\altaffilmark{5}\footnotemark[\textdaggerdbl], and
Drummond~Fielding\altaffilmark{6}
}
\footnotetext[*]{E-mail: jonathan.stern@northwestern.edu}
\footnotetext[\textdagger]{CIERA Fellow}
\footnotetext[\textdaggerdbl]{Hubble and Carnegie-Princeton Fellow}
\altaffiltext{1}{Department of Physics and Astronomy and CIERA, Northwestern University, Evanston, IL, USA}
\altaffiltext{2}{Department of Physics, University of California, Santa Barbara,
CA 93106, USA}
\altaffiltext{3}{Max Planck Institut f\"{u}r Astronomie, K\"{o}nigstuhl 17, D-69117, Heidelberg, Germany} 
\altaffiltext{4}{Department of Astrophysical Sciences, Princeton University, Princeton NJ 08544, USA}
\altaffiltext{5}{The Observatories of the Carnegie Institution for Science, 813 Santa Barbara Street, Pasadena, CA 91101, USA}
\altaffiltext{6}{Astronomy Department and Theoretical Astrophysics Center, University of California Berkeley, Berkeley, CA 94720, USA}

\begin{abstract} 
Large \ovip\ columns are observed around star-forming, low-redshift \sLstar\ galaxies, with a dependence on impact parameter indicating that most $\opv$ particles reside beyond half the halo virial radius ($\gtrsim100\kpc$). 
In order to constrain the nature of the gas traced by \ovip, we analyze additional~observables of the outer halo, namely \hi\ to \ovip\ column ratios of $1-10$, an absence of~low-ion~absorption,~a mean differential extinction of $\ebv\approx10^{-3}$,  and a linear relation between \ovip\ column and velocity width. 
We contrast these observations with two physical scenarios:
(1) \ovip\ traces high-pressure ($\sim30\cm^{-3}\K$) collisionally-ionized gas cooling from a virially-shocked phase, 
and 
(2) \ovip\ traces low-pressure ($\lesssim1\cm^{-3}\K$) gas beyond the accretion shock, where the gas is in ionization and thermal equilibrium with the UV background. 
We demonstrate that the high-pressure scenario requires multiple gas phases to explain the observations, and a large deposition of energy at $\gtrsim100\kpc$ to offset the energy radiated by the cooling gas. 
In contrast, the low-pressure scenario can explain all considered observations with a single gas phase in thermal equilibrium, provided that the baryon overdensity is comparable to the dark-matter overdensity, and that the gas is enriched to $\gtrsim\zsun/3$ with an ISM-like dust-to-metal ratio. 
The low-pressure scenario implies that \ovip\ traces a cool flow with mass flow rate of $\sim5\msun\yr^{-1}$, comparable to the star formation rate of the central galaxies. 
The \ovip\ line widths are consistent with the velocity shear expected within this flow. 
The low-pressure scenario predicts a bimodality in absorption line ratios at $\sim100\kpc$, due to the pressure jump across the accretion shock.
\end{abstract} 


\section{Introduction}\label{s:intro}

Recent observations with the cosmic origin spectrograph (COS) onboard \HST\ have detected a high incidence of \ovip\ $\lambda\lambda1031,1037$ absorption around blue, low-redshift ($z\sim0.2$) galaxies with luminosity \sLstar. The detection fractions of absorbers with columns $\Novi\sim10^{14.5}\cm^{-2}$ are near unity out to impact parameters of order 
$\rvir$, where $\rvir$ is the virial radius  of the dark matter halo (\citealt{ChenMulchaey09,Prochaska+11,Tumlinson+11,Johnson+15}). 
The ubiquitous detection of \ovip\ in the circumgalactic medium (CGM) of blue galaxies is in stark contrast to the CGM of red galaxies with similar redshift and luminosity, in which strong \ovip\ is rarely detected (\citealt{Tumlinson+11}). 
This reflection of the specific star formation rate (sSFR) in the properties of the $100\kpc$-scale CGM holds the potential to provide insight into the long-standing problem of why galaxies have a bimodal color distribution (\citealt{Strateva+01}). 

Broadly speaking, oxygen can be ionized into the $\opv$ state either via collisions with electrons in `warm' gas with temperatures $T\sim10^{5.5}\K$, or photoionized by the UV background in cooler gas with a density of $\nH \lesssim 10^{-4}\cm^{-3}$, assuming current estimates of the UV background are not too far from the correct value. 
While local ionization sources in the galaxy or CGM can in principle also be the source of $\ovip$ (e.g.\ \citealt{OppenheimerSchaye13}), these sources are unlikely to be dominant at distances $>100\kpc$ (\citealt{McQuinnWerk17}) which we show below is where most of the gas traced by $\ovip$ resides. Hence, in order to understand the implication of the observed bimodality in \ovip\ absorption for the physical conditions of the CGM, one must first understand whether \ovip\ is produced by collisional ionization or via photoionization by the UV background.

The question of the \ovip\ ionization mechanism around low-redshift \sLstar\ galaxies has been addressed by a considerable number of studies which employed cosmological simulations, all of which have found that \ovip\ at $\lesssim\rvir$ is produced mainly by collisionally ionized gas rather than by photoionized gas (\citealt{Stinson+12, Hummels+13, Cen13,Oppenheimer+16, Liang+16, Gutcke+17,Suresh+17}). Though, these simulations typically underestimate the observed $\Novi$ by factors of $3-10$. 
The simulations disfavor a photoionization origin since photoionization of \ovip\ by the background is significant only in gas with a low thermal pressure of $\nH T\lesssim5\cm^{-3}\K$. If the photoionized gas is also in thermal equilibrium with the UV background, the implied gas pressure is even lower, of order $1\cm^{-3}\K$ (as we demonstrate later in this paper).
For comparison, the halos in simulations are filled with hot gas shocked to the virial temperature $\Tvir\sim 10^{6}\K$ and a mass similar to the baryon closing fraction, implying significantly larger pressures of $\nH T\approx 30\cm^{-3}\K$ near half the virial radius. Hence, even if some fraction of the halo gas cools to the temperatures of $<10^{5}\K$ where photoionization dominates over collisional ionization, the gas will be compressed by the hot gas to densities in which the oxygen ionization level is well below $\opv$. 

The argument against photoionization as the source of \ovip\ is therefore based on the assumption that the \ovip-gas is within the accretion shock surrounding \sLstar\ galaxies, where the gas pressures are relatively high.  
Despite the general agreement among current simulations that the accretion shock is at $\gtrsim\rvir$, and therefore beyond the distances where large $\Novi$ are observed, this conclusion is both uncertain theoretically and has not been confirmed by observations. 
Observationally, directly detecting virially shocked gas at $\sim\rvir$ is currently beyond the capabilities of X-ray telescopes (see e.g.\ \citealt{Li+17}, and further discussion below). 
Theoretically, the \sLstar\ halo mass of $\sim10^{12}\msun$ inferred from abundance matching (e.g.\ \citealt{Moster+13}) is near the nominal critical halo mass of $\sim10^{11.5}\msun$ required to support a stable virial shock. 
This threshold halo mass is determined by the ratio of the cooling time to the free-fall time (\citealt{ReesOstriker77, WhiteFrenk91}; \citealt{BirnboimDekel03, Keres+05,Keres+09, DekelBirnboim06,  Fielding+17}), which is somewhat uncertain. If, for example, halo gas cooling rates in simulations are modified by metal enrichment (as suggested by the high gas metallicities inferred in recent UV absorber studies, e.g.\ \citealt{Stern+16} and \citealt{Prochaska+17}), the properties of the virial shock could differ from current predictions. It is also worth noting that even for identical physics, different hydrodynamic solvers sometimes predict shock radii differing by a factor up to $\sim2$ (e.g., \citealt{Nelson+13}).

Observational evidence that $\opv$ is collisionally ionized would hence support the picture suggested by cosmological simulations, where \sLstar\ halos are filled with a massive hot gas phase out to $\gtrsim\rvir$. On the other hand, observational evidence that $\opv$ is photoionized by the UV background would suggest a different picture in which the accretion shock is at smaller distances, and \ovip\ traces cool gas beyond the shock. Previous observational work has not been able to distinguish between photoionization and collisional ionization as the source of strong \ovip\ absorption around \sLstar\ galaxies. 
While the observed \ovip\ absorption is inconsistent with the photoionization models of low ions that assume a single density (e.g.\ \citealt{Werk+14}), more flexible photoionization models with two phases or with a density profile are consistent with the \ovip\ absorption (\citealt{Stern+16}). 
The goal of this paper is to confront these two possible ionization scenarios with existing observations, and derive the implied challenges and successes imposed by the observations for each scenario.

Our approach is similar to the approach of \citeauthor{McQuinnWerk17} (2017, hereafter MQW17), who derived physical constraints on the conditions in the CGM directly from observations. The main difference is that MQW17 focused on discriminating between different types of collisional ionized $\opv$ models, while here we focus on the more basic question of whether $\opv$ is collisionally ionized or photoionized by the UV background. 
While the collisional ionization scenario has received the most attention so far, we demonstrate that a low-pressure scenario in which the \ovip\ is in ionization and thermal equilibrium with the UV background is consistent with observational constraints and therefore warrants further investigation. 

This paper is organized as follows. In \S\ref{s:observations} we review available observations that can constrain the pressure of gas in the outer halo and the ionization mechanism of $\opv$. In \S\ref{s:implications} we contrast these observations with the low-pressure and high-pressure scenarios mentioned above. We discuss our results in \S\ref{s:discussion},  focusing on the low-pressure scenario and highlighting predictions that can be used to further test this scenario. We summarize our analysis and conclusions in \S\ref{s:summary}. 

A flat $\Lambda$CDM cosmology with $H_0=68\kms\Mpc^{-1}$, $\OmegaM=0.31$, and $\Omegab/\OmegaM=0.158$ is assumed throughout (\citealt{Planck16}).

\section{Observed properties}\label{s:observations}

As we show below, most of the \ovip\ observed around low-redshift \sLstar\ galaxies originates in gas in the outer halo, at a distance $R\gtrsim100\kpc$. We therefore start by reviewing observations that can be used to constrain the physical conditions in the outer halo.

\subsection{\ovip\ columns}\label{s:ovi}

\begin{figure*}
 \includegraphics{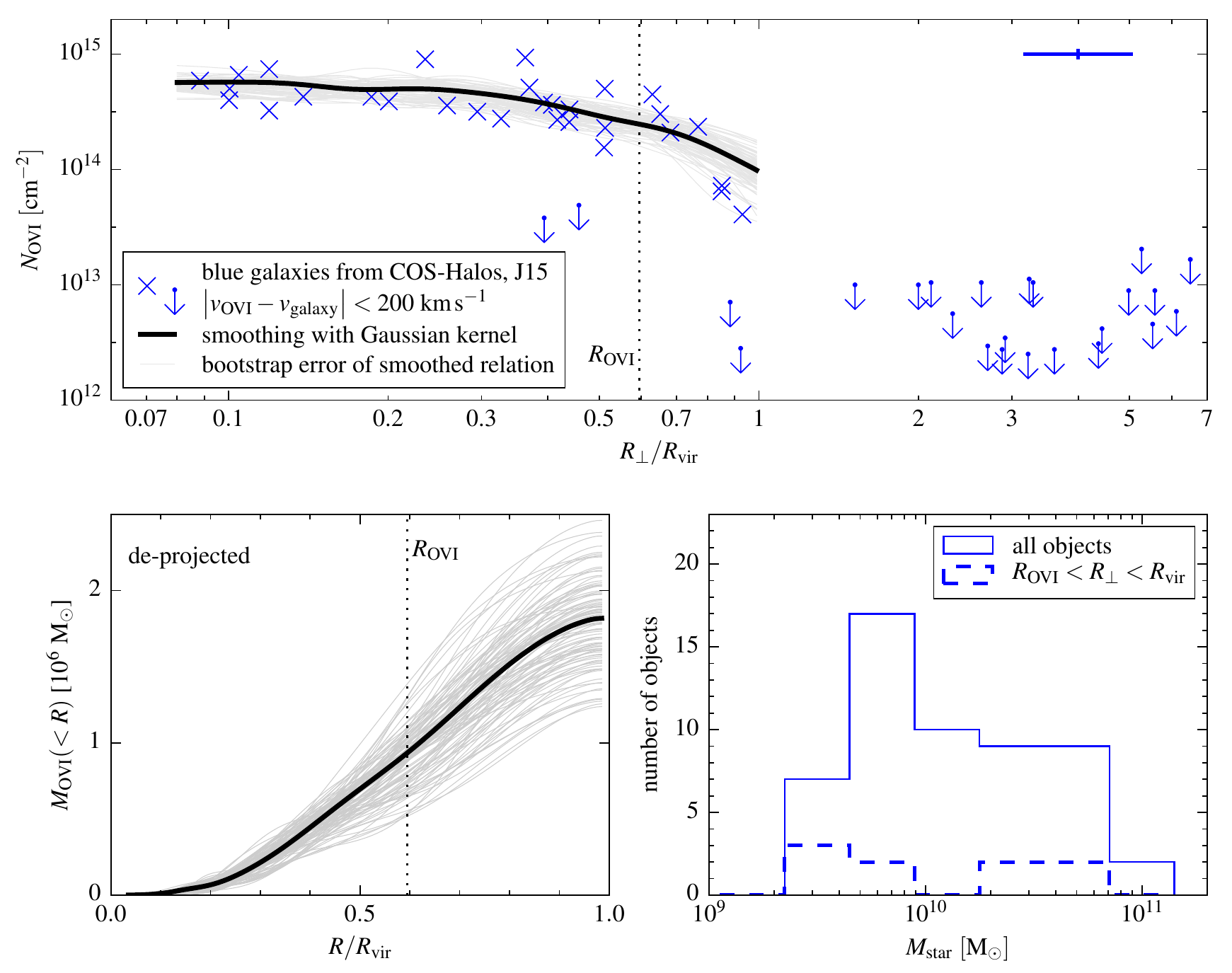}
\caption{
\textbf{Top:}
Observed \ovip\ columns in the CGM of \sLstar\ star-forming galaxies with $0.1<z<0.4$. Data points and upper limits are from the COS-Halos survey and from \cite{Johnson+15}. Impact parameters are normalized by $\rvir$, derived from the \cite{Moster+13} $\Mhalo-\Mstar$ relation. The median $\rvir$ in the sample is $190\kpc$. The typical errors on $\Novi$ ($0.03\dex$) and on the $\rvir$ estimate ($0.1\dex$) are noted in the top-right corner. The thick black line denotes a smoothing of the observations using a Gaussian kernel with width $0.1\dex$. 
The thin lines estimate the error on the smoothed relation using the bootstrap method. 
\textbf{Bottom-left:}
The cumulative mass of \ovip\ within a 3D-distance $R$ from the galaxy, derived using an inverse Abel transform on the smoothed observations. The median distance of the \ovip-gas is $\Rovi=0.59\rvir$ (dotted vertical line), with a $16-84$ percentile range of $0.35\rvir-0.8\rvir$.
\textbf{Bottom-right:}
The stellar mass distribution of the entire sample, and of objects with $\Rovi<\Rimp<\rvir$. 
}
\label{f:ovi observations}
\end{figure*}

The top panel of Figure~\ref{f:ovi observations} shows the $\Novi$ measurements around $z\sim0.2$, \sLstar\ star-forming galaxies (specific SFR $>10^{-11}\yr^{-1}$). We use the Voigt profile fit measurements listed in \citeauthor{Johnson+15} (2015, hereafter J15), which include 27 objects from the COS-Halos sample in which \ovip\ and \hi\ were observed (\citealt{Werk+13}), and 27 additional objects mainly at larger impact parameters, observed by J15. 
To facilitate comparison with previous studies which utilized only the COS-Halos sample, we use only objects from J15 with a stellar mass $\Mstar>10^{9.5}\msun$, the minimum stellar mass of blue COS-Halos galaxies. Also, we limit our analysis to absorption features within $\pm200\kms$ of the galaxy velocity, which includes practically all the \ovip\ absorption observed by COS-Halos (\citealt{Tumlinson+11, Werk+16}). 
The typical $\Novi$ measurement error of $0.03\dex$ is noted in the top-right corner. 
Upper limits are noted by down-pointing errors.

The horizontal axis is the impact parameter of the background quasar normalized by $\rvir$. To derive $\rvir$, we first use the \cite{Moster+13} relation to derive $\Mhalo$ from $\Mstar$. The values of $\Mstar$ are taken from \cite{Werk+13} and J15, and plotted in the bottom-right panel of Fig.~\ref{f:ovi observations}. For consistency, stellar masses from \cite{Werk+13} are corrected from the Salpeter initial mass function (IMF) used in that study to the Chabrier IMF used by J15 and by \cite{Moster+13}.
The implied median $\Mhalo$ in the combined sample is $6\cdot 10^{11}\msun$. 
The value of $\rvir$ is then determined from $\Mhalo$ via
\begin{equation}\label{e:rvir}
 \Mhalo = \frac{4\pi}{3}\Delta_{\rm c}{\bar \rho_{\rm c}}\rvir^3
\end{equation}
where ${\bar \rho_{\rm c}}$ is the critical density at the redshift of each object and $\Delta_{\rm c}$ is derived from \cite{BryanNorman98}\footnote{\cite{Moster+13} use the `200c' definition of the halo mass, which we convert to the eqn.~(\ref{e:rvir}) definition assuming an NFW concentration parameter of $10$.}. 
The implied median $\rvir$ in the sample is $190\kpc$, with a dispersion of $0.12\dex$. 
The error on $\rvir$ can be estimated by adding in quadrature the typical error of the $\Mstar$ measurements of $\pm50\%$ (\citealt{Werk+12}) to the expected error in the \citeauthor{Moster+10} relation of $\approx0.2\dex$. Dividing this implied error on $\Mhalo$ by three yields an error of $0.1\dex$ on $\rvir$, which is also noted in the corner of the top panel.

\begin{figure*}
 \includegraphics{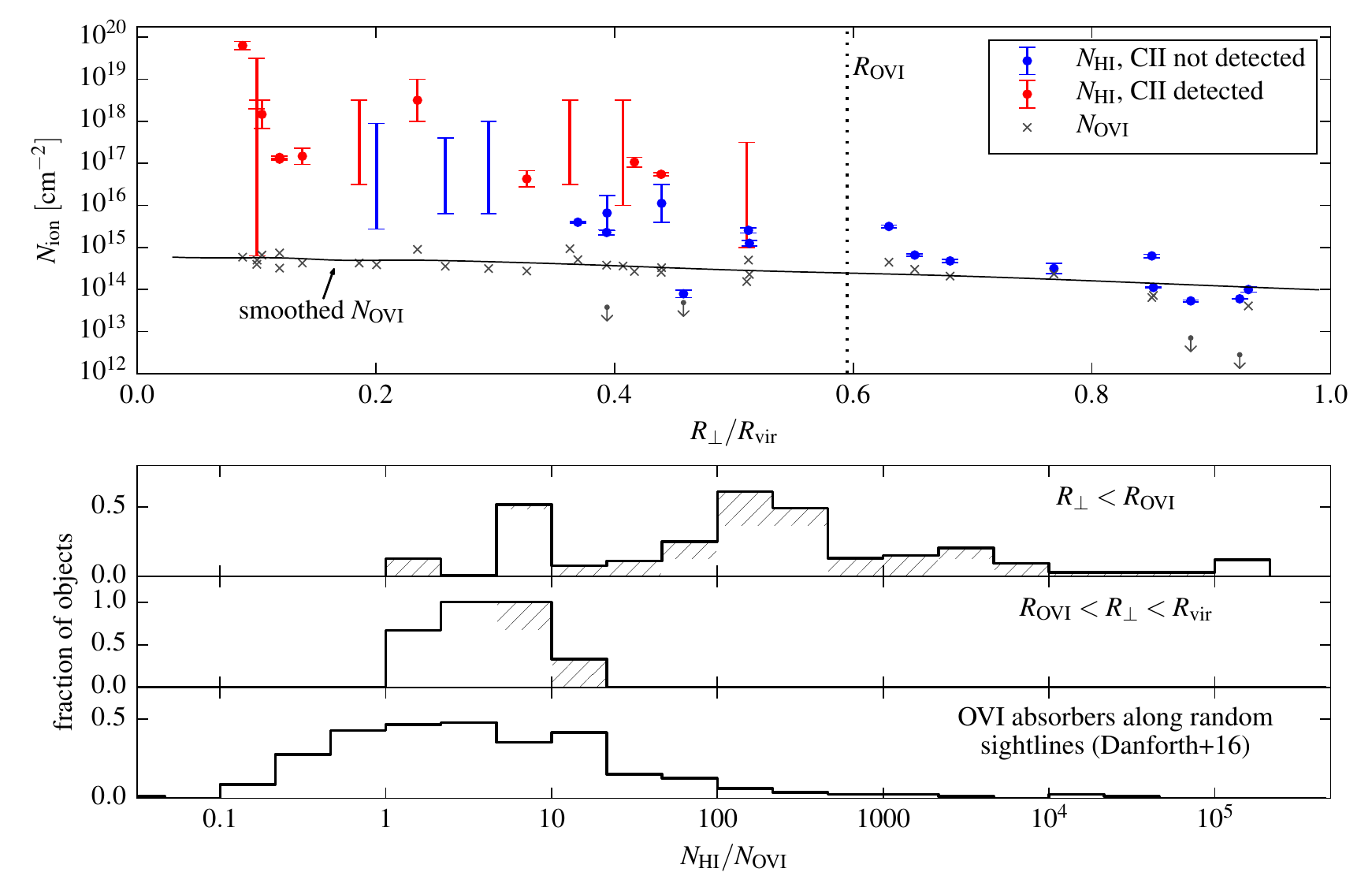}
\caption{
{\bf Top panel:}
\hi\ columns in the halos of \sLstar\ star-forming galaxies with $0.1<z<0.4$. Markers are colored red and blue to denote objects with and without \ciip\ detections, respectively. Measurements of $\NHI$ are based on Lyman transitions and Lyman-limit measurements from \cite{Prochaska+17} and J15. Large errorbars with no symbols indicate objects in which all observed Lyman transitions are saturated and the Lyman-limit is unconstrained. In these objects $\NHI$ is likely $>10^{16.5}\cm^{-2}$ if \ciip\ is detected and $<10^{16.5}\cm^{-2}$ if \ciip\ is not detected (see \S\ref{s:hi}).  
Also plotted are $\Novi$ measurements (grey crosses), the smoothed $\Novi(\Rimp)$ relation (black solid line), and the median \ovip-galaxy distance (dotted line) derived in Fig.~\ref{f:ovi observations}. 
Note that $\NHI>10^{16}\cm^{-2}$ and \ciip\ detections are only observed at impact parameters smaller than $\Rovi$, suggesting that they originate in gas which is distinct in physical radius from most of the \ovip-gas. 
{\bf Bottom panels:} Histograms of $\NHI/\Novi$ ratios at impact parameters below and above $\Rovi$, and in \ovip\ absorbers along random sightlines. 
Hatched regions mark uncertain ratios, either because \ovip\ is not detected or because the error on $\NHI$ is large. 
At $\Rimp>\Rovi$, where \hi\ and \ovip\ most likely trace gas at similar radii,  the column ratios are $\NHI/\Novi = 1-10$ similar to typical ratios along random sightlines.
}
\label{f:hi2ovi}
\end{figure*}

The top panel in Fig.~\ref{f:ovi observations} demonstrates that \ovip\ columns are roughly independent of impact parameter out to about $0.8\rvir$, and drop quickly at larger impact parameters. Beyond $\rvir$ there are no \ovip\ detections within $200\kms$ of the galaxy velocity. 
To derive the implied distribution of the \ovip-gas in physical space, we de-project the observed relation between $\Novi$ and $\Rimp$ assuming the distribution of $\opv$ ions in normalized 3D distance $R/\rvir$ is the same in all galaxies.
To this end, we first smooth the observed $\Novi$ vs.\ $\Rimp/\rvir$ using a Gaussian kernel. We use a Gaussian with width of $0.1\dex$, equal to the error on $\rvir$ noted above. We assume $\Novi=0$ for the four objects within $\rvir$ without an \ovip\ detection. We stop the calculation at $\rvir$, beyond which there are no detections. The derived smoothed relation is plotted as a thick black line in the top panel of Fig.~\ref{f:ovi observations}. To estimate the error in this mean, we repeat the process one hundred times, where in each iteration we choose with replacements 34 objects with $\Rimp<\rvir$, where 34 is the total number of objects that satisfy this criterion. These calculated means are shown as thin lines in Fig.~\ref{f:ovi observations}.

The distribution of the $\opv$ ion in physical space is then derived by deprojecting the smoothed observations using an inverse Abel transform. The result is shown in the bottom-left panel of Fig.~\ref{f:ovi observations}. 
The implied mass of $\opv$ ions is 
\begin{equation}\label{e:Movi}
\Movi(<\rvir)= (1.8 \pm 0.3) \times 10^6 \msun ~,
\end{equation}
where we assumed the median $\rvir=190\kpc$ to calculate the mass, and the quoted error is the $16-84$ percentile range in the bootstrap calculations of $\Movi$. 
The median distance of the \ovip-gas from the galaxy is (dotted line)
\begin{equation}\label{e:Rovi}
 \Rovi = 0.59 \rvir \approx 110\kpc ~,
\end{equation}
where $\Rovi$ satisfies $\Movi(<$$\Rovi)=0.5\Movi(<$$\rvir)$. 
The $16-84$ percentile distance range of \ovip\ is $0.35\rvir-0.8\rvir$. A similar deprojection of the \ovip\ columns was recently done by \cite{MathewsProchaska17}, with similar results.

We check for systematics on the derived $\Rovi$ and $\Movi$ by repeating the above process assuming errors on $\rvir$ in the range $0.075-0.15\dex$. The derived $\Rovi$ changed by less than $10\%$ and the derived $\Movi$ changed by less than $20\%$. Similar small changes occur when we calculate the geometric mean $\Novi$ rather than the arithmetic mean.

\subsection{\hi\ and low-ion columns}\label{s:hi}

In this section we estimate the amount of \hi\ and low-ions associated with \ovip. 
To this end, in the top panel of Figure~\ref{f:hi2ovi} we plot the relation between \hi\ column $\NHI$ and impact parameter in the COS-Halos+J15 sample used above. The objects are colored by whether \ciip\ is detected (red) or not detected (blue) in the same sightline. We focus on sightlines with impact parameters smaller than $\rvir$, beyond which \ovip\ is not detected (see Fig.~\ref{f:ovi observations}).  The values of $\NHI$ are derived by summing the Voigt profile fits in \cite{Tumlinson+13} and J15, using only features within $\pm200\kms$ of the galaxy redshift, as done above for \ovip. \ciip\ detections are taken from \cite{Werk+13} and J15\footnote{In one COS-Halos object \ciip\ has not been observed (J1009+0713\_170\_9). This object is colored in red since \mgip\ absorption is detected.}. When available, we use additional constraints on $\NHI$ based on Lyman limit measurements from \cite{Prochaska+17}. In some objects the error on $\NHI$ is large ($>1\dex$) since all observed Lyman transitions are saturated and the Lyman-limit is unconstrained. In these objects the detection of \ciip\  can be used as a rough constraint on $\NHI$, since \ciip\ is detected in all objects where $\NHI$ is constrained to  $>10^{16.5}\cm^{-2}$, and is not detected in all objects where $\NHI$ is constrained to $<10^{16.5}\cm^{-2}$. We also mark in the panel the $\Novi$ measurements, the smoothed $\Novi$ vs.\ $\Rimp/\rvir$ relation, and the median physical radius of the \ovip-gas $\Rovi$ inferred in the previous section.

At impact parameters larger than $\Rovi$, the values of $\NHI$ are typically $10^{14}-10^{15}\cm^{-2}$, with a tendency to increase towards smaller impact parameters (though note the small number statistics). At impact parameters smaller than $\Rovi$ the values of $\NHI$ exhibit a larger spread, ranging from $10^{14}\cm^{-2}$ to $10^{20}\cm^{-2}$, with the characteristic $\NHI$ generally increasing towards lower impact parameters. 
Specifically, the observed dispersion of $>$$2\dex$ in $\NHI$ at a given impact parameter is substantially larger than the dispersion in $\Novi$ of $0.3\dex$ at $\Rimp<\Rovi$.
These properties of $\Novi$ and $\NHI$ are further demonstrated in the second and third panels of Fig.~\ref{f:hi2ovi}, where we plot the distributions of the column ratio $\NHI/\Novi$ at $\Rimp<\Rovi$ and at $\Rimp>\Rovi$. Uncertain ratios due to a large error on $\NHI$ or a lack of \ovip\ detection are marked with hatched histograms. 
As suggested by the top panel, at $\Rimp<\Rovi$ the column ratio $\NHI/\Novi$ spans a large dynamical range of $1-2\cdot10^5$. In contrast, at $\Rimp>\Rovi$ the column ratios span a significantly smaller range, with the seven \ovip\ detections spanning $1<\NHI/\Novi<10$.

Fig.~\ref{f:hi2ovi} also shows that \ciip\ is not detected beyond $\Rovi$, commensurate with the absence of sightlines with $\NHI>10^{16.5}\cm^{-2}$. In Appendix~\ref{a:low-ions} we show that other ions observed by COS-Halos and J15 are also not detected beyond $\Rovi$, with the exception of relatively weak \ciiip\ and \Siiiip. This correspondence between the lack of large \hi\ columns and absence of low-ion absorption is not a coincidence. Unless the metallicity is highly
super-solar, photoionization equilibrium requires that detectable amounts of low-ion absorption are associated with
a large $\NHI$. We can therefore conclude that the `low-ion' gas, which produces low-ion absorption and $\NHI>10^{16.5}\cm^{-2}$, does not extend beyond $\Rovi$, i.e.\ it does not overlap in radius with the $50\%$ of the \ovip-gas that resides outside $\Rovi$. 
Furthermore, Fig.~\ref{f:hi2ovi} suggests that even at $\Rimp<\Rovi$ the low-ion gas and \ovip-gas are at least partially radially distinct. This follows since the \ovip\ gas is distributed mainly at $0.35-0.8\rvir$ (bottom-left panel of Fig.~\ref{f:ovi observations}), while the \hi\ and low-ion columns tend to increase towards lower impact parameters (top panel of Fig.~\ref{f:hi2ovi} and Appendix~\ref{a:low-ions}), suggesting a centrally-peaked distribution in physical space. Hence the large $\NHI/\Novi$ ratios observed at $\Rimp<\Rovi$ are at least partially due to \hi\ produced in gas at smaller physical radii than \ovip, rather than \hi\ in gas which is co-spatial with \ovip.

We show below that measurements of $\NHI/\Novi$ at $\Rimp>\Rovi$, where \hi\ is not potentially `contaminated' by gas at smaller physical radii than \ovip, provide a useful diagnostic. As the number of sightlines at these impact parameters is small, more measurements would be useful. 

It is also interesting to compare the ratios of $\NHI/\Novi$ in the galaxy-selected sample used here with the ratios observed along random quasar sightlines. We utilize the sample from \cite{Danforth+16}, which detected 255 \ovip\ absorbers with redshifts $0.15-0.4$ ($16-84$ percentiles), comparable to COS-Halos. The range of $\Novi$ in the random sightline sample is $(0.3-1.3)\times10^{14}\cm^{-2}$ (also $16-84$ percentiles). Such absorbers are likely dominated by gas within $\sim2\rvir$ from galaxies with a somewhat lower luminosity than COS-Halos (\citealt{Prochaska+11, McQuinn16}), with a non-negligible contribution from gas around galaxies at luminosities as low as $\sim0.01 \Lstar$ (\citealt{Johnson+17}). 
Almost all (93\%) of the \ovip\ absorbers in the \citeauthor{Danforth+16}\ sample are also detected in \hi, so the $\NHI$ measurements are almost complete. 
The distribution of $\NHI/\Novi$ is shown in the bottom panel of Fig.~\ref{f:hi2ovi}. The median column ratio is $2.9$, with 80\%\ of the objects in the range $0.2-20$ and a tail to larger values. This distribution is consistent with the distribution in the smaller STIS-based random sightline sample of \cite{ThomChen08}, which utilized a different absorber detection technique.

The typical $\NHI/\Novi$ observed along random sightlines is similar to the $\NHI/\Novi=1-10$ seen at $\Rimp\gtrsim\Rovi$ in the galaxy-selected sample, albeit with a larger spread and a tail to higher values. This similarity may suggest similar physical conditions in \ovip\ absorbers along random sightlines and in the outer halo of \sLstar\ galaxies.

\subsection{Differential extinction by dust grains}

Another clue on the nature of the gas traced by \ovip\ can be derived from observations of dust grains in the CGM. \citeauthor{Menard+10}\ (2010, hereafter MSFR) measured the mean differential extinction $\ebv$ of background SDSS quasars, as a function of their angular separation from foreground SDSS galaxies with magnitude $i<21$ and median redshift of $0.36$. They detected an extinction signal of $\ebv \sim \Rimp^{0.8}$ spanning angular separations
of $0.1 -100'$ which corresponds to impact parameters of $20\kpc-20\Mpc$ at the median redshift.
Since the MSFR sample and the COS-Halos+J15 sample are both dominated by \sLstar\ galaxies at similar redshifts, it is plausible that these two samples trace similar galaxies. 
At the characteristic \ovip-galaxy distance $\Rovi\approx100\kpc$ (eqn.~\ref{e:Rovi}), MSFR found $\ebv \approx 2\cdot10^{-3}\,{\rm mag}$. A comparable $\ebv$ at $100\kpc$ was later deduced by \cite{Peek+15}, using galaxies as background sources instead of quasars, though with a somewhat steeper dependence of $\ebv$ on impact parameter.

Assuming that the extinction signal at $\gtrsim 100\kpc$ is dominated by the CGM of the central galaxy rather than by the CGM of neighboring galaxies, as argued by MSFR and as suggested by the analysis of \cite{MasakiYoshida12}, indicates the existence of dust grains at similar radii as \ovip. We discuss the implications of this possibility below.

\begin{figure*}
 \includegraphics{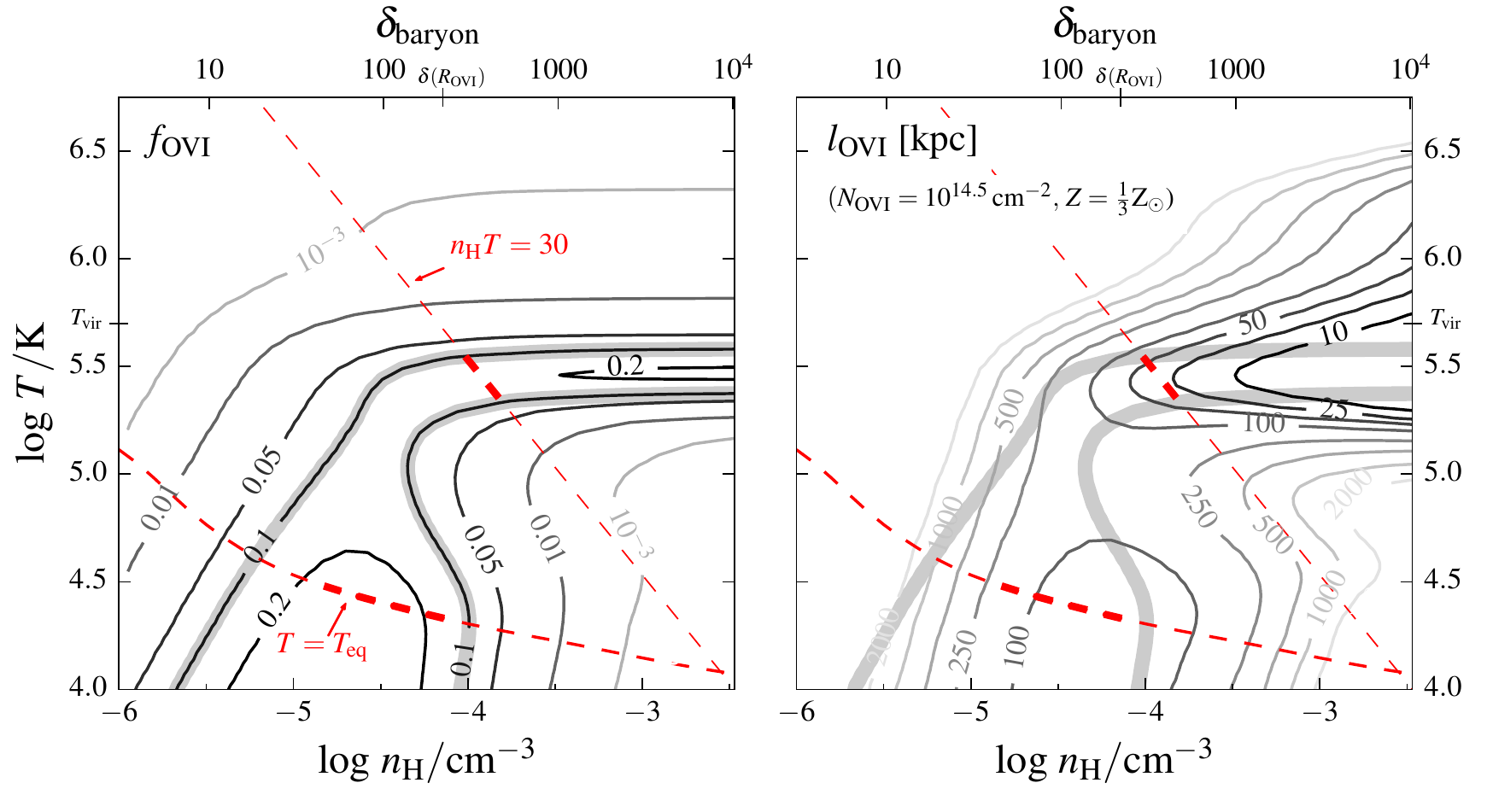}
\caption{
Gas properties as a function of density and temperature, assuming ionization equilibrium conditions. All calculations are done with \cloudy\ assuming the UV background described in \S\ref{s:basic properties}. 
\textbf{(Left)} 
The $\opv/{\rm O}$ fraction $\fovi$ is denoted by thin contours. The region where $\fovi\geq 0.1$, near its peak value, 
is emphasized by thick grey stripes. 
The two `arms' at $\nH\gtrsim10^{-4.5}\cm^{-3},\ T \approx 10^{5.5}\K$ and $\nH \approx 10^{-4.5} \cm^{-3},\ T \lesssim 10^{5}\K$ correspond to \ovip\ being collisionally ionized and photoionized, respectively. 
The red dashed lines mark the two possible scenarios discussed in this work: 
(1) that the typical gas pressure at the median \ovip\ distance $\Rovi\approx0.6\rvir$ is $\nH T \sim 30\cm^{-3}\K$, as expected if the outer halo is filled with virially-shocked gas with an overdensity similar to the dark matter overdensity ($\Tvir$ and $\delta$ are marked on the axes). In this scenario \ovip\ is collisionally ionized; 
(2) that \ovip\ traces gas in ionization and thermal equilibrium with the UV background, as expected if \ovip\ traces metal-enriched gas beyond the accretion shock. 
The intersections of the red dashed lines with the peak of $\fovi$ (emphasized by thick lines) gives the range of densities and temperatures likely traced by \ovip\ in each scenario. 
The two scenarios differ by a factor of $\sim40$ in the implied gas pressure in the outer halo, so we name them the `high-pressure' scenario and the `low-pressure' scenario, respectively. 
\textbf{(Right)} 
Contour values denote the pathlength through the \ovip-gas for the observed $\Novi = 10^{14.5} \cm^{-2}$ and a third-solar oxygen abundance. 
The thick gray stripes and the red lines are as in the left panel. 
The implied pathlengths are $\approx 30\,(3Z/\zsun)^{-1}\kpc$ in the high-pressure scenario and at least $70\,(3Z/\zsun)^{-1}\kpc$ in the low-pressure scenario. Densities of $\nH\lesssim10^{-5}\cm^{-3}$ are ruled out since they would imply a pathlength larger than $2\Rovi$, which is unphysical.
}
\label{f:nTplots OVI}
\end{figure*}

\subsection{Other CGM observations}\label{s:other obs}

Several additional observations have been used in the literature to constrain the conditions in the CGM of \sLstar\ galaxies, including detections of X-ray emission (\citealt{AndersonBregman11, Dai+12, Bogdan+13a,Bogdan+13b,Anderson+16,Li+17}), modelling the ram pressure stripping of local group satellites (\citealt{BlitzRobishaw00, GrcevichPutman09, Gatto+13}) and the Large Magellanic Cloud (LMC, \citealt{Salem+15}), the dispersion measure of pulsars in the Magellanic Clouds (\citealt{AndersonBregman10}), and $z=0$ \oviip\ and \oviiip\ absorption features in the X-ray (\citealt{Wang+05,Fang+06,BregmanLloydDavies07,Faerman+17}). 
However, direct X-ray emission from the hot gas is detected only out to $\approx 50\kpc$ and hence does not directly constrain the conditions at $0.5-1\rvir$ where most of the \ovip-gas resides. Similarly, sightlines to the Magellanic clouds which are located at $50-60\kpc$ can only constrain the physical conditions within this distance. The lack of \hi\ in local group satellites has been attributed to ram pressure stripping by a volume filling hot phase within $\lesssim 120\kpc$ (\citealt{GrcevichPutman09}), which is equivalent to $\lesssim0.4\rvir$ for a MW virial radius derived from eqn.~(\ref{e:rvir}) assuming $\Mstar=6\cdot10^{10}\msun$. 
This constraint hence also does not directly constrain the conditions at larger $R/\rvir$ which are the focus of this work. 
Last, the physical distance of the $z=0$ X-ray \oviip\ and \oviiip\ absorption features is unclear, and may well be limited to $<100\kpc$. 
Hence, since all the observations mentioned above primarily constrain the conditions at radii $\lesssim0.5\rvir<\Rovi$, we do not consider them further in the analysis below, instead giving predictions for future observations.

\section{Implications for the physical conditions in the outer halo}\label{s:implications}

Above we showed that most of the \ovip-gas resides in the outer halo, beyond $0.5\rvir$.
At these distances available observations indicate typical column ratios of $\NHI/\Novi=1-10$, $\ebv\approx2\cdot10^{-3}\,{\rm mag}$, weak \ciiip\ and \Siiiip\ absorption, and a lack of other low-ion absorption. 
In this section we derive several physical properties of the \ovip-gas based on these observations. We start with general physical implications (\S\S\ref{s:basic properties}--\ref{s:Z}), and continue with physical implications under two assumed physical scenarios (\S\S\ref{s:pressure}--\ref{s:summary PIE}). 

\subsection{Mass, pathlength, and the \ovip\ fraction}\label{s:basic properties}

\newcommand{\highP}{{\rm high}\text{-}P}
\newcommand{\lowP}{{\rm low}\text{-}P}

We assume that the gas traced by \ovip\ is irradiated by the UV background from \citeauthor{HaardtMadau12} (2012, hereafter HM12) at the median $z=0.24$ of the COS-Halos+J15 sample, after multiplying the intensity at $1\ryd$ by a factor of two ($J_\nu = 3.2\cdot10^{-23}\erg\s^{-1}\cm^{-2}\Hz^{-1}\,{\rm sr}^{-1}$). This factor of two is suggested by comparing models of the \Lya\ forest with observations (\citealt{Shull+15, Gaikwad+17}, cf.\ \citealt{Kollmeier+14}), and by the \Ha\ fluorescence of UGC 7321 (\citealt{Fumagalli+17}). This factor of two is also consistent with models in which the UV background is dominated by quasars (\citealt{HaardtMadau15}). The spectrum of \cite{FaucherGiguere+09} is comparable to HM12 at low redshift, so our assumed spectrum is also a factor of $\sim2$ stronger than this estimate. To avoid over-predicting the X-ray background which is a lower factor of $1.2$ above HM12, we multiply the HM12 spectrum by $2(h\nu/1\ryd)^{-0.1}$. Based on the analysis of MQW17 and \cite{UptonSanderbeck+17}, we do not include local sources in the galaxy and CGM, as they are unlikely to exceed the background at the characteristic \ovip\ scale of $\Rovi=0.6\rvir\approx110\kpc$. The implications of uncertainties in the UV background on our results are addressed in the discussion. 

The left panel of Figure~\ref{f:nTplots OVI} plots $\fovi$, the fraction of oxygen particles in the $\opv$ ionization state, as a function of gas density $\nH$ and temperature $T$. The ionization calculations are done with \cloudy\ (\citealt{Ferland+13}), assuming optically thin gas, third-solar abundances (where solar is defined by \citealt{Asplund+09}), and ionization equilibrium conditions. The assumed abundances have a negligible effect on $\fovi$, while the ionization equilibrium assumption is addressed below. 
Fig.~\ref{f:nTplots OVI} shows that $\fovi$ peaks at a value of $\approx0.2$ in two `branches' of $\nH-T$ space: at $\nH>10^{-4.5}\cm^{-3}$ and $T \approx 10^{5.5}\K$, where $\opv$ is collisionally ionized, and at $\nH \sim 10^{-4.5} \cm^{-3}$ and $T<10^{5}\K$, where $\opv$ is photoionized by the UV background. 
We argue next that the large observed $\Novi$ columns imply that the mass and pathlength of \ovip-gas would be too large if $\fovi$ were significantly below its peak value (see also MQW17). So, we delineate the region in phase space where $\fovi$ peaks by marking the $\fovi=0.1$ contours with thick gray lines. These gray lines are plotted for reference also in the right panel and in following figures. 

The total mass of gas traced by \ovip\ is equal to 
\begin{equation}\label{e:Mgasovi}
 \Mgasovi =  5\times10^{9} \left(\frac{\Movi}{2\cdot10^{6}\msun}\right) 
\left(\frac{\fovi}{0.2}\right)^{-1} \left(\frac{Z}{\frac{1}{3}Z_\odot}\right)^{-1}\msun ~,
\end{equation}
where $\Movi$ is derived above (Fig.~\ref{f:ovi observations}) and we used an oxygen mass fraction of $0.006\,(Z/\zsun)$ based on \cite{Asplund+09}. 
This gas mass can be compared to the total baryon budget of the halo  $(\Omegab/\OmegaM)\Mhalo$:
\begin{equation}\label{e:Mgasovi frac}
 \frac{\Mgasovi}{(\Omegab/\OmegaM)\Mhalo} \approx 0.05\left(\frac{\Movi}{2\cdot10^{6}\msun}\right) \left(\frac{\fovi}{0.2}\right)^{-1} \left(\frac{Z}{\frac{1}{3}Z_\odot}\right)^{-1} ~,
\end{equation}
where we assumed the median $\Mhalo=0.6\cdot10^{12}\msun$ in the COS-Halos+J15 sample. 
Similarly, the pathlength through the gas traced by \ovip\ is equal to
\begin{eqnarray}\label{e:lovi}
\lovi &=& \frac{\Novi}{\novi} = 99\, \left(\frac{\Novi}{10^{14.5}\cm^{-2}}\right)\times \nonumber\\
      & & \left(\frac{\nH}{10^{-4.5}\cm^{-3}}\right)^{-1} 
	  \left(\frac{\fovi}{0.2}              \right)^{-1}
 	  \left(\frac{Z}{\frac{1}{3}\zsun}                \right)^{-1} \kpc ~,\nonumber\\
\end{eqnarray}
where $\novi$, the volume density of $\opv$ ions, is equal to
\begin{equation}\label{e:novi}
\novi=4.9\cdot10^{-4}(Z/\zsun)\fovi\nH ~.
\end{equation}
The right panel of Fig.~\ref{f:nTplots OVI} plots $\lovi$ as a function of density and temperature. The numerical values for $\nH$ and $Z$ are justified below. 

Since $\Mgasovi$ likely does not exceed $(\Omegab/\OmegaM)\Mhalo$, and the pathlength cannot be larger than the size of the system (say $\sim2\Rovi\approx200\kpc$), equations~(\ref{e:Mgasovi frac}) and (\ref{e:lovi}) suggest that $\fovi$ cannot be far below $0.2$. This constraint become stronger with decreasing metallicity, and with the mass of the non-\ovip\ gas assumed to exist in the halo. Furthermore, the right panel in Fig.~\ref{f:nTplots OVI} shows that $\nH\lesssim10^{-5}\cm^{-3}$ is ruled out due to the large implied pathlength, even if $\fovi$ is near its peak value.

Our conclusion that $\fovi$ is near its peak value of $0.2$, i.e.\ that \ovip\ traces either collisionally ionized gas with $T\approx10^{5.5}\K$ or photoionized gas with $\nH\sim10^{-4.5}\cm^{-3}$, is based on the assumption of equilibrium ionization fractions. Is this assumption reasonable?
The ${\rm O}^{4+}\rightarrow\opv$ photoionization timescale is $\approx300\Myr$ (\citealt{VernerYakovlev95}) for our assumed background spectrum, while the ${\rm O}^{6+}$ $+\ e\rightarrow\opv$ recombination timescale is $\sim 40(\nH/10^{-4.5}\cm^{-3})^{-1}(1+T/10^{5.5}\K)\Myr$ (\citealt{Colgan+04}). These timescales are shorter than the dynamical timescale of $\sim1.5\Gyr$, supporting our equilibrium assumption. However, the recombination timescale is comparable to the cooling timescale of gas with $T\lesssim10^{5}\K$. \cite{GnatSternberg07} calculated the non-equilibrium ionizations fractions in radiatively cooling gas, and found that even in solar enriched gas, $\fovi$ is significantly enhanced only at $T<10^5\K$, to values which are an order of magnitude below the near-peak value $\sim0.2$. Hence, for the range of parameters where $\fovi$ peaks, rapid cooling will not cause the gas to depart significantly from ionization equilibrium. Our assumption of equilibrium is also unlikely to be affected by time-dependent local ionizing sources (e.g.\ \citealt{Vasiliev+15, Segers+17, Oppenheimer+17}), since such sources are expected to be sub-dominant to the background at $\gtrsim100\kpc$ (MQW17, \citealt{UptonSanderbeck+17}).

\begin{figure*}
\includegraphics[width=0.48\textwidth]{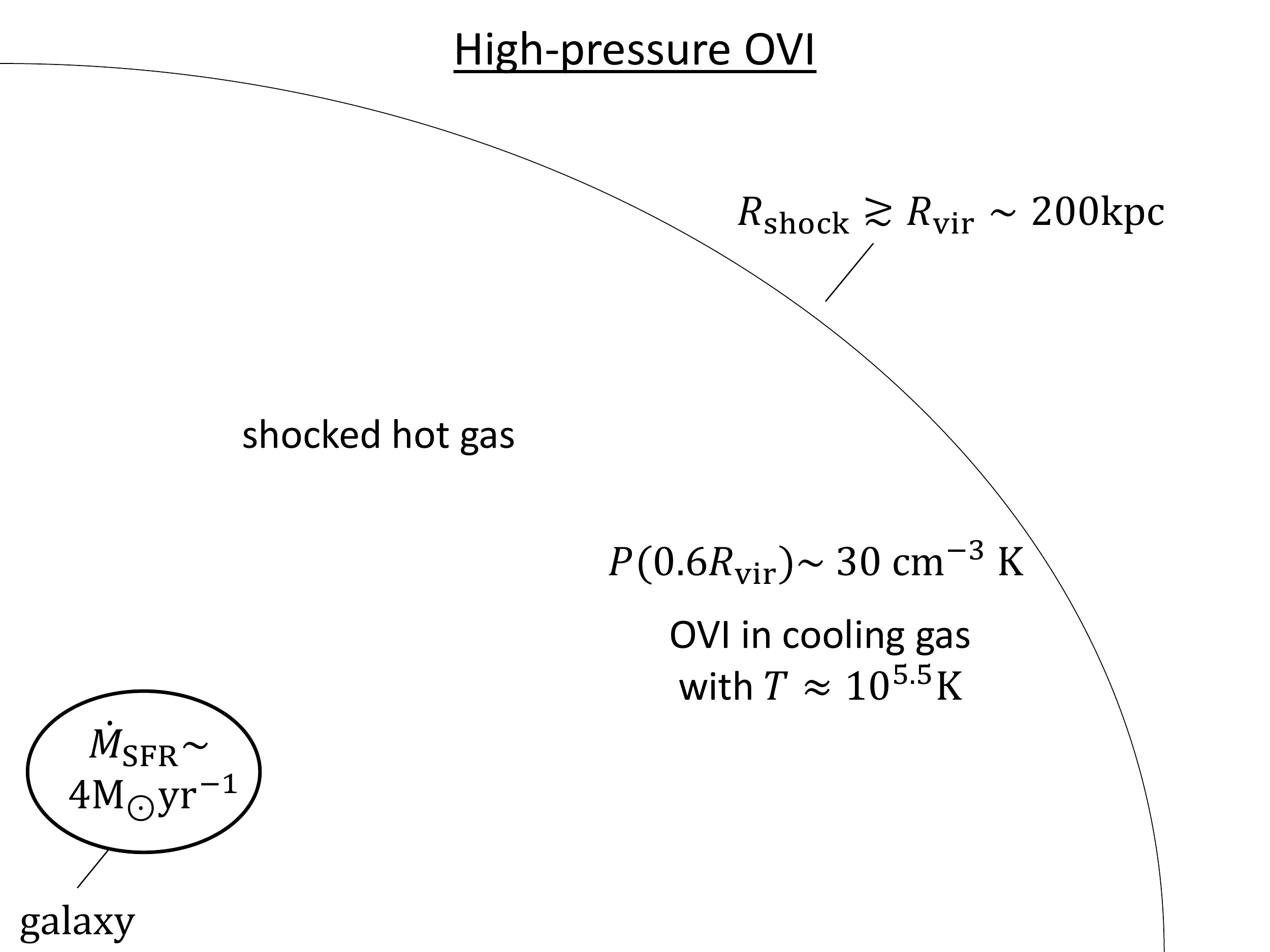}
\includegraphics[width=0.48\textwidth]{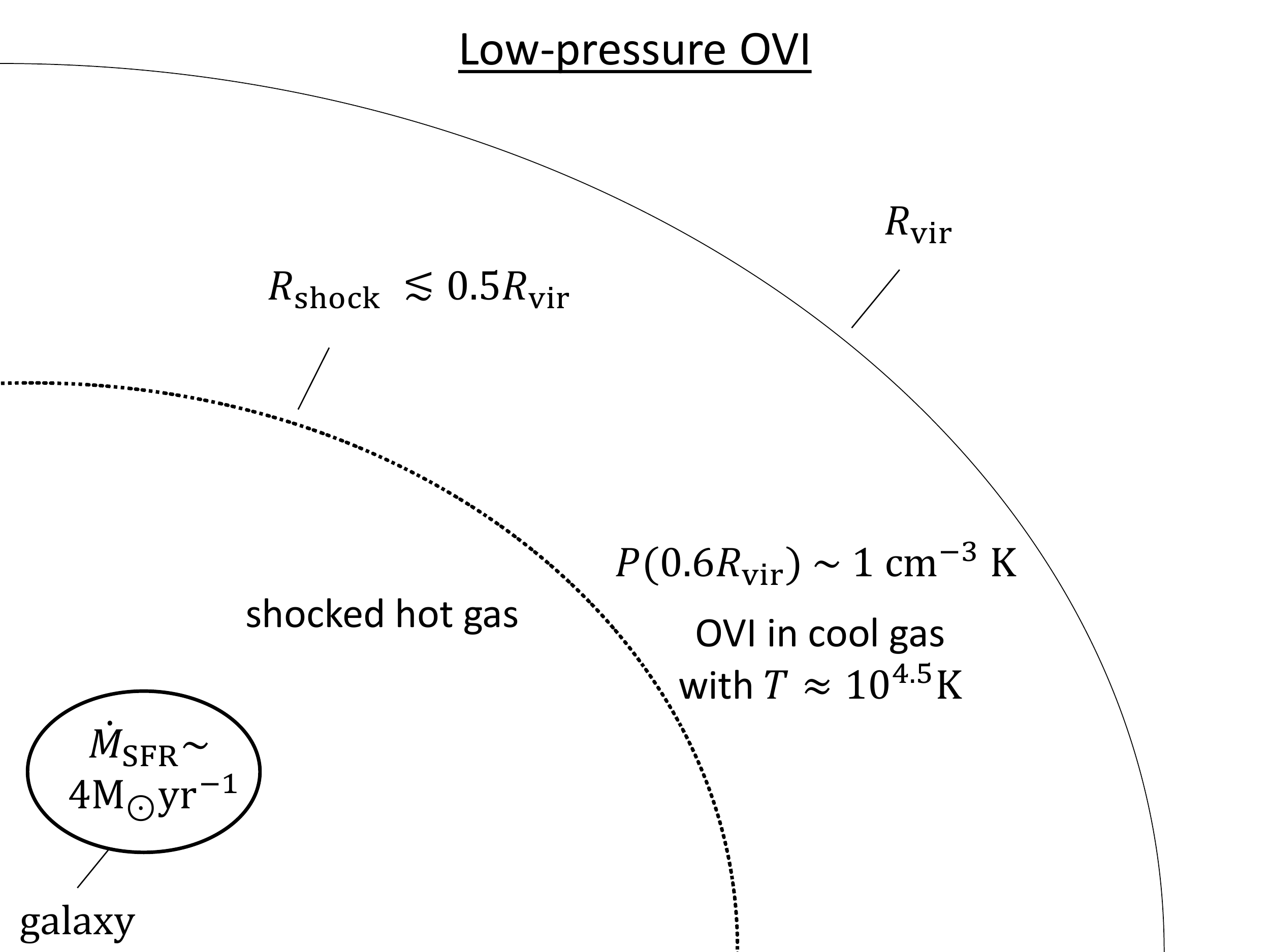}
\caption{
Schemas of the two scenarios discussed in this work for the halo gas of blue \sLstar\ galaxies at low redshift.
\textbf{(Left)} The `high-pressure \ovip' scenario, where a virially-shocked hot phase extends out to distances $\gtrsim\rvir$. The gas pressure at $R=\Rovi=0.6\rvir$ is $\nH T \sim 30\cm^{-3}\K$. \ovip\ is produced via collisional ionization in gas cooling from the hot phase. 
\textbf{(Right)} The `low-pressure \ovip' scenario, where the accretion shock is at $\lesssim0.5\rvir$. \ovip\ traces low-pressure gas beyond the shock in ionization and thermal equilibrium with the UV background. 
The high-pressure scenario is favored by current cosmological simulations. The low-pressure scenario may however be possible if the cooling time of the postshock gas is shorter than in current simulations (e.g., due to a more metal enriched CGM), resulting in an unstable virial shock at $\sim\rvir$.
}
\label{f:schema}
\end{figure*}

\subsection{Metallicity}\label{s:Z}

What is the metallicity of CGM absorbers? 
Photoionization modeling of COS-Halos absorption data (excluding \ovip) using single-density models suggest $0.02<Z/\zsun<3$ with a median of $Z=0.3\zsun$ (\citealt{Prochaska+17}). 
Alternative multi-density photoionization models of all COS-Halos ions suggest a somewhat higher median metallicity of $Z=0.6\zsun$ with a smaller dispersion of $0.3\dex$ (\citealt{Stern+16}, hereafter S16), and a similar $Z=0.5\zsun\pm0.4\dex$ if \ovip\ is excluded from the modelling (see \S5.3 in S16). The smaller metallicity dispersion suggested by the multi-density models is favored by the small dispersion seen in $\Novi$ at $\Rimp<\rvir$ ($0.3\dex$, see Fig.~\ref{f:ovi observations}). The results of S16 are further addressed in the discussion.  
Metallicities of order solar are also suggested by the ISM-like dust-to-gas ratios found in \mgiip\ absorbers (\citealt{MenardChelouce09, MenardFukugita12}), derived from comparing the measured \hi\ column and the observed $\ebv$. 

Hence, analysis of CGM absorbers suggests relatively high metallicities of $Z\gtrsim\zsun/3$.  These metallicities are consistent with the lower limit on $Z$ implied by the mass and pathlength constraints discussed in the previous section (eqns.~\ref{e:Mgasovi frac} and \ref{e:lovi}). 
For consistency with previous studies we use a fiducial $Z=\zsun/3$, though we note that a factor of $\sim 2$ higher metallicity is favored by some observational analyses.

\subsection{Thermal pressure and cooling luminosity}\label{s:pressure}

In this section we calculate the typical gas pressures $P$ at a distance $\Rovi\approx0.6\rvir$ (eqn.~\ref{e:Rovi}). We employ two distinct physical assumptions, plotted schematically in Figure~\ref{f:schema}. In the first scenario (left panel) the virially-shocked phase extends beyond $\Rovi\approx0.6\rvir$, and \ovip\ traces gas in pressure equilibrium with the hot phase.
In the second scenario (right panel) the accretion shock occurs at $R<\Rovi$, and \ovip\ traces low-pressure cool gas outside the shock. 
We denote the two scenarios by `high-$P$' and `low-$P$', respectively. 
The high-$P$ scenario, which is favored by current cosmological simulations, has been discussed previously in several papers (e.g., \citealt{Faerman+17, McQuinnWerk17, MathewsProchaska17,Armillotta+17,Bordoloi+17}). The low-$P$ scenario, which is possible if the virial shock around blue \sLstar\ galaxies is unstable (see discussion), has not been systematically investigated before. 

Given that \ovip\ is produced over a range of radii (Fig.~\ref{f:ovi observations}), it is also possible that some of the \ovip\ is produced in high-pressure gas within the shock, while some of the \ovip\ is produced in low-pressure gas outside the shock. In this work we focus on the median \ovip\ radius $\Rovi$, and hence on the question what are the conditions where {\it most} of the \ovip\ is produced.

\subsubsection{High-pressure scenario for the \ovip-gas}\label{s:highP}

In the high pressure scenario, one can derive a rough estimate of the gas pressure in the outer halo by assuming the hot gas over-density follows the dark matter over-density, and the hot gas temperature is equal to the virial temperature (e.g.\ \citealt{MallerBullock04}). For our median halo with $\Mhalo=6\cdot10^{11}\msun$ at $z=0.2$ this estimate gives a pressure of
\begin{eqnarray}\label{e:P CI}
(\nH T)_{\highP} &\sim& 200\left(\frac{R}{\Rovi}\right)^{-2.7} {\bar \nH}T_{\rm vir} \nonumber\\
&\approx& 30 \left(\frac{R}{\Rovi}\right)^{-2.7}\cm^{-3}\K
\end{eqnarray}
where ${\bar \nH}=3.3\cdot10^{-7}\cm^{-3}$ is the cosmic baryon mean density at $z=0.2$, $200(R/\Rovi)^{-2.7}$ is the over-density of an NFW profile near $\Rovi=0.6\rvir$ assuming a concentration parameter of 10 (\citealt{DuttonMaccio14}), and $T_{\rm vir}\equiv\mu\mp G\Mhalo / 2\rvir k = 5\cdot10^5\K$ is the virial temperature (we assume a mean molecular weight $\mu=0.6$ throughout). 
The pressure estimate in eqn.~(\ref{e:P CI}) and its dependence on $R$ is comparable to the average pressure seen in the FIRE m12i simulation (\citealt{Hopkins+17}) at $z=0$, in which we find $\nH T \approx 25 (R/0.6\rvir)^{-3}\cm^{-3}\K$. Comparable pressures are also seen in the EAGLE simulation (fig.~11 at \citealt{Oppenheimer+16}). The idealized CGM simulations of \cite{Fielding+17} with $\Mhalo=10^{12}\msun$ show a somewhat lower $\nH T\approx 10\cm^{-3}\K$ (see fig.~A1 there, note they use the `200m' definition for $\Mhalo$ and $\rvir$).

The characteristic pressure in eqn.~(\ref{e:P CI}) is marked with red dashed lines in both panels of Fig.~\ref{f:nTplots OVI}. This pressure crosses the collisional ionization `arm' of the peak in $\fovi$ (the intersection is emphasized with a thick line). Hence, if the \ovip-gas pressure is similar to the estimate in eqn.~(\ref{e:P CI}), we can conclude that \ovip\ originates in collisionally ionized gas with $T\approx10^{5.5}\K$. Fig.~\ref{f:nTplots OVI} shows that for photoionization to allow $T\lesssim10^5\K$ gas to have $\fovi$ near its peak value, the characteristic pressure needs to be $\sim5\cm^{-3}\K$ or lower, a factor of six less than estimated in eqn.~(\ref{e:P CI}). 

The implied volume density scale for the gas traced by \ovip\ is hence
\begin{equation}\label{e:n CI}
 n_{\rm H,\, \highP} \approx \frac{(\nH T)_{\highP}}{10^{5.5}\K} = 10^{-4}\left(\frac{R}{\Rovi}\right)^{-2.7}\cm^{-3}~.
\end{equation}
Using eqns.~(\ref{e:novi}) and (\ref{e:n CI}), the implied $\opv$ volume density is
\begin{equation}\label{e:novi CI}
 n_{\ovip,\, \highP} = 3.3\cdot10^{-9}\left(\frac{\fovi}{0.2}\right)\left(\frac{Z}{\frac{1}{3}\zsun}\right)\left(\frac{R}{\Rovi}\right)^{-2.7}\cm^{-3} 
\end{equation}

\begin{figure}
 \includegraphics{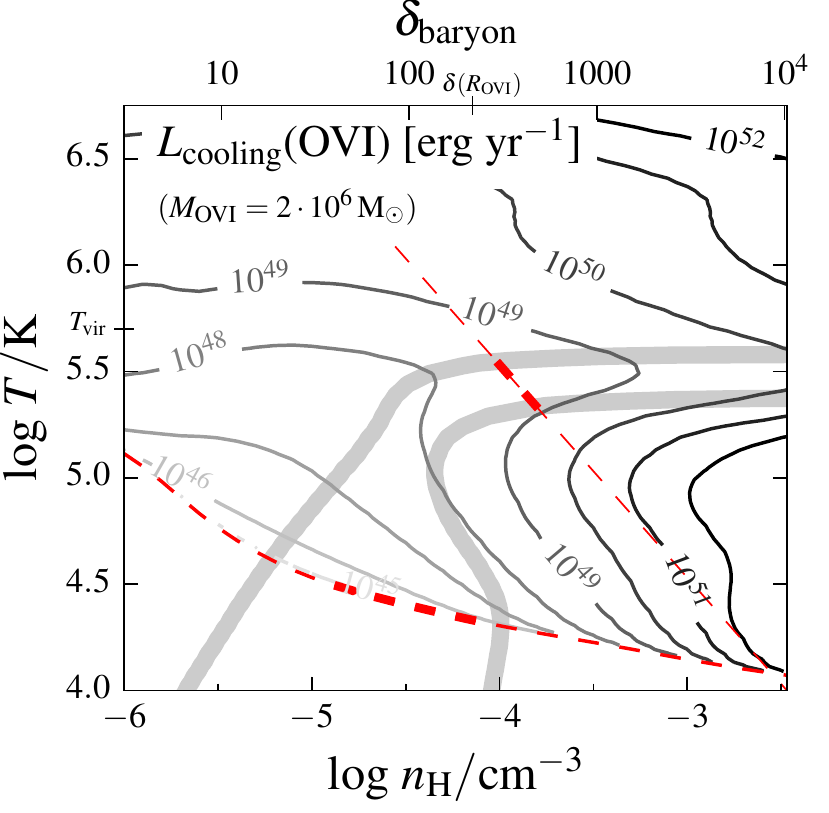}
\caption{
Cooling luminosity of the gas traced by \ovip\ as a function its density and temperature, calculated from the $\ovip$ mass of $2\cdot 10^{6}\msun$ derived in Fig.~\ref{f:ovi observations}. Thick grey stripes and red dashed lines are as in Fig.~\ref{f:nTplots OVI}. 
Negative cooling values in the lower-left corner (net heating) are not shown. 
In the high-pressure scenario ($\nH\approx10^{-4}\cm^{-3}, T\approx10^{5.5}\K$) the cooling luminosity is $\approx3\times 10^{48}\erg\yr^{-1}$, which requires efficient deposition of energy in the outer halo to avoid collapse (see \S\ref{s:highP}).
In the low-pressure scenario the net $L_{\rm cool}$ is zero by construction, since the gas is in thermal equilibrium with the UV background.
}
\label{f:nTplots cooling}
\end{figure}

Figure~\ref{f:nTplots cooling} plots the cooling luminosity from gas traced by \ovip, $L_{\rm cool}^{(\ovip)}$, assuming it originates from a specific gas density and temperature. We derive $L_{\rm cool}^{(\ovip)}$ using
\begin{eqnarray}\label{e:Lcool}
 L_{\rm cool}^{(\ovip)}(\nH,T) & = & V^{(\ovip)}\nH^2\Lambda(\nH,T,Z)\nonumber\\
 		& \approx & \frac{X\Mgasovi}{\mp}\nH\Lambda(\nH,T,Z)  ~,
\end{eqnarray}
where $V^{(\ovip)}$ is the volume occupied by the gas traced by \ovip, $X=0.75$ is the hydrogen mass fraction, and we define $\Lambda\nH^2$ as the net cooling per unit volume after accounting for heating by the UV background (calculated using \cloudy). 
For $\Mgasovi$, we use eqn.~(\ref{e:Mgasovi}) with $\Movi=2\cdot10^6\msun$ and $\fovi$ appropriate for the assumed $\nH$ and $T$ (left panel of Fig.~\ref{f:nTplots OVI}). 
Note that eqn.~(\ref{e:Lcool}) implies that $L_{\rm cool}^{(\ovip)}$ is independent of $Z$ at temperatures $T\sim 10^{5.5}\K$, since the metals dominate the cooling and $\Lambda\propto Z$, while the gas mass $\Mgasovi$ is inversely proportional to $Z$ (eqn.~\ref{e:Mgasovi}).

Fig.~\ref{f:nTplots cooling} demonstrates that the high-pressure \ovip\ scenario has a cooling luminosity of $3\times10^{48}\erg\yr^{-1}$, in the limit that all the \ovip\ gas has $\nH=10^{-4}\cm^{-3}$ and $T=10^{5.5}\K$ (a more accurate calculation can be derived by convolving $L_{\rm cool}^{(\ovip)}(\nH,T)$ with an assumed $\nH-T$ distribution). 
For comparison, the entire thermal energy of the halo is $\approx\frac{3}{2}kT_{\rm vir}\frac{\Omegab}{\OmegaM}\Mhalo/\mu\mp\approx 2\cdot10^{58}\erg$. Hence, without a source of heating, the cooling from the \ovip\ gas would have radiated away all the thermal energy of the halo gas during the last $\approx 7\Gyr$. This `cooling problem' is exacerbated by the fact that our cooling luminosity estimate includes only gas which produces observable \ovip\ absorption, and hence is a lower limit on the total cooling luminosity of the halo gas. MQW17 showed that the total luminosity of a cooling flow which reproduces the observed $\Novi$ is $6\times10^{48}\erg\yr^{-1}$, if we assume $\nH T=30\cm^{-3}\K$ and an initial flow temperature of $5\cdot10^{5}\K$ in their calculation. The MQW17 estimate is a factor of two higher than our estimate which includes only the \ovip-gas. 

Energy sources which are potentially large enough to compensate the cooling from the \ovip-gas include supernovae feedback (see \citealt{Cen13} and MQW17), feedback from the central massive black hole (see \citealt{MathewsProchaska17}), and IGM gas accretion (see \citealt{Cen13}).

\subsubsection{Low-pressure scenario for the \ovip-gas}\label{s:lowP}

In the low-$P$ scenario, we assume that \ovip-traces cool gas outside the accretion shock. In this scenario the gas temperature will be set by the competition between radiative cooling and heating sources beyond the shock, including radiative heating, adiabatic compression, stirring by satellites, mechanical heating from large scale structure formation, and winds from neighboring galaxies. However, the cooling time of metal-enriched gas with $T\sim10^{4.5}-10^5\K$ is 
\begin{eqnarray}\label{e:tcool lowP}
 t_{\rm cool} &=& \frac{2.2\cdot\frac{3}{2}\,kT}{\nH\Lambda(\nH,T,Z)} = 410 \left(\frac{Z}{\frac{1}{3}\zsun}\right)^{-0.87}\cdot  \nonumber \\ 
&\cdot& \left(\frac{T}{10^{5}\K}\right)^{-1.15} \left(\frac{n_{\rm H}}{10^{-4.5}\cm^{-3}}\right)^{-1.5}\Myr ~,\nonumber\\
\end{eqnarray}
where we assumed the thermal energy per unit volume is $2.2\cdot\frac{3}{2}\nH k T$ (the number of particles per hydrogen particle is $2.2=(\mu X)^{-1}$), isochoric cooling beyond the shock, and we remind the reader that in our notation $\nH^2\Lambda$ is the net cooling per unit volume calculated by \cloudy. The power-law dependence on the parameters are approximations applicable at $10^{4.5}<T<10^5\K$ and near the stated numerical values of $\nH$ and $Z$. This cooling time is generally shorter than the dynamical time scale of $\approx 1.5\Gyr$, which is the heating timescale of e.g.\ adiabatic compression. Hence, if heating sources outside the accretion shock occur on a dynamical timescale and heat the gas to temperatures $\lesssim10^{5}\K$, radiative cooling will dominate and the gas would roughly be in thermal equilibrium with the UV background. 

The equilibrium temperature $T_{\rm eq}$ can be calculated using \cloudy\footnote{This calculation is similar to the calculations of the models shown in Figs.~\ref{f:nTplots OVI} and \ref{f:nTplots cooling}, but instead of fixing $T$ to a certain value, \cloudy\ derives $T=T_{\rm eq}$ such that  the gas is in thermal equilibrium.}   and approximated as
\begin{equation}\label{e:T_eq}
  T_{\rm eq} = 2.6\times 10^{4}\left(\frac{\nH}{10^{-4.5}\cm^{-3}}\right)^{-0.23}\K ~, 
\end{equation}
which is accurate to 2\%\ at $10^{-5}<\nH<10^{-4}\cm^{-3}$. 
This solution is plotted as dashed red curves in both panels of Fig.~\ref{f:nTplots OVI} and in Fig.~\ref{f:nTplots cooling}. By definition, this solution is where the net cooling $L_{\rm cool}^{(\ovip)}$ equals zero. We henceforth assume that in the low-pressure \ovip\ scenario, the temperature of the \ovip-gas is $T_{\rm eq}$. 

Given the mass and pathlength arguments above, we expect  $\fovi$ to be near its peak value, so we emphasize the intersection of the thermal equilibrium solution with the region where $\fovi>0.1$, excluding low densities where the pathlength is larger than the size of the system (see right panel of Fig.~\ref{f:nTplots OVI}). 
The expected gas density is hence in the range $\nH\sim0.1-1\times10^{-4}\cm^{-3}$, i.e.\ an overdensity of $30-300$. This overdensity is comparable on the high end to the dark matter overdensity of 200 at $\Rovi=0.6\rvir$, and hence the hydrogen densities in both scenarios are comparable (see eqn.~\ref{e:n CI}). Based on eqn.~(\ref{e:novi}), the implied $\novi$ in the low-$P$ scenario is
\begin{equation}\label{e:novi PI}
 n_{\ovip,\, \lowP}= 10^{-9}\left(\frac{\nH}{10^{-4.5}\cm^{-3}}\right)\left(\frac{\fovi}{0.2}\right)\left(\frac{Z}{\frac{1}{3}\zsun}\right)\cm^{-3}~,
\end{equation}
while the gas pressure is
\begin{equation}\label{e:P PI}
 (\nH T)_{\lowP} = \nH T_{\rm eq}=0.78\left(\frac{\nH}{10^{-4.5}\cm^{-3}}\right)^{0.77} \cm^{-3}\K ~,
\end{equation}
where we used eqn.~(\ref{e:T_eq}) for $T_{\rm eq}$. This characteristic pressure is a factor of $\approx40$ lower than the characteristic pressure in the high-$P$ scenario discussed above (eqn.~\ref{e:P CI}).

How far inward can the gas pressure estimated in eqn.~(\ref{e:P PI}) remain so low? 
The typical gas pressures at distances smaller than $\Rovi$ can be estimated from the pressure of the low-ionization cool clouds observed at small impact parameters. 
Fig.~\ref{f:hi2ovi} shows that \hi\ columns of $\NHI\gtrsim10^{16.5}\cm^{-2}$ are observed in roughly half the objects with $0.2<\Rimp/\rvir<0.5$, and in all objects with $\Rimp/\rvir<0.2$. Detections of \ciip\ and other low-ions follow a similar trend (see Fig.~\ref{f:hi2ovi} and Appendix~\ref{a:low-ions}). Single-density photoionization modeling of these low-ions and \hi\ columns deduce that they originate in gas with a typical density of $\gtrsim 10^{-3}\cm^{-3}$ (\citealt{Prochaska+17}, see table~3 there). Multi-density photoionization modeling of the same data also suggests that gas with $\NHI=10^{16.5}\cm^{-2}$ originates in gas with $\nH\gtrsim10^{-3}\cm^{-3}$ (S16). 
These derived densities should be multiplied by two (i.e., $\nH\gtrsim2\cdot10^{-3}\cm^{-3}$), since both \cite{Prochaska+17} and S16 assumed a HM12 ionizing spectrum, which has a factor of two lower intensity than the spectrum assumed here (see \S\ref{s:basic properties}).
A similar gas density of $3\cdot10^{-3}\cm^{-2}$ is also directly deduced from the $\NHI\approx10^{16.5}\cm^{-2}$ and $N_\ciiip\gtrsim 10^{14}\cm^{-2}$ observed in these sightlines, using the analytic formula of MQW17 (eqns.~22-23 there). Hence, for our assumed UV background, the gas pressure at distances $R\lesssim0.5\rvir$ is constrained to be $\nH T\gtrsim2\cdot10^{-3}\cdot10^4=20\cm^{-3}\K$, at least in sightlines which exhibit large \hi\ columns and low-ions. This estimated pressure is at least a factor of $\sim25$ higher than the pressure at $\Rovi \approx 0.6\rvir$ estimated in eqn.~(\ref{e:P PI}).
Such relatively high pressures in the inner halo are also suggested by the hot gas densities of $\sim10^{-4}\cm^{-3}$ estimated from the extended X-ray emission in nearby spirals  (\citealt{AndersonBregman11, Bogdan+13a, Bogdan+13b}) and the similar densities inferred from ram pressure stripping of Local Group satellites and the LMC (\citealt{GrcevichPutman09,Salem+15}). 
Therefore, observations of low-ions, \hi\ columns, X-ray emission and Local Group satellites all indicate the existence of a high-pressure hot phase in the inner halo, i.e.\ within roughly half the virial radius. The low-$P$ scenario for \ovip\ suggests that this hot phase does not extend beyond this distance, as pictured in the right panel of Fig.~\ref{f:schema}.

\subsection{$\NHI/\Novi$ ratio}\label{s:hi res}

\begin{figure*}
 \includegraphics{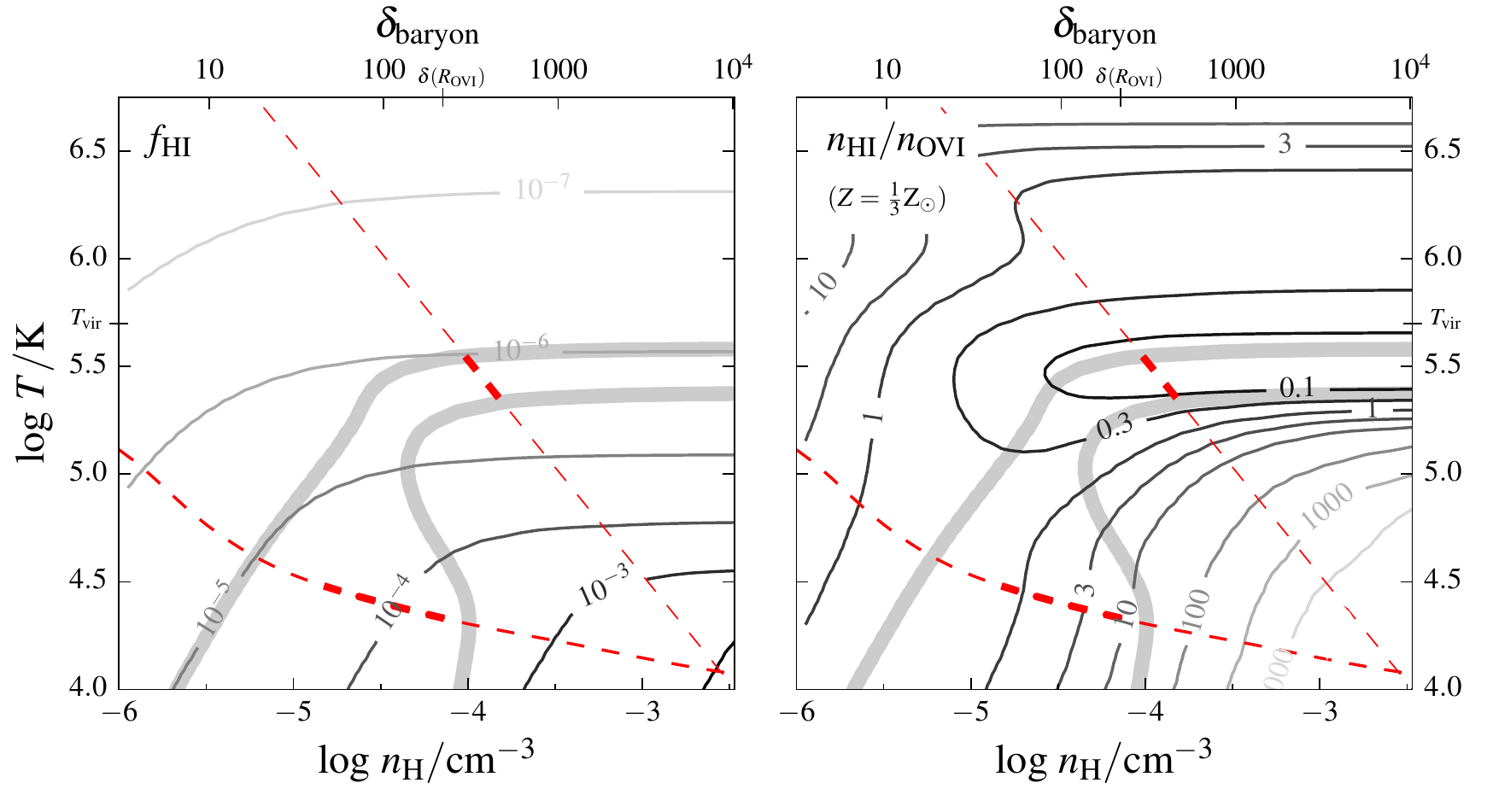}
\caption{
Contours denote the neutral hydrogen fraction $\fhi$ (\textbf{left panel}) and the ratio of \hi\ to \ovip\ density (\textbf{right panel}), as a function of gas density and temperature. The corresponding baryon overdensity is noted on top. Third-solar abundances are assumed in the right panel ($\nhi/\novi$ scales inversely with abundance). 
The thick grey and red dashed lines are as in Fig.~\ref{f:nTplots OVI}. 
In the low-pressure \ovip\ scenario (thick lower red line), the predicted range of $\nhi/\novi = 0.5-20$ is consistent with the observed $\NHI/\Novi$ in sightlines with $\Rimp>\Rovi=0.6\rvir$, where \hi\ is not dominated by gas on scales smaller than \ovip\ (Fig.~\ref{f:hi2ovi}). Therefore, in the low-pressure scenario the \hi\ observed at $\Rimp>0.6\rvir$ originates in the same gas phase as \ovip.
In the high-pressure \ovip\ scenario (thick upper red line) the predicted range of $\nhi/\novi \lesssim0.1$ is too low to explain the observations, so \hi\ must originate in a different gas phase than \ovip.
}
\label{f:nTplots HI}
\end{figure*}

What is the predicted $\NHI/\Novi$ in each of the two scenarios discussed in the previous section? 
To address this question, in the left panel of Figure~\ref{f:nTplots HI} we plot the neutral hydrogen fraction $\fhi$ as a function of $\nH$ and $T$. The value of $\fhi$ decreases to higher $T$ due to increased collisional ionization, and to lower $\nH$ due to increased photoionization. The right panel shows the implied neutral hydrogen to \ovip\ ratio $\nhi/\novi$ assuming $Z=\zsun/3$.

Fig.~\ref{f:nTplots HI} shows that in the high-pressure \ovip\ scenario, we expect $\nhi/\novi \lesssim 0.1(Z/\frac{1}{3}\zsun)^{-1}$ in the gas traced by \ovip. For comparison, the observed ratios are $\NHI/\Novi>1$ in all sightlines (Fig.~\ref{f:hi2ovi}).
A metallicity of $Z\lesssim0.03\zsun$ would resolve this difference for the lowest values of $\NHI/\Novi$, but would imply a mass for the \ovip-gas which is larger than the baryon budget (eqn.~\ref{e:Mgasovi frac}) and a pathlength larger than the size of the system (eqn.~\ref{e:lovi}, right panel of Fig.~\ref{f:nTplots OVI}). Hence, a low $Z<0.03\zsun$ is ruled out. Moreover, the \hi\ absorption features in objects with $\Rimp>0.5\rvir$ show a median line width of $b_\hi=40\kms$, compared to the $b_\hi\geq\sqrt{2kT/\mp}=70\kms$ expected in gas with $T=10^{5.5}\K$ ($b_\hi$ can be larger than this estimate if non-thermal broadening is significant). The typical line width hence implies that \hi\ originates in gas with $T<10^5\kms$, cooler than the gas which produces \ovip. We conclude that in any scenario where \ovip\ is collisionally ionized, \hi\ must originate from a different gas phase than \ovip. 

In contrast, in the low-pressure \ovip\ scenario, Fig.~\ref{f:nTplots HI} shows we expect $\nhi/\novi$ between $0.8(Z/\frac{1}{3}\zsun)^{-1}$ and $20(Z/\frac{1}{3}\zsun)^{-1}$. Given the likely range of metallicities of the \ovip\ gas of  $0.3<Z/\zsun<1$ (\S\ref{s:Z}), this calculation correctly predicts the range of $1<\NHI/\Novi<10$ observed at $\Rimp\gtrsim\Rovi$ (with the caveat that there are only a small number of $\NHI/\Novi$ measurements at outer radii). 
At smaller impact parameters the low-pressure model falls short of the observed $\NHI/\Novi$ in most sightlines, though as we argued in \S\ref{s:hi} at these impact parameters \hi\ may be dominated by gas at smaller radii than \ovip. 
Hence, in the low pressure scenario the \ovip\ and \hi\ absorption observed at $\Rimp\gtrsim\Rovi$ originate in the same phase. This conclusion is further explored below.

The same conclusions for the low and high-pressure scenarios also apply to \ovip\ absorbers along random sightlines, which likely originate in the vicinity of galaxies with a somewhat lower mass than in the galaxy-selected sample used above (\citealt{Prochaska+11}). These absorbers exhibit a typical column ratio of $0.2<\NHI/\Novi<20$ (lower panel of Fig.~\ref{f:hi2ovi}), similar to the values observed at $\Rimp>\Rovi$ in the galaxy-selected sample. Hence if these \ovip\ absorbers are collisionally-ionized in $\sim10^{5.5}\K$ gas, then they are always associated with a cooler phase which produces on average three \hi\ particles per \ovip\ particle, with a factor of three dispersion around this average. Alternatively, if \ovip\ absorbers along random sightlines are photoionized by the background, then their $\NHI/\Novi$ ratios suggest that \ovip\ and \hi\ originate in a single phase with metallicity of $\sim1/3\zsun$, as derived above for galaxy-selected absorbers.

An additional constraint on the relation between \ovip\ and \hi\ can be derived from the velocity profiles of the two absorption features. The two features exhibit kinematic alignments in their velocity profiles, though this alignment is not perfect. This characteristic behavior is observed both in random IGM sightlines (e.g.\ \citealt{ThomChen08, Fox11}), and in galaxy selected samples such as COS-Halos, especially in absorbers with $\NHI\lesssim10^{15}\cm^{-2}$ that do not exhibit low-ionization absorption (see fig.~5 in \citealt{Werk+16}). 
The observed kinematic alignment argues for a physical connection between the gas traced by \hi\ and \ovip. 
In the high-pressure scenario, this alignment may be explained if \hi\ and \ovip\ trace gas at different temperatures in the same cooling flow (e.g.\ MQW17). 
In the low-pressure scenario, the kinematic alignment is consistent with our conclusion that \hi\ and \ovip\ traces the same gas phase. 

If \hi\ and \ovip\ originate in the same gas, as suggested by the low-pressure scenario, why then is their kinematic alignment not perfect? 
A possible explanation is the high sensitivity of the \hi\ to \ovip\ ratio to the gas density.
The right panel of Fig.~\ref{f:nTplots HI} demonstrates that $\nhi/\novi$ can change by a factor of $\approx10$ as a result of a relatively mild change of a factor of two in $\nH$. 
Given that the pathlength through \ovip\ absorbers spans at least tens of kpc (eqn.~\ref{e:lovi}), a change of a factor of $\sim 2$ in $\nH$ across a single absorber is plausible. In this case the velocity-resolved $\NHI(v)/\Novi(v)$ ratio would change across the absorber, and the kinematics of the two absorption features would be correlated, but not fully aligned.

\begin{figure*}
 \includegraphics{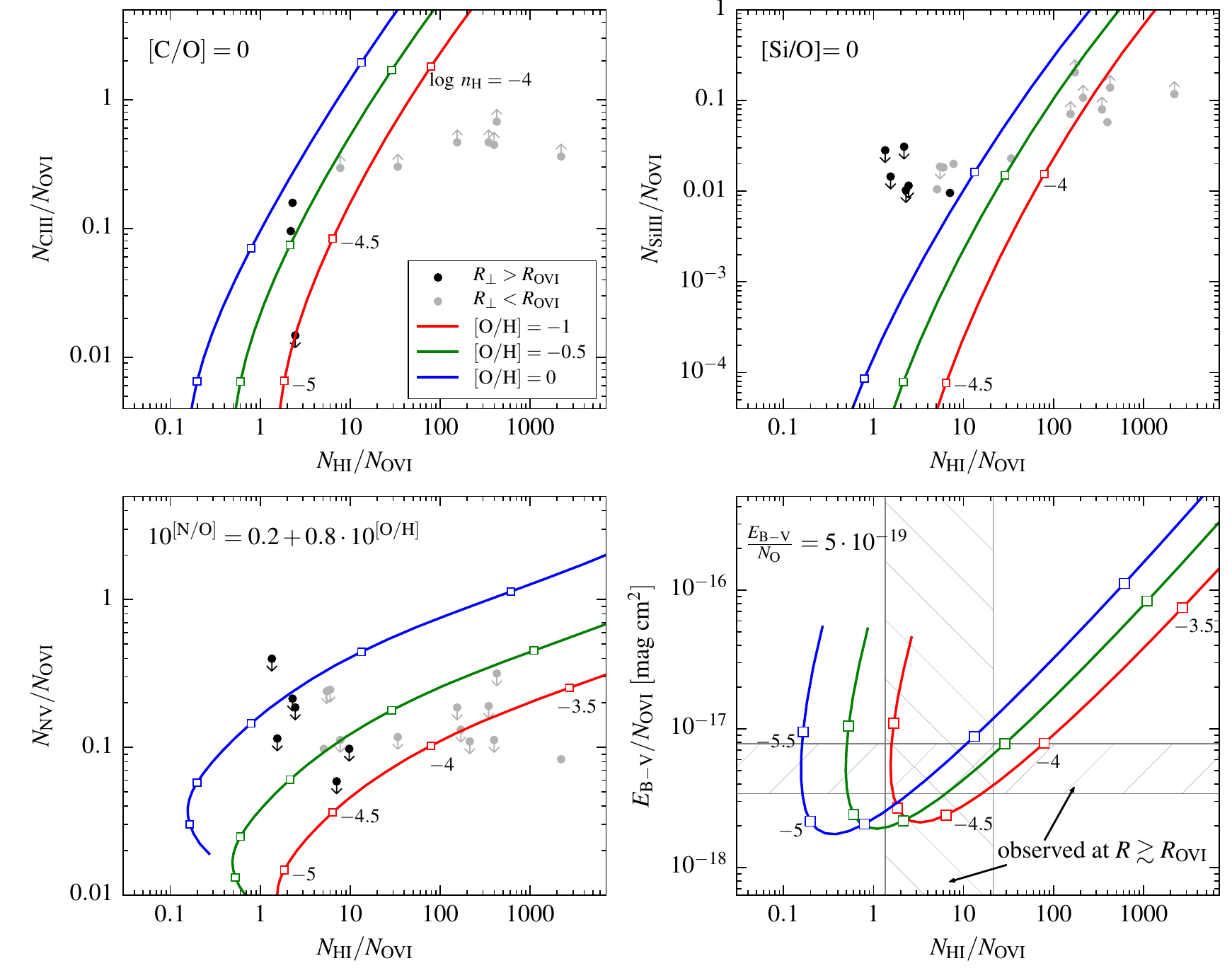}
\caption{
Expected absorption line ratios and $\ebv$ in single-density PIE models, in which the gas is in ionization and thermal equilibrium with the UV background. These models are applicable in the low-pressure \ovip\ scenario for sightlines with $\Rimp > \Rovi$, where the single-density assumption is plausible. 
Models with the same metallicity are connected by colored lines according to the legend in the top-left panel.
Every half-decade in $\nH$ is marked by a large white square, and noted in the panels. 
\textbf{(Top and bottom-left panels)}
The models assume solar C/O and Si/O abundance ratios, and an ISM-like N/O noted in the bottom-left panel. 
Observations of ion columns are from COS-Halos and J15, where we use black markers to emphasize sightlines with $\Rovi<\Rimp<\rvir$. 
Single-density PIE models with $10^{-4.5}\lesssim\nH\lesssim10^{-4}\cm^{-3}$ are consistent with most observations at $\Rimp>\Rovi$, assuming the actual \nvp\ columns are not significantly below the measured upper limits.  
\textbf{(Bottom-right panel)}
The models assume an ISM-like $\ebv$-to-oxygen ratio (noted in the panel). 
The horizontal stripe marks the average $\ebv$ at $\Rimp=\Rovi$ measured by MSFR divided by the average observed $\Novi=10^{14.5}\cm^{-2}$. The vertical stripe marks the range of $\NHI/\Novi$ observed at $\Rimp\gtrsim\Rovi$. The observed $\ebv/\Novi$ is also consistent with $\nH\lesssim10^{-4}\cm^{-3}$.
}
\label{f:PIE}
\end{figure*}

\subsection{Other ions}\label{s:other ions}

In this section we use the detections and non-detections of the `intermediate ions' (\ciiip, \Siiiip, \nvp) in the outer halo to further constrain the two scenarios for \ovip.

In the low-pressure scenario, the expected ion columns relative to \ovip\ can be calculated using photoionization equilibrium models (PIE models), where the gas is assumed to be in ionization and temperature equilibrium with the UV background. We use \cloudy\ to calculate the expected line ratios as a function of gas metallicity and density, assuming as above that the gas is irradiated by the UV background (\S\ref{s:basic properties}).  Figure~\ref{f:PIE} plots the predicted $N_\ciiip/\Novi$ vs.\ $\NHI/\Novi$ (top-left panel), $N_\Siiiip/\Novi$ vs.\ $\NHI/\Novi$ (top-right), and $N_\nvp/\Novi$ vs.\ $\NHI/\Novi$ (bottom-left). Models with the same metallicity are connected using a colored line, as noted in the legend. The carbon and silicon abundances are scaled with the oxygen abundance, while nitrogen, which is a secondary nucleosynthesis element (e.g.\ \citealt{Henry+00}), 
is scaled as in \hii\ regions in the ISM of galaxies (eqn.~2 in \citealt{Groves+06}):
\begin{equation}\label{e:N2O}
 10^{\rm [N/O]} = 0.19 + 0.81\cdot10^{\rm [O/H]} ~.
\end{equation}
Also plotted in the panels are observations of ion column ratios in the COS-Halos+J15 sample. We show only objects in which the relevant intermediate ion has been observed, and at least one ion in each axis has been detected. We focus on impact parameters $\Rimp>\Rovi$ (black markers) where the single-density assumption is plausible. At smaller impact parameters \hi, \ciiip, and \Siiiip\ are potentially dominated by gas on smaller scales than \ovip\ (see \S\ref{s:hi}), in which case multi-density PIE models are required (e.g.~S16). These latter objects are plotted with gray markers (objects with $>$$1\dex$ error on $\NHI$ are not shown to avoid clutter).

In the previous sections we deduced a gas density of $\nH\sim 10^{-4.5}\cm^{-3}$ and a metallicity of $Z\gtrsim\zsun/3$ for the \ovip-gas in the context of the low-pressure scenario, based on the \ovip-gas mass and pathlength. 
Fig.~\ref{f:PIE} demonstrates that these gas parameters are generally consistent also with the observational constraints on $\NHI/\Novi$, $N_\ciiip/\Novi$, $N_\Siiiip/\Novi$ and $N_\nvp/\Novi$ at impact parameters $\Rimp>\Rovi$. Specifically, in the \ciiip\ panel (top-left) two of the three objects suggest $0.3<Z/\zsun<1$ and $\nH\gtrsim10^{-4.5}\cm^{-3}$, while the third object\footnote{This object has peculiar \ovip\ kinematics, see below.} suggests a lower metallicity of $Z\lesssim0.1\zsun$ and a lower density of $\nH\approx10^{-4.8}\cm^{-3}$. 
The single detection of \Siiiip\ suggests $\nH\lesssim10^{-4}\cm^{-2}$ and $Z\approx\zsun$ (top-right), while the upper limits on $N_{\Siiiip}/\Novi$ in the remaining objects are also consistent with similar parameters.
In the bottom-left panel, four of the upper limits on $N_{\nvp}/\Novi$ are consistent with $Z\gtrsim0.3\zsun$ (bottom-left), while two objects require somewhat lower metallicities. Last, in one object in the sample \civp\ has been observed (not shown in the figure). Comparing its $N_\civp/\Novi=0.5$ and $\NHI/\Novi=9.8$ with the PIE models suggests $Z\approx\zsun/3$ and $\nH\approx10^{-4.4}\cm^{-3}$, again similar to the parameters deduced above.

Fig.~\ref{f:PIE} also shows that at impact parameters smaller than $\Rovi$, the observed line ratios of objects with $\NHI/\Novi<10$ imply a similar gas density and metallicity as implied by objects with $\Rimp>\Rovi$. However, the line ratios of objects with $\NHI/\Novi\gg10$ apparently imply a gas density larger than $10^{-4}\cm^{-3}$. Such a large density is ruled out for the \ovip-gas in the low-pressure model due to the large implied gas mass (\S\ref{s:basic properties}). As mentioned above, in these objects \hi, \ciiip\ and \Siiiip\ are likely dominated by a gas phase which is closer to the galaxy and denser than the gas which produces \ovip. The absence of \nvp\ detections in these objects (bottom-left panel) constrains this additional dense phase not to produce observable quantities of \nvp. We address this constraint in the discussion.

To conclude, existing ion columns observations at $\Rimp>\Rovi$ are generally consistent with a low-pressure scenario for \ovip\ with $\nH\approx10^{-4.5}\cm^{-3}$ and $Z\gtrsim\zsun/3$, though the existing constraints are not very strong. Additional observations of the CGM of galaxies at $z>0.2$ where \ciiip\ is observable with COS can increase the statistics of this ion. Also, somewhat deeper measurements of \nvp\ would be able to verify that \nvp\ is indeed close to the measured upper limits, as implied by the low-pressure \ovip\ model (bottom-left panel of Fig.~\ref{f:PIE}).

We note that \cite{Werk+16} concluded that the observed upper limits on $N_{\nvp}/\Novi$ rule out a photoionization origin for \ovip, because the implied pathlengths assuming PIE are unphysically large. Our analysis does not reach a similar conclusion. In Appendix \ref{a:NV} we list the \ovip\ pathlengths implied by the preferred PIE model for each object. For all objects except one (J0914+2823\_41\_27), the observed upper limits on \nvp\ allow a pathlength smaller than $\rvir$, typically by a factor of $2-10$. That is, the derived pathlengths are consistent with \ovip\ originating in the ambient photoionized medium beyond the shock as suggested by the low pressure scenario, provided that the actual $N_{\nvp}$ are not significantly below the measured upper limits. The disparity in the conclusions is mainly due to the ISM-based scaling of [N/O] used here (eqn.~\ref{e:N2O}) compared to the solar [N/O] used by \cite{Werk+16}.

In the high-pressure scenario, where $T\approx10^{5.5}\K$ and $\nH\approx10^{-4}\cm^{-3}$, the \cloudy\ calculations used in \S\ref{s:basic properties} give $n_{\rm C^{2+}}/\nopv=10^{-4}$, $n_{\rm Si^{2+}}/\nopv=10^{-6}$, and $n_{\rm N^{4+}}/\nopv=0.03$. These values are a factor of $1000$ too low to explain the two detections of \ciiip\ at $\Rimp>\Rovi$ (top-left panel of Fig.~\ref{f:PIE}) and a factor of $10^4$ too low to explain the single detection of \Siiiip\ (top-right panel) at the same impact parameters. Therefore, the high-pressure scenario requires a multi-phase solution to explain the observed \Siiiip\ and \ciiip\ absorption at $\Rimp>\Rovi$. This conclusion is similar to our conclusion for \hi\ in \S\ref{s:hi res}, and in contrast with the conclusion for the low-pressure scenario, where only a single phase is required at $\Rimp>\Rovi$.

\subsection{Differential extinction}

What is the expected $\ebv$ in each of the two scenarios discussed above? 
In the low-pressure scenario, the hot gas phase does not extend beyond $R \sim 0.5\rvir$, so the \ovip-gas is the dominant gas phase in the outer halo. We hence expect dust embedded in the \ovip-gas to dominate the extinction in the outer halo. We can estimate the expected $\ebv$ from the \ovip-gas by assuming the dust-to-oxygen ratio in the CGM is similar to that seen in the diffuse ISM of the Milky-Way (MW). Using $\ebv/\NH=1.7\cdot10^{-22}\,{\rm mag}\cm^2$ (\citealt{Bohlin+78,Rachford+09}) and an oxygen abundance relative to hydrogen of $3.19\cdot10^{-4}$ (\citealt{Meyer+98}) we get
\begin{equation}\label{e:ebv to N_O}
 \left.\frac{\ebv}{N_{\rm O}}\right|_{\rm CGM} = D\left.\frac{\ebv}{N_{\rm O}}\right|_{\rm ISM} = 5.3 \times10^{-19}\,D\, {\rm mag} \cm^2
\end{equation}
where $N_{\rm O}$ is the oxygen column, and the factor $D$ represents any differences between $\ebv/N_{\rm O}$ in the CGM and in the MW ISM. 
Since the dust-to-oxygen ratio in the ISM of local galaxies appears to be independent of ISM metallicity at $Z_{\rm ISM}\gtrsim0.2\zsun$ and shows a relatively small factor of $\sim2$ dispersion between individual galaxies (\citealt{Issa+90, Draine+07, RemyRuyer+14}), 
we expect $D\approx1$ in the simplest scenario where grains and metals are coupled when ejected from the ISM, and the grains do not experience further growth or destruction in the CGM. 

To derive the expected ratio of $\ebv$ to $\Novi$, we divide eqn.~(\ref{e:ebv to N_O}) by $\fovi$:
\begin{equation}\label{e:ebv to N_OVI}
 \frac{\ebv}{N_{\ovip}} = \frac{\ebv}{\fovi N_{\rm O}} = 2.7 \times10^{-18}\  \left(\frac{\fovi}{0.2}\right)^{-1}\,D\,{\rm mag} \cm^2~.
\end{equation}
We calculate $\fovi$ using the PIE models presented in the previous section, and plot the implied $\ebv/\Novi$ as a function of $\nH$ and $Z$ in the bottom-right panel of Fig.~\ref{f:PIE}. A value of $D=1$ is assumed.
The panel shows that the models reach a minimum $\ebv/\Novi=2\cdot10^{-18}\,{\rm mag}\cm^2$ at $\nH=10^{-4.5}\cm^{-3}$, which is the gas density where $\fovi$ peaks. 
For comparison, the horizontal hatched stripe marks the $\ebv$ measurement at $\Rimp=\Rovi\approx110\kpc$ from MSFR (calculated from their eqn.~28 at the mean redshift of the MSFR sample), divided by the average $\Novi=10^{14.5}\cm^{-2}$. 
The observed range of $\NHI/\Novi$ at $\Rimp\gtrsim\Rovi$ (Fig.~\ref{f:hi2ovi}) is shown as a vertical hatched stripe. 
Fig.~\ref{f:PIE} demonstrates that PIE models with $0.5<\nH<1\cdot 10^{-4}\cm^{-2}$ and $0.3<Z/\zsun<1$ reproduce the observed $\ebv/\Novi$. These parameters are similar to the parameters deduced above based on the \ovip-gas mass, \ovip-gas pathlength, and ion column ratios.

Calculating the expected $\ebv$ in the high-pressure \ovip\ scenario is less straightforward than in the low-pressure scenario, due to the possible contribution to $\ebv$ from dust embedded in the unconstrained hot phase, and due to grain sputtering by this hot phase. Sputtering occurs on a timescale of (\citealt{Draine11})
\begin{equation}\label{e:t_sputtering}
t_{\rm sput} = 3.3\frac{a}{0.05\mic}\left(\frac{\nH}{10^{-4.5}\cm^{-3}}\right)^{-1}\left(\frac{1+T_6^{-3}}{2}\right)\Gyr ~,
\end{equation}
where $a$ is the grain size, and $T\equiv10^6 T_6 \K$. The normalization of $a$ is the maximum grain size which can produce the differential extinction between the SDSS-$u$ band and SDSS-$g$ band observed by MSFR ($\approx4000\AA/(1+z_{\rm MSFR})/2\pi=0.05\mic$, where $z_{\rm MSFR}=0.36$ is the median redshift in the MSFR sample). 
The timescale in eqn.~(\ref{e:t_sputtering}) may be lower than the Hubble time, in which case the factor $D$ in eqn.~(\ref{e:ebv to N_OVI}) would be lower than unity. Furthermore, this sputtering by the hot phase at $\sim100\kpc$ is on top of any sputtering that the dust experienced on its way out of the galaxy, an effect which could also be relevant to the low-pressure scenario. \cite{McKinnon+16, McKinnon+17} included the effect of sputtering on the grains in the galaxy formation model of \cite{Vogelsberger+13}.
They found that the simulation underpredicts the dust mass deduced by MSFR at $R=\Rovi=110\kpc$ by a factor of ten (see figure~3 in \citealt{McKinnon+17}).

\subsection{\ovip\ line widths}\label{s:tidal}

\begin{figure}
 \includegraphics[width=0.48\textwidth]{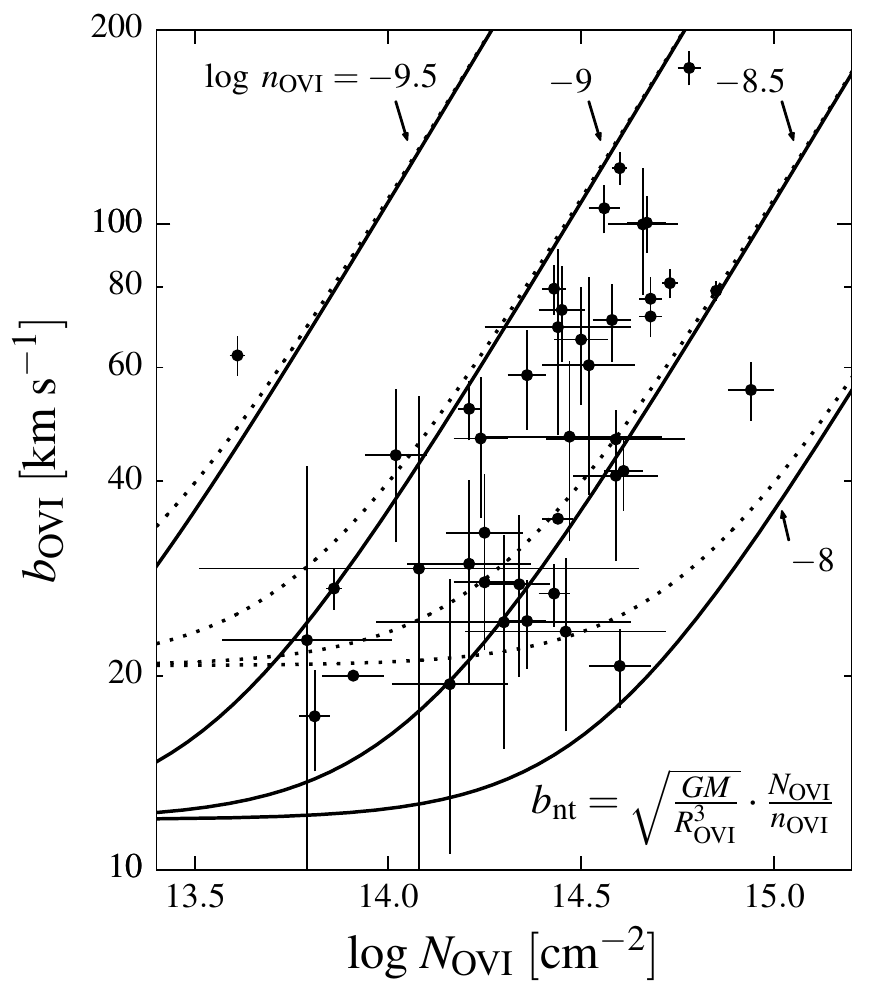}
\caption{Line width versus column density for \ovip\ absorption features. Black lines plot the expected relation for gravitational broadening of \ovip\ absorbers, for different assumed volume densities of the $\opv$ ion as noted in the figure. The equation for gravitational broadening is noted in the lower-right and discussed in \S\ref{s:tidal}. Solid lines assume an \ovip-gas temperature of $10^{4.5}\K$ (thermal equilibrium with the UV background), while dotted lines assume $T=10^{5.5}\K$ (collisionally ionized \ovip). Errorbars plot the observed values in the COS-Halos+J15 sample. The observed relation between $\bovi$ and $\Novi$ is consistent with gravitational broadening of the \ovip-gas if $10^{-9}\lesssim\novi\lesssim10^{-8.5}\cm^{-3}$. These values of $\novi$ are expected in both the high-pressure and low-pressure scenarios (see eqns.~\ref{e:novi CI} and \ref{e:novi PI}).
}
\label{f:b vs N}
\end{figure}

\begin{figure*}
 \includegraphics{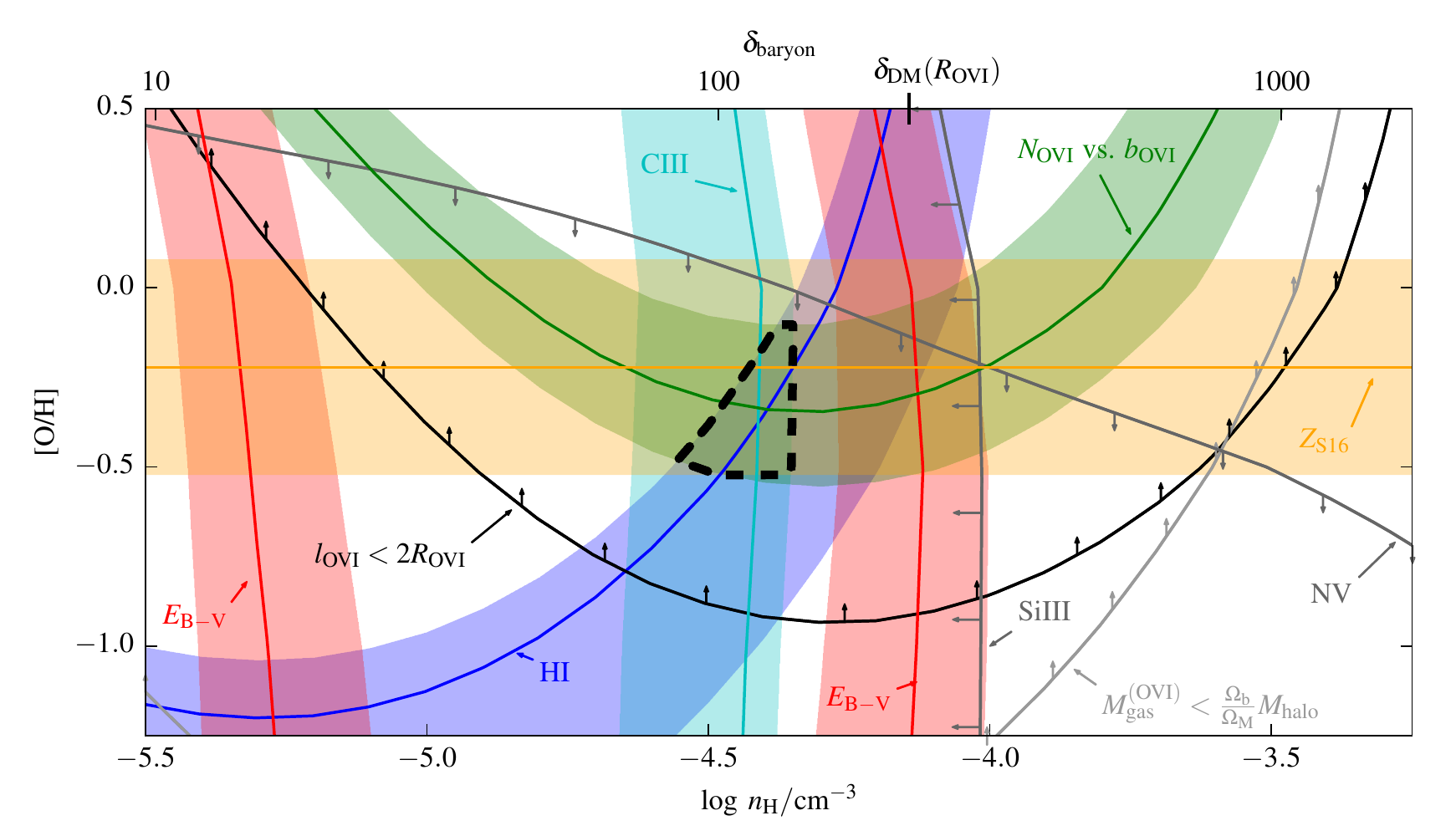}
\caption{Summary of observational and physical constraints on the \ovip-gas in the context of the low-pressure scenario. Stripes show the allowed regions in $\nH-Z$ space implied by different observables: the observed ratios of $\NHI/\Novi$ ({\it blue}) and $N_{\ciiip}/\Novi$ ({\it cyan}) at $\Rimp>\Rovi$; the $\novi$ suggested by the $\Novi$ vs.\ $\bovi$ relation ({\it green}, assuming $g=1$ in eqn.~\ref{e:btidal func}); and the observed $\ebv/\Novi$ ({\it red}, allows two distinct densities, see Fig.~\ref{f:PIE}). Black and grey lines with arrows mark one-sided constraints: that the pathlength is smaller than the size of the system, that the mass of the gas traced by \ovip\ is lower than the halo baryon budget, and that the models conform with observed upper limits on $N_{\nvp}/\Novi$ and $N_\Siiiip/\Novi$ at $\Rimp>\Rovi$. The yellow horizontal stripe marks the metallicities derived by modelling sightlines at all impact parameters with multi-density PIE models (S16).  
The top x-axis shows the baryon overdensity at the sample redshift corresponding to each $\nH$, and the dark matter overdensity at $\Rovi$ is marked. The dashed polygon marks the region in parameter space which satisfies all constraints except the constraint based on $\ebv$, where the latter suggests densities a factor of $\approx2$ higher. Note that PIE models with $\delta_{\rm baryon} \approx \delta_{\rm DM}$ and $Z\gtrsim1/3$ are consistent with all considered constraints to within a factor of two, which provides observational support for the low-pressure scenario.}
\label{f:PIE_allConstraints}
\end{figure*}

The pathlength $\lovi$ we derive for the \ovip-gas spans at least tens of kpc in both scenarios (right panel of Fig.~\ref{f:nTplots OVI}), which is a significant fraction of the halo size. 
Hence, if the absorber kinematics are dominated by bulk motions, we expect a significant velocity shear within the absorbers. This shear is expected to broaden the observed absorption profile, and therefore can be tested against observations. In the limit of coherent ballistic gas motions, we expect this broadening to be of order
\begin{equation}\label{e:dvds}
\frac{b_{\rm nt}}{\lovi} \sim \frac{{\rm d} v_{\rm los}}{{\rm d} s} \sim \sqrt{\frac{G\Mhalo(<R)}{R^3}}
\end{equation}
where $b_{\rm nt}$ is the non-thermal broadening component of the absorption profile, and ${\rm d}v_{\rm los}/{\rm d}s$ is the velocity gradient along the line of sight. Eqn.~(\ref{e:dvds}) states that the ratio of the total velocity shear across the absorber to the absorber pathlength should roughly equal the reciprocal of the dynamical time. 

Since $\lovi=\Novi/\novi$, equation~(\ref{e:dvds}) can be converted to a relation between \ovip\ column $\Novi$ and \ovip\ line width $\bovi$:
\begin{eqnarray}\label{e:btidal func}
 b_{\rm nt}(\ovip) &=& g\sqrt{\frac{G\Mhalo(<R)}{R^3}}\cdot\frac{\Novi}{\novi} \nonumber \\
               &=& g\sqrt{\frac{4\pi\Delta_{\rm c}\rho_{\rm c}G}{3}}\left(\frac{R}{\rvir}\right)^{-1.15}\cdot\frac{\Novi}{\novi}
\end{eqnarray}
where we introduced the order-unity geometric factor $g$ to account for projection effects, possible differences between $b_{\ovip}/\lovi$ and ${\rm d}v_{\rm los}/{\rm d}s$, and for the inaccuracy of approximating the broadened absorption profile as a Gaussian. In the second equality in eqn.~(\ref{e:btidal func}) we approximate the NFW halo mass as $\Mhalo(<R)=(4\pi\Delta_{\rm c}\rho_{\rm c}\rvir^3/3)\cdot(R/\rvir)^{0.7}$, which is accurate to $10\%$ at $0.3<R/\rvir<2$ for a concentration parameter of $10$.
Plugging in eqn.~(\ref{e:btidal func}) the characteristic radius of an \ovip\ absorber of $R=\Rovi=0.6\rvir$ (eqn.~\ref{e:Rovi}), and using $\Delta_{\rm c}\rho_{\rm c}$ appropriate for $z=0.2$ we get 
\begin{eqnarray}\label{e:btidal num}
	  b_{\rm nt}(\ovip)& & =  34\, g\left(\frac{\Novi}{10^{14.5}\cm^{-2}}\right)\cdot \nonumber\\
  & & \cdot\left(\frac{\novi(0.6\rvir)}{10^{-8.5}\cm^{-3}}\right)^{-1}\left(\frac{R}{0.6\rvir}\right)^{\beta-1.15} \kms ~, \nonumber\\
\end{eqnarray}
where we define $\beta$ such that $\novi\propto (R/\rvir)^{-\beta}$. Since $\beta$ is plausibly in the range $1-3$, eqn.~(\ref{e:btidal num}) suggests that $b_{\rm nt}$ scales as $\sim R^{0}-R^{1.85}$. Hence, the factor of $\sim2$ dispersion of the \ovip-gas in physical radius (lower-left panel of Fig.~\ref{f:ovi observations}) introduces only a modest factor of $\lesssim3$ dispersion in the expected $b_{\rm nt}$ vs.\ $\Novi$ relation.
Note that the a broadening of $34\kms$ as suggested by eqn.~(\ref{e:btidal num}) is detectable with COS, which has a spectral resolution of $b_{\rm res}\approx {\rm FWHM}_{\rm res}/1.66=11\kms$.

Our assumption of ballistic motions essentially assumes an \ovip-gas temperature which is $\ll T_{\rm vir}$, a condition which may not hold in the high-pressure \ovip\ scenario where $T\approx10^{5.5}\K$. Also, the derivation above neglects possible drag on the gas, which may be significant especially in the high-pressure scenario where the \ovip\ gas is embedded in a relatively dense medium. 
Both effects will decrease $b_{\rm nt}$ relative to the estimate in eqn.~(\ref{e:btidal num}). 

To compare eqn.~(\ref{e:btidal num}) with observations we calculate the total line broadening of
\begin{equation}\label{e:bovi}
 \bovi = \sqrt{b_{\rm nt}^2(\Novi,\novi) + b_{\rm th}^2(T) + b_{\rm res}^2}
\end{equation}
where $b_{\rm th}=5.7\,(T/10^{4.5}\K)^{1/2}\kms$ is the thermal broadening. 
Figure~\ref{f:b vs N} plots eqn.~(\ref{e:bovi}) for $\novi(\Rovi)$ in the range $10^{-9.5}-10^{-8}\cm^{-3}$, using eqn.~(\ref{e:btidal num}) for $b_{\rm nt}$ with $g=1$ and $R=\Rovi$. We show the relation for both $T=10^{4.5}\K$ (solid lines) and $T=10^{5.5}\K$ (dotted lines), corresponding to the two scenarios for \ovip. 

The observed values in the COS-Halos+J15 sample are plotted as errorbars in Fig.~\ref{f:b vs N}. Most objects are consistent with the predicted relation for values of $10^{-9}\lesssim\novi\lesssim10^{-8.5}\cm^{-3}$, as expected in both the high-pressure and low-pressure scenarios (eqns.~\ref{e:novi CI} and \ref{e:novi PI}). Deriving $\novi$ from the observed $\Novi$ and $\bovi$ and from eqns.~(\ref{e:btidal num})--(\ref{e:bovi}) we get a 16--84 percentile range of $1.2<\novi<3.3\cdot10^{-9}\cm^{-3}$ for $T=10^{4.5}\K$ and a similar range for $T=10^{5.5}\K$.
This success of eqn.~(\ref{e:btidal num}) in reproducing the observations supports both our estimate of $\novi$ and the assumption that the kinematics of the \ovip-gas are dominated by bulk motions.

The absorber around the galaxy 04:07:50.57-12:12:24.0 from J15 has $\Novi=10^{13.6}\cm^{-2}$ and $\bovi=60\kms$, which suggests $\novi\approx10^{-9.6}\cm^{-3}$, a factor of $5-15$ lower than most other objects. This low $\novi$ is consistent with the low $N_\ciiip/\Novi<0.015$ in this object (see top-left panel of Fig.~\ref{f:PIE}), which suggests a relatively low $\nH<10^{-4.8}\cm^{-2}$ and $Z\lesssim0.1\zsun$ in the context of the low-pressure model. We suspect this absorber originates from gas at distances significantly larger than $\Rovi$, where the characteristic densities are likely lower.  

Our explanation of the observed $\Novi$ vs.\ $\bovi$ relation using the velocity shear expected in gas dominated by bulk motions is qualitatively different from the explanation proposed by \cite{Heckman+02}, who argued that this relation is a general property of radiatively cooling gas. Specifically, the broadening mechanism discussed here is relevant to any gas which kinematics is dominated by gravity, and hence is potentially consistent with both the high-pressure and low-pressure scenarios. In contrast, the \citeauthor{Heckman+02}\ mechanism is applicable only in the high-pressure scenario where \ovip\ is out of thermal equilibrium and hence radiatively cooling. We will further explore the gravitational broadening mechanism in future work.

\subsection{Summary of low-pressure \ovip\ scenario}\label{s:summary PIE}

Figure~\ref{f:PIE_allConstraints} summarizes the observational constraints on the single-density PIE models, which are applicable in the low-pressure \ovip\ scenario for sightlines through the outer halo.
The regions in $\nH-Z$ space allowed given the maximum pathlength and \ovip-gas mass are marked by black and light grey lines with upward-pointing arrows.
The locus of models suggested by the median observed $\NHI/\Novi$ and $N_\ciiip/\Novi$ column ratios at $\Rimp>\Rovi$ are marked by blue and cyan solid lines, respectively, surrounded by same-color stripes to denote the 16\th--84\th\ percentiles of the observed column ratios.
The parameter space region allowed by the median upper limits on $N_{\nvp}/\Novi$ and $N_\Siiiip/\Novi$ are marked with dark gray lines and arrows. Models that give $1.2<\novi<3.3\cdot10^{-9}\cm^{-3}$, as suggested by the observed $\Novi-\bovi$ relation and gravitational broadening (eqn.~\ref{e:btidal num}, assuming $g=1$) are marked with a green line and stripe. Models that reproduce the observed $\ebv/\Novi$ at $\Rimp=\Rovi$ (based on eqn.~\ref{e:ebv to N_OVI}, assuming $D=1$) are marked with a red line and stripe. Note that there are two distinct densities that produce the observed $\ebv/\Novi$, because the predicted ratio depends on $\fovi$ which has the same value at $\nH=10^{-4.2}\cm^{-3}$ and $\nH=10^{-5.3}\cm^{-3}$. The yellow horizontal line and stripe mark the median and dispersion in metallicities derived by S16 by modelling all ions observed in COS-Halos with multi-density PIE models.

The dotted polygon in Fig.~\ref{f:PIE_allConstraints} marks the range of PIE models which satisfy all constraints except the constraint based on the observed $\ebv/\Novi$.
The factor of $\sim2$ higher densities suggested by the $\ebv/\Novi$ ratio compared to other constraints may reflect the fact that $\ebv$ is derived from a different sample (MSFR) than the sample used to derive the other constraints (COS-Halos+J15), or alternatively may suggest that the parameter $D$ defined in eqn.~(\ref{e:ebv to N_O}), which reflects differences between $\ebv/N_{\rm O}$ in the MW ISM and the CGM, has a value of $\approx2$. In general, the fact the all observational constraints are consistent to a factor of two with $\nH\approx0.5\times10^{-4}\cm^{-3}$ and $1/3<Z/\zsun<1$, provides observational support to the low-pressure \ovip\ scenario. Moreover, the preferred density suggests a baryon overdensity within a factor of two of the dark matter overdensity at $R=\Rovi$ (top x-axis), which is plausible for gas outside the accretion shock.

\section{Discussion}\label{s:discussion}

In the previous section, we analyzed constraints on the physical properties of halo gas traced by \ovip. In contrast to the common assumption that \ovip\ around \sLstar\ galaxies is mostly collisionally ionized, we demonstrated that a low-pressure scenario in which \ovip\ is in ionization and thermal equilibrium with the cosmic UV background can explain the observables that we considered. 
Specifically, we showed that cool gas with $\nH\approx10^{-4.5}\cm^{-3}$ and $Z/\zsun \sim 0.3-1$ can simultaneously explain the ionic column ratios at $R\gtrsim0.5\rvir$, the relation between $\Novi$ and $\bovi$, and the observed $\ebv$ to a factor of two. This scenario also has the advantage of invoking an equilibrium phase to explain the ubiquitous \ovip\ absorption, rather than the rapidly cooling phase invoked by the high-pressure \ovip\ scenario (Fig.~\ref{f:nTplots cooling}). In this section we further explore the properties of this low-pressure \ovip\ scenario. Some comments on the high pressure scenario, which has been addressed by other studies (e.g.\ \citealt{Faerman+17, McQuinnWerk17, MathewsProchaska17}), are given in the last subsection.

\subsection{The location of the accretion shock}\label{s:accretion shock}
\newcommand{\Mthres}{M_{\rm thres}}
\newcommand{\Tshock}{T_{\rm h}}
\newcommand{\Rshock}{R_{\rm s}}
\newcommand{\nHhot}{n_{\rm H,\,h}}
\newcommand{\vvir}{v_{\rm vir}}
\newcommand{\vinf}{v_{\rm inf}}

In the low-pressure scenario, \ovip\ traces gas with pressure $\approx1\cm^{-3}\K$ (eqn.~\ref{e:P PI}) at a distance of $\approx\Rovi=0.6\rvir$ (eqn.~\ref{e:Rovi}). In comparison, photoionization modeling of low-ions and \hi-columns of $\gtrsim10^{16}\cm^{-2}$ observed at $R\lesssim0.5\rvir$ suggest gas pressures of $\nH T \approx 20\cm^{-3}\K$ (\S\ref{s:lowP}).
This difference suggests a very steep dependence of pressure on distance ($\nH T \sim R^{-7}$), which indicates the presence of a shock at $R_{\rm shock}\approx0.5\rvir$. The low pressure scenario therefore implies that the accretion shock is located well within $\rvir$. 
Which physical conditions would place a shock at this radius? 

If the cooling time of virially-shocked heated gas is short compared to the dynamical time, than the virial shock would be unstable (e.g.\ \citealt{BirnboimDekel03}). In this regime, the accreting gas would potentially shock when it converges against an outflow from the galaxy, and hence the properties of the shock would depend on the properties of galaxy outflows. This effect was demonstrated by
\cite{Fielding+17}, who calculated the physical conditions in the CGM using idealized hydrodynamic simulations including galaxy outflows. They identified a threshold halo mass of $\Mthres\sim10^{11.5}\msun$ for their assumed CGM metallicity of $\zsun/3$, below which the virial shock is unstable and the location of the shock depends on outflow parameters. Specifically, they demonstrated that below this mass threshold the shock can occur at a radius significantly less than $\rvir$ (see their fig.~2), as suggested for the shock in the low-pressure scenario discussed here. 

Can the halo masses of blue \sLstar\ galaxies be below the threshold mass $\Mthres$ for a stable virial shock? The median $\Mhalo=6\cdot10^{11}\msun$ deduced above for the COS-Halos+J15 sample is above the threshold predicted by \cite{Fielding+17} for $Z=\zsun/3$, though not by a large factor. Hence, the conclusion on virial shock stability around \sLstar\ galaxies may depend on CGM parameters. To gain analytic insight on how $\Mthres$ scales with CGM parameters, we follow the analytic derivation of $\Mthres$ in \cite{DekelBirnboim06}, which deduced that
\begin{equation}\label{e:stability crit}
 t_{\rm cool}(\Mthres,\Rshock) = A\times \frac{\Rshock}{\vinf(\Mthres,\Rshock)}~,
\end{equation}
where $t_{\rm cool}$ is the cooling time, $\Rshock$ is the shock radius, and $\vinf$ is the infall velocity. The dimensionless pre-factor $A$ depends on the shock velocity relative to the inflow velocity, and has a value of $\approx6$ for a shock velocity which is $\ll\vinf$. The cooling time is equal to (see eqn.~\ref{e:tcool lowP})
\begin{equation}
  t_{\rm cool} = \frac{2.2\cdot\frac{3}{2}k\Tshock}{\nHhot\Lambda(\nHhot,\Tshock,Z)}  ~,
\end{equation}
where $\nHhot$ and $\Tshock$ are the density and temperature of the hot post-shock gas. For $\Tshock\sim\Tvir\approx5\cdot10^5\K$, $Z\sim\zsun/3$, and $\nHhot\sim10^{-4}\cm^{-3}$, the cooling time scales as 
\begin{equation}\label{e:tcool prop}
t_{\rm cool} \propto Z^{-0.8}\nHhot^{-1.1}\Tshock^{1.5}~,
\end{equation}
where we derive these scalings using the \cloudy\ calculation of $\Lambda$ mentioned above. 
Now, using the scalings
\begin{equation}
 T_{\rm h}\propto \Mhalo^{2/3}\left(\frac{\vinf}{\vvir}\right)^2 ~~,~~ 
  \frac{\Rshock}{\vinf} \propto \frac{\Rshock/\rvir}{\vinf/\vvir}
\end{equation}
in equations~(\ref{e:stability crit}) and (\ref{e:tcool prop}),
we get a threshold mass that scales as
\begin{equation}\label{e:Mthres}
 \Mthres \propto Z^{0.8} n_{\rm H,hot}^{1.1}\left(\frac{\vinf}{\vvir}\right)^{-4}\frac{\Rshock}{\rvir} ~.
\end{equation}
Equation~(\ref{e:Mthres}) demonstrates that at a given $\Rshock/\rvir$, the threshold mass for a stable virial shock depends roughly linearly on CGM metallicity and post-shock gas density, and to the fourth power on the ratio of the inflow velocity to $\vvir$. 
We now discuss how these scalings affect the threshold halo mass relevant to blue \sLstar\ galaxies.

\subsubsection{What is the CGM metallicity?}

Our analysis above of the \ovip\ metallicity suggests $Z/\zsun\sim 0.3-1$ (Fig.~\ref{f:PIE_allConstraints}). 
Photoionization modeling which includes the low-ions suggests either a median $Z \approx 0.6\zsun$ with a dispersion of $0.3\dex$, or a median $Z\approx 0.3\zsun$ with a larger dispersion of $1\dex$, depending on the modelling method (\S\ref{s:Z}). Hence, if the actual CGM metallicity is at the high end of the range deduced from absorption features, the analysis of \cite{Fielding+17} for $Z=\zsun/3$ may underestimate the threshold mass, and the virial shock in $6\cdot10^{11}\msun$ halos could be unstable. However, if the CGM has a range of metallicities, absorption features will generally be skewed towards the high metallicity phases. This bias holds for both the metal absorption lines and the \hi\ absorption, since low-metallicity gas cools less efficiently, and hence will have a lower \hi\ fraction due to its higher temperature. Therefore, based on absorption features alone one cannot rule out the existence of low-metallicity gas which has not been detected. Since such low-metallicity gas would decrease $\Mthres$ below the halo masses of \sLstar\ galaxies, the low-pressure scenario suggests that this phase is indeed absent, rather than that is has merely avoided detection. 

Enriching a CGM with mass $10^{11}\msun$ (the baryon budget of a $6\cdot10^{11}\msun$ halo) to e.g.\ $Z=\zsun/2$ requires $\approx10^9\msun$ of metals. This metal mass is comparable to estimates of the total metal mass produced by the central galaxy, of which only $10-40\%$ are observed in the galaxy itself (\citealt{Peeples+14}). It is hence possible that most of the metals produced by the galaxy reside in the CGM, and therefore that most of the CGM is enriched to $\approx\zsun/2$, similar to the metallicity deduced for CGM absorbers. 
Such a high level of enrichment is also consistent with phenomenological models of the hot CGM gas around the Milky-Way based on \oviip\ and \oviiip\ emission and absorption features (\citealt{MillerBregman13, MillerBregman15, Miller+16, Faerman+17}), though additional modelling is required to test whether such models are consistent with an accretion shock at $\approx0.5\rvir$. 
Also, such a highly-enriched CGM may cool and increase the SFR above observed values. Additional modeling is required to determine whether such a highly enriched CGM is consistent with constraints on galaxy evolution. 

\subsubsection{How does the inflow velocity compare to $\vvir$?}\label{s:vinf}

Equation.~(\ref{e:Mthres}) demonstrates that $\Mthres$ is highly sensitive to the velocity of the supersonic IGM inflows. At a given halo mass and distance from the galaxy, the inflow velocity depends on the shape of the gravitational potential (which is sensitive to the distribution of neighboring halos), on the inflow trajectory within the gravitational potential, and on whether the potential energy of the flows has been converted into kinetic energy, or rather radiated away by small shocks within the flow which subsequently cool. The velocity of intergalactic inflows is a subject of active research, albeit mainly at $z\gtrsim2$ (\citealt{FaucherGiguere+10,RosdahlBlaizot12,WetzelNagai15, GoerdtCeverino15,Nelson+16,Mandelker+16}). 
The exact deduced inflow velocities can depend on the numerical method used. Moreover, the low-redshift metal-enriched inflows envisioned have properties different from the high-redshift cosmological inflows that have been the focus of most simulation analyses. Given the strong dependence of $\Mthres$ on $\vinf$, the uncertainty in $\vinf$ can strongly affect the stability of virial shocks around low redshift \sLstar\ galaxies.

\subsection{The origin of the flow traced by \ovip\ in the low-pressure scenario}\label{s:flow properties}

In the low-pressure scenario the \ovip\ gas is too cool to be supported by thermal pressure, and is located outside the accretion shock. Since gas outside the accretion shock is expected to be predominantly inflowing, the \ovip\ gas most likely traces pre-shock infall.
The characteristic radial velocity of this inflow $\vr$ can be estimated from the velocity centroids of the \ovip\ absorption profiles, which are typically offset by $\approx100\kms$ from the galaxy velocity (see \citealt{Werk+16}). In the limit that the velocity field is purely radial and that all the \ovip\ gas resides at $R\sim\Rovi=0.6\rvir$, we expect $\vr$ to equal
\begin{equation}
 \vr \approx \frac{\left|v_{\rm los}\right|}{\sqrt{1-\left(\frac{\Rimp}{0.6\rvir}\right)^2}}~,
\end{equation}
where $v_{\rm los}$ is the observed velocity centroid relative to the galaxy velocity. 
Using the component fit to the \ovip\ absorption features from \cite{Werk+13}, and weighting each component by its $\Novi$, we derive an \ovip-weighted average of $|\vr|=91\kms$. Assuming instead $\vr \sim |v_{\rm los}|$ gives a similar $\vr=84\kms$, since most of the \ovip\ absorption in the sample is observed at $\Rimp^2\ll (0.6\rvir)^2$. 

The derived $|\vr|\approx100\kms$ implies a mass inflow rate of
\begin{eqnarray}\label{e:Mff}
 {\dot M}_{\rm in} &\sim& \frac{\Mgasovi \vr}{\Delta\Rovi} = \frac{\left[0.006\left(\frac{\fovi Z}{\zsun}\right)^{-1}\Movi\right] \vr}{0.4\rvir} \nonumber \\
	  &=& 5.4\,v_{100}\left(\frac{Z}{\frac{1}{3}\zsun}\right)^{-1}\left(\frac{\fovi}{0.2}\right)^{-1}\msun\yr^{-1}~,
\end{eqnarray}
where $|\vr|\equiv100\,v_{100}\kms$, $\Delta\Rovi\approx0.4\rvir$ is the $16-84$ percentile radius range of the \ovip-gas  (Fig.~\ref{f:ovi observations}), the oxygen mass fraction is $0.006(Z/\zsun)$, and in the last equality we use $\rvir=190\kpc$ and $\Movi=1.8\cdot10^6\msun$ deduced for the median halo mass in the COS-Halos+J15 sample (\S\ref{s:ovi}). This mass flow rate is similar to the mean SFR of $4.2\msun/\yr$ in the galaxies of the sample. This quantitative correspondence between the deduced $M_{\rm in}$ in the outer halo and the SFR may suggest that they are physically connected, i.e.\ that the inflow traced by \ovip\ supplies the necessary fuel for star formation in steady state where ${\dot M}_{\rm in} \sim {\rm SFR}$. This connection is further supported by the absence of the \ovip-flow around red galaxies (\citealt{Tumlinson+11}, see further discussion below).

If \ovip\ traces pre-shock infall then we expect its ram pressure to be comparable to the thermal pressure of the post-shock gas\footnote{As this discussion applies to the regime where the virial shock is unstable (\S\ref{s:accretion shock}), the temperature and other properties of the post-shock gas depend on the properties of galaxy feedback (e.g.\ \citealt{Fielding+17}). In future work we will further explore the properties of the accretion shock in the context of the low-pressure scenario.}. 
We can calculate the ram pressure $P_{\rm ram}$ of the \ovip-gas by averaging the mass flow rate calculated in eqn.~(\ref{e:Mff}) over the surface area of the shock:
\begin{eqnarray}\label{e:Finfall}
& & \frac{P_{\rm ram}}{k} = \frac{{\dot M}_{\rm in} \vr}{4\pi \Rshock^2 k} = \nonumber \\
 & &  22\, v^2_{100}\left(\frac{Z}{\frac{1}{3}\zsun}\right)^{-1}\left(\frac{\fovi}{0.2}\right)^{-1} \left(\frac{\Rshock}{90\kpc}\right)^{-2} \cm^{-3}\K ~. \nonumber\\
\end{eqnarray}
where we assumed a shock radius of $\Rshock\approx0.5\rvir\approx90\kpc$, as deduced in \S\ref{s:lowP}. The thermal pressure of the gas within the shock can be estimated from the low-ion clouds observed at $R\lesssim\Rshock$, which are plausibly pressure confined by the hot ambient medium within the shock. Photoionization modelling of these clouds suggests $\nH \approx 2\cdot10^{-3}\cm^{-3}$ (\S\ref{s:lowP}), which implies a thermal pressure of $P/k\approx2.2\nH T\approx 40\cm^{-3}\K$, within a factor of two of the value in eqn.~(\ref{e:Finfall}). 
As both derived pressures are uncertain to a factor of at least two, this similarity is consistent with the ram pressure of the \ovip\ gas being equal to the thermal pressure of the gas within the shock, and hence supports the assumption that the \ovip-gas is pre-shocked infall. 

On the other hand, the deduced inflow velocities of $\vr\approx100\kms$ are significantly lower than the velocities of $\approx175\kms$ expected for gas at $\Rovi=0.6\rvir$ which had free-falled from $2\rvir$ -- the `turnaround radius' 
of the spherical collapse model for cosmological structure formation. This indicates that, for the pre-shock infall interpretation to be consistent with the observed \ovip\ velocities, either the inflows are radiating away a large fraction of their potential energy during infall, or alternatively that the spherical collapse model is not an accurate approximation of inflow trajectories.

The inflow interpretation for \ovip\ implies that the region around galaxies is enriched to the deduced $Z \gtrsim 0.3\zsun$ of the \ovip\ gas. Such a level of enrichment could have thus far eluded detection, since the cool gas beyond $\rvir$ would be highly ionized due to its low density of $\lesssim10^{-5}\cm^{-2}$. This high ionization level implies that the gas enrichment would be evident only in absorption features such as $\oviip$, $\oviiip$ and $\neviiip$, which are relatively hard to detect (see next section). The origin of this enrichment may be in outflows in a past epoch when the SFR densities were larger and the halo potential wells of galaxies were shallower. 
Cosmological simulations have also found that \ovip\ absorbers originate in such `ancient outflows', although with different physical properties suggesting collisionally ionized \ovip\ (e.g., \citealt{Ford+16} and \citealt{Oppenheimer+16}).

\subsection{Predictions}\label{s:predictions}

How can we test the low-pressure \ovip\ scenario with additional observations?
In this section we first discuss potential observables of the accretion shock itself, and then discuss observables of low-pressure gas beyond the shock.

\begin{figure}
 \includegraphics[width=0.48\textwidth]{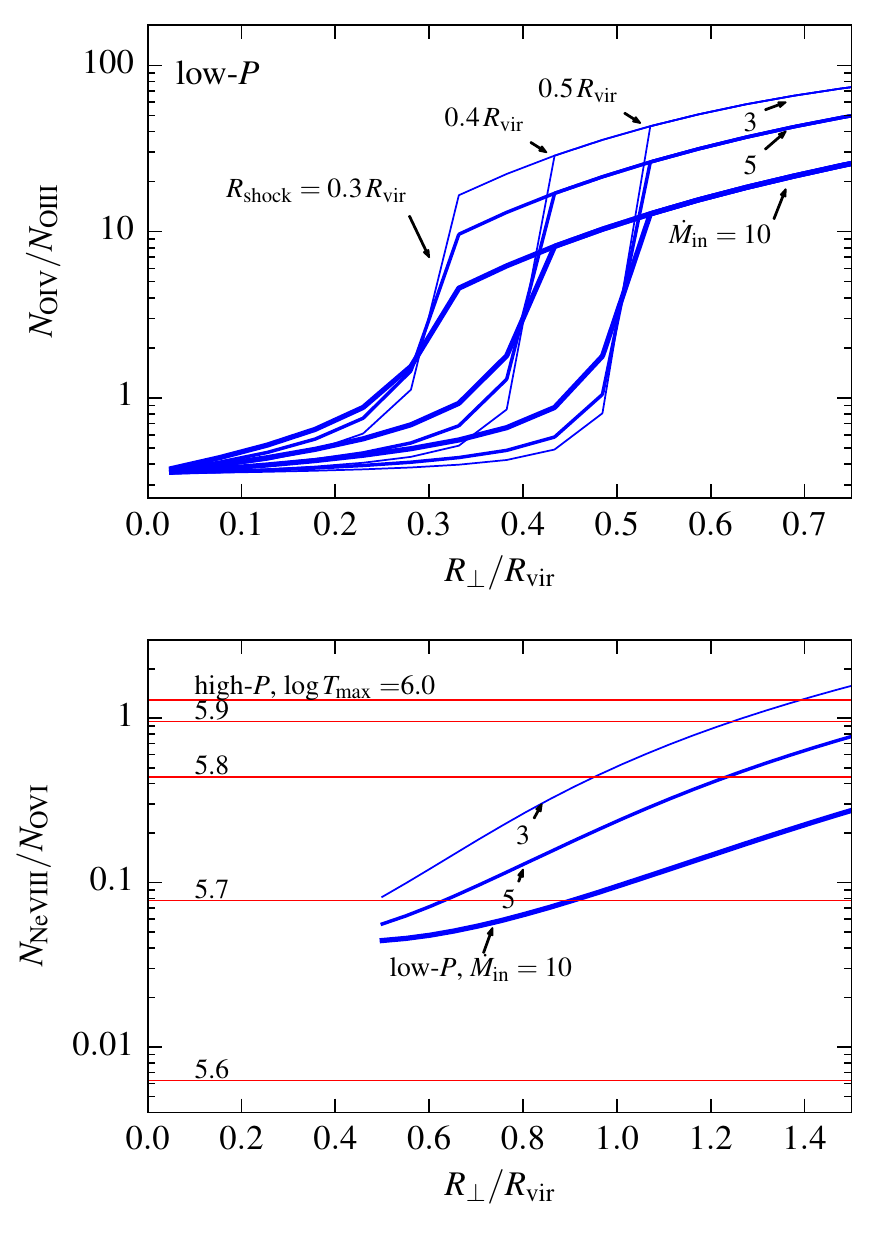}
\caption{
Predictions of the two scenarios discussed in this work for EUV absorption lines. 
\textbf{(Top)} Blue lines plot the predicted $N_\oivp/N_\oiiip$ in the low-pressure scenario, as a function of the shock radius and mass inflow rate (noted in $\msun\yr^{-1}$). The calculation assumes a constant inflow velocity of $100\kms$ and ionization by the UV background at $z=0.2$. The low-pressure scenario predicts an abrupt change in $N_\oivp/N_\oiiip$ and other line ratios at the shock radius, due to the pressure jump across the shock. This feature is potentially detectable as a bimodality in the line ratios distributions. 
\textbf{(Bottom)} The predicted $N_\neviiip/N_\ovip$ in both scenarios. In the low-pressure scenario (blue) the ionization level of the gas increases outward beyond the shock radius. In the high-pressure scenario (red horizontal lines, based on MQW17), $N_\neviiip/N_\ovip$ is mainly determined by the maximum temperature of the cooling flow, as noted in the figure. 
Note that column ratios of $N_\neviiip/N_\ovip\approx0.1$, which are expected in the low-pressure scenario at $\gtrsim0.5\rvir$, are expected only for a narrow range in $T_{\rm max}$ in the high-pressure scenario.
}
\label{f:prediction}
\end{figure}

\subsubsection{The accretion shock}\label{s:predictions virial shock}
\newcommand{\Mdotin}{{\dot M}_{\rm in}}

As mentioned in \S\ref{s:lowP}, photoionization modeling of cool gas immediately inside the inferred shock radius suggests $\nH \approx2\cdot10^{-3}\cm^{-3}$, significantly higher than the $n_{\rm H}\lesssim5\cdot 10^{-5}\cm^{-3}$ deduced for the \ovip-gas outside the shock radius (Fig.~\ref{f:PIE_allConstraints}). The low-pressure \ovip\ scenario hence implies that cool gas with intermediate densities of $10^{-4}\lesssim\nH \lesssim 10^{-3}\cm^{-3}$ is `missing', due to the discontinuous pressure profile across the shock. 
A gap in gas densities is expected to be evident as a bi-modality in line ratios at sightlines with $\Rimp\sim R_{\rm shock}$, corresponding to lines of sight which cross immediately within or immediately outside the shock radius. 
An example of this predicted gap is shown in the top panel of Figure~\ref{f:prediction}, where we plot the expected $N_{\oivp}/N_{\oiiip}$ in the low-pressure model as a function of impact parameter. The predicted line ratios are calculated using a simple constant-velocity inflow model, where the inflow is in thermal and ionization equilibrium with the UV background. We assume a constant mass inflow rate (i.e.\ $\nH\propto R^{-2}$) in the range $\Mdotin=3-10\msun/\yr$, and a volume filling factor of $1/3$ outside the shock radius. For $\Mdotin=5\msun/\yr$, these parameters reproduce the deduced gas density of $\nH=10^{-4.5}\cm^{-3}$ at $R=\Rovi$. Changing the assumed volume filling factor has the same effect on the line ratios as changing the assumed $\Mdotin$. At the shock radius we assume isothermal shock conditions (e.g.\ \citealt{LamersCassinelli99}), which implies that the cool gas pressure and density increase by a factor of $\scriptM^2$ ($\scriptM$ is the Mach number) at the shock, while the volume filling factor decreases by the same factor. Such conditions are expected in the low-pressure scenario where the cooling time of the post-shocked gas is short.  The ion fractions at each distance are then calculated with \cloudy, and summed along the line of sight to produce the predicted column ratios at a given impact parameter. We neglect absorption due to collisionally-ionized gas cooling from the hot phase within the shock radius, and due to galaxy outflows. In future work we will test the predictions plotted in Fig.~\ref{f:prediction} using full hydrodynamic models.

The top panel of Fig.~\ref{f:prediction} shows that the line ratios differ by a factor of $3-30$ immediately inside and immediately outside the shock. A direct detection of this gap in line ratios, independent of any photoionization modeling, would provide support to the low-pressure model in which the accretion shock is at $R\lesssim0.5\rvir$. Moreover, in all plotted models $N_\oivp/N_\oiiip>3$ outside the shock radius and $N_\oivp/N_\oiiip<1$ inside the shock radius, suggesting that a gap in line ratios may be detectable even in the likely presence of variance in inflow rate and accretion shock radius.

A bi-modality in line ratios would be best detectable with transitions of abundant ions that are expected to be produced in observable quantities both outside and inside the shock. For the flow parameters estimated for the COS-Halos sample, these include the `intermediate ions' $\oivp~\lambda788$, $\oiiip~\lambda833$, $\ciiip~\lambda977$, $\Siiiip~\lambda1206$, and $\civp~\lambda1550$, and also the \hi\ Lyman-series. Is this expected bi-modality observed in the COS-Halos+J15 sample?
In sightlines with $0.3\rvir<\Rimp<0.5\rvir$, six objects have a column ratio of $N_\Siiiip/\Novi\approx0.02$ (including one upper limit and one lower limit), while the other five objects have a column ratio of $N_\Siiiip/\Novi = 0.05-0.2$, four of which are lower limits (see $N_\Siiiip$ plotted in Appendix~\ref{a:low-ions}). There hence may be a gap in intermediate values of $N_\Siiiip/\Novi=0.02-0.05$, as expected in the above inflow models, though the sample size is not large enough to test its significance.
The column ratio $N_\ciiip/\Novi$ is measurable in only four objects with $0.3<\Rimp/\rvir<0.5$ in the sample, in all of which \ciiip\ is saturated.
We advocate for additional observations of $\ciiip$ (observable with COS at $z>0.2$), $\civp$, $\Siiiip$, $\oiiip$ ($z>0.4$), and $\oivp$ ($z>0.5$), to test this prediction. 
A bi-modality may also be detectable in \hi\ absorption features, since the density discontinuity across the shock implies that the \hi\ fraction is also discontinuous. Observations of Ly$\zeta$ (observable at $z>0.25$) and higher-series Lyman lines which are not saturated at the characteristic \hi\ columns of $10^{16}\cm^{-2}$ near $0.5\rvir$ (Fig.~\ref{f:hi2ovi}) would allow to detect or rule out such a bi-modality. 
We note that galaxy-selected samples with a relatively small range in $\Mstar$ and $z$ (such as COS-Halos) are required to detect the bi-modality, since these parameters will affect $\Mdotin$ and the predicted column ratios, and hence the expected `missing' gas density.

The expected bi-modality in line ratios at $\Rimp\sim R_{\rm shock}$ is a general prediction of the pressure jump across the accretion shock, regardless of the distance of the shock from the galaxy. Thus, a bi-modality in line ratios may be expected also in the high-pressure \ovip\ scenario, albeit at larger impact parameters of $\gtrsim\rvir$. 

The `missing' gas densities of $10^{-4}\lesssim\nH \lesssim 10^{-3}\cm^{-3}$ in the low-pressure scenario could also explain the paucity of \nvp\ detections in the COS-Halos sample (see bottom-left panel in Fig.~\ref{f:PIE}), since in photoionization equilibrium conditions the $\nvp/{\rm N}$ fraction is significant only at these missing densities (see figure~1 in S16). This explanation for the lack of \nvp\ differs from the model of \cite{Bordoloi+17}, in which high ions including \ovip\ and \nvp\ originate in radiatively cooling collisionally ionized gas, and the paucity of \nvp\ is a result of the short cooling time of gas at temperatures in which $\nvp/{\rm N}$ peaks.

An accretion shock at $R=0.5\rvir$ also implies that observables of the hot phase, such as X-ray emission, X-ray absorption of \oviip\ and \oviiip\ lines, and the thermal Sunyaev-Zeldovich effect (see \S\ref{s:other obs}), should exhibit drops in emission / absorption beyond impact parameters of $\approx0.5\rvir$. 
Additional modeling of the hot phase within the shock radius is required to estimate the expected signal.

\subsubsection{Beyond the accretion shock}

In gas photoionized by the UV background, the ionization level scales as $\nH^{-1}$, so ionization is expected to increase outwards. This property predicts that column ratios of high-ionization lines such as $N_{\neviiip}/\Novi$ and $\Novi/N_{\ovp}$, which are sensitive to the gas ionization level outside the shock, should increase with increasing impact parameter. The bottom panel of Fig.~\ref{f:prediction} plots the predicted $N_{\neviiip}/\Novi$ in the inflow models discussed above (blue lines). We plot the predicted line ratios only outside the assumed shock radius of $0.5\rvir$, since within the shock radius these ions are not produced in the cool phase due to its relatively high density, but might be produced via collisional ionization in the hot phase.

For comparison, we also plot in the bottom panel the expected $N_\neviiip/\Novi$ in the high-pressure scenario where \ovip\ traces collisionally-ionized gas in a cooling flow (red horizontal lines). For isobaric cooling, the predicted line ratio in this scenario depends mainly on the maximum temperature of the cooling flow $T_{\rm max}$ (see MQW17 and \citealt{Bordoloi+17}). The plotted ratios are based on the calculation described in MQW17 for the characteristic gas pressure of $\nH T=30\cm^{-3}\K$ derived above (M.~McQuinn, private communication). 
Note that ratios of $N_\neviiip/N_\ovip\approx0.1$, which are expected in the low-$P$ scenario at $\gtrsim0.5\rvir$, are expected only for a narrow range in $T_{\rm max}$ in the high-$P$ scenario. Also note that if $T_{\rm max}$ decreases outwards as expected in the CGM, the expected trend of $N_\neviiip/\Novi$ with impact parameter in the high-pressure scenario is opposite to the expected trend in the low-pressure scenario. However, at sufficiently large distances the densities would be low enough such that photoionization will become important even in the high-pressure scenario, which might reverse the trend of $N_\neviiip/\Novi$ vs.\ $\Rimp$. Additional modeling is required to check whether the trend of $N_\neviiip/\Novi$ vs.\ $\Rimp$ can be used to discriminate between the two scenarios.


\subsection{Red galaxies}

\cite{Tumlinson+11} demonstrated that sightlines around red \sLstar\ galaxies do not exhibit the strong and ubiquitous \ovip\ absorption observed around blue galaxies at similar luminosity and redshift. What is the source of this difference in the context of the low pressure \ovip\ scenario? Possibly, the accretion shock is further out in the halos of red galaxies compared to the halos of blue galaxies. If the shock around red galaxies is beyond $\rvir$, then the implied high pressures in the outer halo would suppress the low-pressure phase which gives rise to strong \ovip\ in the low-pressure scenario. Photoionized \ovip\ would then only be produced beyond the shock, where the lower densities and gas columns would create weaker \ovip\ absorption.

Why would red galaxies have an accretion shock further out than blue galaxies? 
Red galaxies in the COS-Halos sample are somewhat more massive than blue galaxies in the sample (\citealt{Tumlinson+11, Oppenheimer+16}), which suggests a larger halo mass. 
A larger halo mass implies that it is more likely to be above the threshold mass for a stable virial shock, in which case the shock is expected to be at $\gtrsim\rvir$.
In addition, the halos of red galaxies may be subject to stronger and more effective feedback from supermassive black holes (e.g., \citealt{Suresh+17,MathewsProchaska17}), since their central black holes are more than order of magnitude more massive (e.g.\ \citealt{ReinesVolonteri15}). Since in halos with an unstable virial shock the location of the accretion shock depends on feedback parameters (e.g.\ \citealt{Fielding+17}), stronger feedback in red galaxies may drive the accretion shock outward relative to blue galaxies, even for the same halo mass. We note that the effects of quasar feedback on surrounding gas can also be constrained directly from observations (e.g., \citealt{Liu+13, Stern+16b}).

\subsection{Dwarf galaxies}

\cite{Johnson+17} detected \ovip\ absorption in the CGM of $z\sim0.2$ star-forming dwarf galaxies with $\Mstar=10^8-10^9\msun$. They find an $\Novi$ vs.\ $\Rimp$ relation similar in shape to the relation around \sLstar\ galaxies shown in Fig.~\ref{f:ovi observations}, with a drop in the \ovip\ detection rate at impact parameters beyond $\rvir$. This drop indicates that the \ovip\ gas resides at physical distances $\lesssim\rvir$, similar to \ovip\ around \sLstar\ galaxies.

The low halo masses of the dwarf galaxies in the \cite{Johnson+17} sample  ($\lesssim10^{11}\Mhalo$) suggest they are below the threshold for a stable virial shock. It is hence likely that in these galaxies the accretion shock is also at $<0.5\rvir$, as inferred for \sLstar\ galaxies in the low-pressure scenario. 
Is this scenario consistent with the $\Novi$ profile observed by \citeauthor{Johnson+17}?
We showed above that \ovip\ around \sLstar\ galaxies is observed at a distance with an overdensity of $\delta\sim200$, where $\fovi$ peaks in gas photoionized by the UV background. Since the overdensity is roughly independent of $\Mhalo$ at a given $R/\rvir$, we expect $\fovi$ to peak at the same $R/\rvir$ also in the halos of dwarf galaxies, consistent with the similar shapes of the $\Novi$ vs.\ $\Rimp/\rvir$ relations in the two samples. 

\subsection{Comparison with \cite{Stern+16}}

S16 modeled the observed ion columns in COS-Halos with multi-density PIE models, under two assumptions: 
(1) that different galaxy halos in the COS-Halos sample have a similar relation between gas density and gas column, 
and (2) that dense gas which produces low ions and large \hi\ columns is located within low-density gas which produces higher ionization ions and relatively small \hi\ columns. 
They derived the following relation between gas column and gas density:
\begin{equation}\label{e:AMD}
 \frac{{\rm d}\NH}{{\rm d}\log \nH} = 3.6\times10^{18}\left(\frac{\nH}{10^{-4.5}\cm^{-2}}\right)^{0.05}\cm^{-2}~.
\end{equation}
S16 also derived a median metallicity of $Z=0.6\zsun$, with a dispersion of $0.3\dex$ between different objects (Fig.~\ref{f:PIE_allConstraints}).  

The assumption of S16 that dense gas is embedded in low-density gas can be interpreted either in a `local' density gradient picture or a `global' density gradient picture.
In the local picture (which S16 focuses on), the distribution of gas densities is independent of distance from the galaxy, and the low-ions form in dense pockets in larger low-density clouds which produce high-ions. This picture would be expected for example for self-gravitating clouds. 
In the alternative `global' picture, dense gas which produces low-ions resides at smaller distances from the galaxy than low-density gas which produces high-ions. The analysis in Figs.~\ref{f:ovi observations}--\ref{f:hi2ovi} suggests that \ovip\ is indeed located at larger distances than the large \hi\ columns and low ions, and hence this global picture seems preferred, at least for the relation between \ovip\ and the low-ions. We hence utilize only the results of S16 which are independent of the assumption of a local density gradient, namely eqn.~(\ref{e:AMD}) and the metallicity distribution. 

We now compare eqn.~(\ref{e:AMD}) deduced by S16 with the single-density PIE modeling of gas at $R\gtrsim\Rovi$ done above in the context of the low-pressure scenario. 
The hydrogen column in the single-density model is:
\begin{eqnarray}
 \NH &=& \left(5\cdot10^{-4}\frac{Z}{\zsun}\right)^{-1}\frac{\Novi}{\fovi} \nonumber\\
     &=& 9.5 \times10^{18}\frac{\Novi}{10^{14.5}\cm^{-2}}\left(\frac{\fovi}{0.2}\right)^{-1}\left(\frac{Z}{\frac{1}{3}\zsun}\right)^{-1} \cm^{-2} ~. \nonumber\\
\end{eqnarray}
which is comparable to the characteristic $\NH=3.6\cdot 10^{18}\cm^{-2}$ in eqn.~(\ref{e:AMD}), especially if the metallicity is somewhat higher than $\zsun/3$ as suggested by the S16 analysis. A similar characteristic $\NH$ was also deduced by \cite{Prochaska+04} for \ovip\ absorbers along the sightline to PKS~0405--123.

Equation~(\ref{e:AMD}) also shows that the column density distribution $\d\NH/\d\log\nH$ found by S16 is nearly flat (same total column per $\log\nH$). How does this flat distribution relate to the cool inflow interpretation? If the \ovip\ and low-ions trace inflowing gas with a constant mass flow rate and velocity, then $\nH\propto R^{-2}$ and hence $\NH\propto\Rimp^{-1}$. In this scenario, one would expect ${\rm d} \NH/{\rm d} \log \nH\propto \nH^{1/2}$, in contrast with the power-law index of $0.05$ found by S16. 
This difference might be explained by the `missing' gas densities $10^{-4}\lesssim\nH\lesssim10^{-3}\cm^{-3}$ implied by the accretion shock in the low-pressure scenario, in which the density is discontinuous at the shock (see \S\ref{s:predictions virial shock}). Such a density jump implies a relation between $\nH$ and $R$ which is steeper than $R^{-2}$, and hence a flatter ${\rm d} \NH/{\rm d} \log \nH$. S16 assumed a continuous power-law distribution in density, so they could not detect a jump over specific densities. Indeed, the S16 model predicts \nvp\ columns above the observed upper limits in many objects (see their fig.~4). The \nvp\ fraction peaks at the `missing' gas density (see fig.~1 in S16), and hence removing this phase from the S16 model would make the model more consistent with the $\nvp$ observations.

\subsection{Uncertainties in the ionizing spectrum}

How would our conclusions for the low pressure scenario change for a different assumed ionizing spectrum?
Most of the constraints shown in Fig.~\ref{f:PIE_allConstraints} depend on $\nH$ and the background intensity $J_\nu$ only through their ratio, which is proportional to the ionization parameter.  The deduced $\nH$ in these constraints would hence change proportionally with the assumed $J_\nu$.
The exceptions to this rule are the constraints based on the pathlength $\lovi$, which include the requirement that $\lovi<2\Rovi$ (black line with lower limits in Fig.~\ref{f:PIE_allConstraints}), and the constraint based on the observed $\Novi$ vs.\ $\bovi$ relation (green stripe). Since $\lovi$ is proportional to $\novi^{-1}$, and $\novi\propto\nH Z\fovi(J_\nu/\nH)$ (see eqn.~\ref{e:novi}), then keeping $\lovi$ and $J_\nu/\nH$ constant requires $Z\propto \nH^{-1}$, or equivalently $Z\propto J_\nu^{-1}$. Thus, a different assumed $J_\nu$ will change the density and metallicity implied by the $\Novi$ vs.\ $\bovi$ constraint proportionally to $J_\nu$ and $J_\nu^{-1}$, respectively. 
Fig.~\ref{f:PIE_allConstraints} suggests that the location of the intersection of all (non-dust) constraints will change in a similar fashion, i.e.\ the implied gas density in the low-pressure scenario is proportional to the assumed background, while the implied metallicity is inversely proportional to it. If $J_\nu(1\ryd)$ is more than three times weaker than the assumed value of $3\cdot10^{-23}\erg\s^{-1}\cm^{-2}\Hz^{-1}\,{\rm sr}^{-1}$, no region in parameter space would satisfy all non-dust constraints, which would be a challenge for the low-pressure scenario.

\subsection{The high-pressure \ovip\ scenario}

In Figure~\ref{f:nTplots HI} we demonstrated that the amount of \hi\ in the $T=10^{5.5}\K$ gas traced by \ovip\ in the high pressure scenario is insufficient to explain the observed column ratios of $\NHI/\Novi=1-10$ at $\Rimp\gtrsim\Rovi$ (Fig.~\ref{f:hi2ovi}). The high pressure scenario hence requires a cooler gas phase to explain the observed \hi. A similar conclusion holds for the observed \ciiip\ (\S\ref{s:other ions}). In principle, these two phases could be unassociated with one another, for example if \hi\ originates in gas that is more distant than \ovip. This possibility appears unlikely, for two reasons. First, the typical $\NHI$ at impact parameter $\approx\Rovi$ is $\sim10^{15}\cm^{-2}$, an order of magnitude larger than at $\Rimp\sim\rvir$ (top panel of Fig.~\ref{f:hi2ovi}), suggesting that most of the \hi\ observed at $\Rimp\approx\Rovi$ originates at a 3D distance $R\approx\Rovi$, similar to the 3D distance of the \ovip-gas. Second, \hi\ and \ovip\ are correlated in velocity space (\S\ref{s:hi res}), which also argues for a physical connection.  

A more likely possibility in the context of the high pressure scenario is that \hi\ and \ciiip\ trace cooler phases of the same cooling flow traced by \ovip, hence explaining both the spatial and kinematic alignments of these phases with \ovip. While this seems plausible, the observed $\NHI$ is {orders of magnitude lower than naively expected in the \ovip-traced cooling flow. A cooling flow would result in $M_{\rm cool} \approx t_{\rm dyn}\Mdot_{\rm cooling}\sim3\cdot10^{10}\msun$ of cool gas at the same location as \ovip, where $\Mdot_{\rm cooling}\sim 30\,(\Novi/10^{14.5}\cm^{-2})\msun\yr^{-1}$ is the mass flow rate of the cooling flow traced by \ovip\ (see MQW17), and $t_{\rm dyn}\approx10^9\yr$ is the dynamical time required for the cool gas to traverse a significant distance in the halo. This mass of cool gas would yield a characteristic \hi\ column of $\NHI\sim3\cdot10^{10}\msun\cdot \fhi/ (\pi\Rovi^2\mp)=7\cdot10^{17}(\fhi/0.01)\cm^{-2}$, where $\fhi\approx0.01$ is the \hi\ fraction expected for $T\approx10^4\K$ gas with $\nH T\approx30\cm^{-3}\K$ (the value at the intersection of the two red dashed lines in the left panel of Fig.~\ref{f:nTplots HI}). This expected \hi\ column suggests $\NHI/\Novi\approx2000$, almost three orders of magnitude larger than the typical values observed at $\Rimp\gtrsim\Rovi$. The cooled gas therefore must be heated up (or mixed) on a timescale significantly shorter than $t_{\rm dyn}$ to avoid producing large \hi\ columns, in the right amount to produce the observed $\NHI/\Novi\sim1-10$.

Furthermore, if the $\NHI/\Novi\approx1-10$ observed in the outer halos of \sLstar\ galaxies is determined by the hot gas cooling rate and the cool gas mixing rate as suggested above, it is somewhat surprising that \ovip\ absorbers along random sightlines show similar values of $\NHI/\Novi$ (bottom panel of Fig.~\ref{f:hi2ovi}), since the mentioned timescales are likely different in these systems.

\section{Summary}\label{s:summary}

In this work we analyzed the physical properties of gas traced by \ovip\ absorption around low-redshift, \sLstar, star-forming galaxies. We demonstrated that the characteristic $\opv$-galaxy distance is $\Rovi\approx0.6\rvir$, and that gas at this distance also exhibits  $\NHI/\Novi\sim 1-10$, $\ebv\approx10^{-3}\,{\rm mag}$, weak  $N_\ciiip\approx10^{13.5}\cm^{-2}$ and $N_\Siiiip \lesssim10^{12.5}\cm^{-2}$ absorption, and a lack of other low-ion absorption. Larger $\NHI/\Novi$ of $\gtrsim100$ and stronger low-ion absorption are only seen at $\Rimp\lesssim0.5\rvir$, suggesting that the gas which produces \hi\ columns of $\gtrsim 10^{16.5}\cm^{-2}$ and low-ion absorption features resides at smaller physical radii than the \ovip\ gas.  

We compared these observations with two distinct physical scenarios.
In the `high-pressure \ovip' scenario, the accretion shock is at $\gtrsim\rvir$, and \ovip\ traces collisionaly ionized gas within the shock radius with a characteristic pressure of $\nH T\approx30\cm^{-3}\K$.
In the alternative `low-pressure \ovip' scenario, the accretion shock is at $\lesssim0.5\rvir$, and \ovip\ traces gas beyond the shock radius, which is in ionization and thermal equilibrium with the UV background. The gas pressure that maximizes the \ovip\ fraction in such conditions is $\nH T\sim 1\cm^{-3}\K$. 
The high-pressure scenario is favored by current cosmological simulations, though a low-pressure scenario is possible if the cooling time of the postshock gas is shorter than in simulations, e.g.\ due to a higher metallicity or lower preshock inflow velocity (see \S\ref{s:accretion shock}). Such a scenario may be realized if low redshift inflows originate in outflows from a previous epoch that are currently falling back onto the central galaxy. 

Using simplified analytic derivations, \cloudy\ numerical calculations, and the available observational constraints, we deduce the following:
\begin{enumerate}

 \item The low-pressure \ovip\ scenario can explain the observed \ovip, \ciiip, \hi\ and dust absorption at $R\gtrsim\Rovi$ with a single gas phase, provided that the circumgalactic \ovip-gas has a density of $\nH\approx10^{-4.5}\cm^{-3}$, a metallicity of $\zsun/3-\zsun$, and a dust-to-metal ratio similar to that in the ISM. This phase is also consistent with existing upper limits on $N_{\nvp}$ and $N_\Siiiip$. The implied baryon overdensity of $\approx100$ is comparable to the dark-matter overdensity at the location of the \ovip-gas.
 In contrast, the high-pressure \ovip\ scenario requires multiple gas phases to explain the coincident \ovip, \hi\ and \ciiip\ absorption. 

 \item The high-pressure \ovip\ scenario implies that the \ovip-gas is rapidly cooling, at a rate roughly equal to the halo gas thermal energy per Hubble time. A mechanism that efficiently dissipates energy in the outer halo is hence required to avoid collapse. A similar conclusion has also recently been reached by \cite{Faerman+17} and MQW17. This cooling `problem' is mitigated in the low-pressure scenario where the \ovip-gas is in thermal equilibrium with the UV background. 

 \item The expected density of $\opv$ ions in both scenarios is $\novi=1-3\cdot 10^{-9}\cm^{-3}$, suggesting an absorber pathlength of $30-100\kpc$ for the typical $\Novi=10^{14.5}\cm^{-2}$. This pathlength is a significant fraction of the halo size. The observed \ovip\ line widths are consistent with the velocity shear expected within absorbers of this size, provided that their kinematics are dominated by bulk motions.

\item When the halo mass is below the threshold mass for a stable virial shock, the accretion shock may be located at $\lesssim0.5\rvir$, as assumed in the low-pressure \ovip\ scenario. The median halo mass of $6\cdot10^{11}\msun$ of the COS-Halos+J15 sample (based on the \citealt{Moster+13} relation) is a factor of $\sim2$ higher than the threshold of $10^{11.5}\msun$ deduced by the idealized CGM simulations of \cite{Fielding+17} for a CGM metallicity of $\zsun/3$. The actual threshold mass could however be higher if the CGM metallicity is $>\zsun/3$, 
or alternatively if inflows reach $\rvir$ at a velocity somewhat lower than the virial velocity.
\end{enumerate}

The following conclusions are applicable in the context of the low-pressure scenario, in which \ovip\ is assumed to trace cool gas beyond the accretion shock:

\begin{enumerate}\setcounter{enumi}{4}


 \item The location of the \ovip-gas outside the shock suggests that it is infalling. The centroid offsets of the \ovip\ absorption profiles relative to the galaxy imply a typical infall velocity of $\approx100\kms$.

\item We derive a mass flow rate of $\sim5\msun\yr^{-1}$ for the flow traced by \ovip, comparable to the mean star formation rate of $4.2\msun\yr^{-1}$ in the central galaxies in the sample. This suggests that low-redshift \sLstar\ galaxies form stars at a rate comparable to the large-scale ($>100\kpc$) inflow rate.

\item We derive a ram pressure of $P_{\rm ram}/k\approx20\cm^{-3}\K$ for the flow traced by \ovip, comparable to the thermal pressure within the shock radius estimated from the low-ion absorption observed at $\lesssim 0.5\rvir$. This similarity supports the pre-shock infall interpretation for the \ovip\ gas.

 \item The lack of strong \ovip\ absorption in the halos of red galaxies suggests the accretion shock is farther out in these halos, which would cause the low pressure \ovip\ phase to disappear. The more extended accretion shock around red galaxies could be due to their higher halo masses, which implies the virial shock is stable over a larger range of CGM parameters, or due to different feedback properties (e.g., AGN feedback).

 \item The low-pressure scenario predicts the existence of a bimodality in line ratios at $\Rimp\approx0.5\rvir$,
due to the factor of $\sim40$ difference between the density of $\sim 10^4\K$ gas immediately within and immediate
outside the shock radius. This scenario also predicts that $N_\neviiip/\Novi$ and $\Novi/N_\ovp$ should increase with impact parameter beyond the shock radius, due to the increase in ionization with distance. Both predictions can be tested using COS observations of the CGM at $0.5\lesssim z\lesssim1$.

\end{enumerate}

Although it is often assumed, largely based on theoretical models of galaxy formation, that the \ovip\ observed around low-redshift \sLstar\ galaxies arises in a high-pressure, collisionally ionized medium, we have demonstrated that in many respects CGM observations can instead be explained by \ovip\ originating primarily in a low-pressure, photoionized medium. 
Going forward, it will therefore be important to test in more detail the predictions of the low- and high-pressure scenarios against observations. In particular, our analysis suggests that it would be instructive to compare the $\NHI/\Novi$ ratio and circumgalactic dust extinction predicted by simulations with observations. Moreover, it would be useful to develop more quantitative predictions for the low-pressure scenario (e.g., for the metal-enriched inflows interpretation mentioned above), since this scenario may not be realized in standard simulations.

\acknowledgements
We thank the anonymous referee for a thorough report which significantly improved the paper. We also wish to thank Matthew McQuinn and Hsiao-Wen Chen for reading and sending insightful comments on a draft version of this manuscript. We thank M.~McQuinn also for providing the $N_\neviiip/\Novi$ ratios in the high-pressure scenario. JS thanks J.\ X.\ Prochaska, William Mathews, Amiel Sternberg, Yakov Faerman, Ben Oppenheimer, Avishai Dekel, Frank van den Bosch, and Jose O{\~n}orbe for useful discussions.
JS acknowledges support from the Alexander von Humboldt foundation in the form of the Humboldt
Postdoctoral Fellowship. The Humboldt foundation
is funded by the German Federal Ministry for
Education and Research.
CAFG and ZH were supported by NSF through grants AST-1412836, AST-1517491, AST-1715216, and CAREER award AST-1652522, by NASA through grant NNX15AB22G, by STScI through grants HST-GO-14681.011, HST-GO-14268.022-A, and HST-AR-14293.001-A, and by a Cottrell Scholar Award from the Research Corporation for Science Advancement. 
SDJ is supported by a NASA Hubble Fellowship (HST-HF2-51375.001-A).

\bibliographystyle{apj}

\begin{thebibliography}{100}
\bibitem[Anderson \& Bregman(2010)]{AndersonBregman10} Anderson, M.~E., \& Bregman, J.~N.\ 2010, \apj, 714, 320 
\bibitem[Anderson \& Bregman(2011)]{AndersonBregman11} Anderson, M.~E., \& Bregman, J.~N.\ 2011, \apj, 737, 22 
\bibitem[Anderson et al.(2016)]{Anderson+16} Anderson, M.~E., Churazov, E., \& Bregman, J.~N.\ 2016, \mnras, 455, 227 
\bibitem[Armillotta et al.(2017)]{Armillotta+17} Armillotta, L., Fraternali, F., Werk, J.~K., Prochaska, J.~X., \& Marinacci, F.\ 2017, \mnras, 470, 114 
\bibitem[Asplund et al.(2009)]{Asplund+09} Asplund, M., Grevesse, N., Sauval, A.~J., \& Scott, P.\ 2009, \araa, 47, 481 
\bibitem[Birnboim \& Dekel(2003)]{BirnboimDekel03} Birnboim, Y., \& Dekel, A.\ 2003, \mnras, 345, 349 
\bibitem[Blitz \& Robishaw(2000)]{BlitzRobishaw00} Blitz, L., \& Robishaw, T.\ 2000, \apj, 541, 675 
\bibitem[Bogd{\'a}n et al.(2013a)]{Bogdan+13a} Bogd{\'a}n, {\'A}., Forman, W.~R., Vogelsberger, M., et al.\ 2013a, \apj, 772, 97 
\bibitem[Bogd{\'a}n et al.(2013b)]{Bogdan+13b} Bogd{\'a}n, {\'A}., Forman, W.~R., Kraft, R.~P., \& Jones, C.\ 2013b, \apj, 772, 98 
\bibitem[Bohlin et al.(1978)]{Bohlin+78} Bohlin, R.~C., Savage, B.~D., \& Drake, J.~F.\ 1978, \apj, 224, 132 
\bibitem[Bordoloi et al.(2017)]{Bordoloi+17} Bordoloi, R., Wagner, A.~Y., Heckman, T.~M., \& Norman, C.~A.\ 2017, \apj, 848, 122 
\bibitem[Bregman \& Lloyd-Davies(2007)]{BregmanLloydDavies07} Bregman, J.~N., \& Lloyd-Davies, E.~J.\ 2007, \apj, 669, 990 
\bibitem[Bryan \& Norman(1998)]{BryanNorman98} Bryan, G.~L., \& Norman, M.~L.\ 1998, \apj, 495, 80 
\bibitem[Cen(2013)]{Cen13} Cen, R.\ 2013, \apj, 770, 139 
\bibitem[Chen \& Mulchaey(2009)]{ChenMulchaey09} Chen, H.-W., \& Mulchaey, J.~S.\ 2009, \apj, 701, 1219 
\bibitem[Colgan et al.(2004)]{Colgan+04} Colgan, J., Pindzola, M.~S., \& Badnell, N.~R.\ 2004, \aap, 417, 1183 
\bibitem[Dai et al.(2012)]{Dai+12} Dai, X., Anderson, M.~E., Bregman, J.~N., \& Miller, J.~M.\ 2012, \apj, 755, 107 
\bibitem[Danforth et al.(2016)]{Danforth+16} Danforth, C.~W., Keeney, B.~A., Tilton, E.~M., et al.\ 2016, \apj, 817, 111 
\bibitem[Dekel \& Birnboim(2006)]{DekelBirnboim06} Dekel, A., \& Birnboim, Y.\ 2006, \mnras, 368, 2 
\bibitem[Draine et al.(2007)]{Draine+07} Draine, B.~T., Dale, D.~A., Bendo, G., et al.\ 2007, \apj, 663, 866 
\bibitem[Draine(2011)]{Draine11} Draine, B.~T.\ 2011, Physics of the Interstellar and Intergalactic Medium by Bruce T.~Draine.~Princeton University Press, 2011.~ISBN: 978-0-691-12214-4,  
\bibitem[Dutton \& Macci{\`o}(2014)]{DuttonMaccio14} Dutton, A.~A., \& Macci{\`o}, A.~V.\ 2014, \mnras, 441, 3359 
\bibitem[Faerman et al.(2017)]{Faerman+17} Faerman, Y., Sternberg, A., \& McKee, C.~F.\ 2017, \apj, 835, 52 
\bibitem[Fang et al.(2006)]{Fang+06} Fang, T., Mckee, C.~F., Canizares, C.~R., \& Wolfire, M.\ 2006, \apj, 644, 174 
\bibitem[Faucher-Gigu{\`e}re et al.(2009)]{FaucherGiguere+09} Faucher-Gigu{\`e}re, C.-A., Lidz, A., Zaldarriaga, M., \& Hernquist, L.\ 2009, \apj, 703, 1416 
\bibitem[Faucher-Gigu{\`e}re et al.(2010)]{FaucherGiguere+10} Faucher-Gigu{\`e}re, C.-A., Kere{\v s}, D., Dijkstra, M., Hernquist, L., \& Zaldarriaga, M.\ 2010, \apj, 725, 633 
\bibitem[Ferland et al.(2013)]{Ferland+13} Ferland, G.~J., Porter, R.~L., van Hoof, P.~A.~M., et al.\ 2013a, \rmxaa, 49, 137 
\bibitem[Fielding et al.(2017)]{Fielding+17} Fielding, D., Quataert, E., McCourt, M., \& Thompson, T.~A.\ 2017, \mnras, 466, 3810 
 \bibitem[Ford et al.(2016)]{Ford+16} Ford, A.~B., Werk, J.~K., Dav{\'e}, R., et al.\ 2016, \mnras, 459, 1745 
\bibitem[Fox(2011)]{Fox11} Fox, A.~J.\ 2011, \apj, 730, 58 
\bibitem[Fumagalli et al.(2017)]{Fumagalli+17} Fumagalli, M., Haardt, F., Theuns, T., et al.\ 2017, arXiv:1702.04726 
\bibitem[Gaikwad et al.(2017)]{Gaikwad+17} Gaikwad, P., Khaire, V., Choudhury, T.~R., \& Srianand, R.\ 2017, \mnras, 466, 838 
\bibitem[Gatto et al.(2013)]{Gatto+13} Gatto, A., Fraternali, F., Read, J.~I., et al.\ 2013, \mnras, 433, 2749 
\bibitem[Gnat \& Sternberg(2007)]{GnatSternberg07} Gnat, O., \& Sternberg, A.\ 2007, \apjs, 168, 213 
\bibitem[Goerdt \& Ceverino(2015)]{GoerdtCeverino15} Goerdt, T., \& Ceverino, D.\ 2015, \mnras, 450, 3359 
\bibitem[Grcevich \& Putman(2009)]{GrcevichPutman09} Grcevich, J., \& Putman, M.~E.\ 2009, \apj, 696, 385-395 
\bibitem[Groves et al.(2006)]{Groves+06} Groves, B.~A., Heckman, T.~M., \& Kauffmann, G.\ 2006, \mnras, 371, 1559 
\bibitem[Gutcke et al.(2017)]{Gutcke+17} Gutcke, T.~A., Stinson, G.~S., Macci{\`o}, A.~V., Wang, L., \& Dutton, A.~A.\ 2017, \mnras, 464, 2796 
\bibitem[Haardt \& Madau(2012)]{HaardtMadau12} Haardt, F., \& Madau, P.\ 2012, \apj, 746, 125 (HM12)
\bibitem[Heckman et al.(2002)]{Heckman+02} Heckman, T.~M., Norman, C.~A., Strickland, D.~K., \& Sembach, K.~R.\ 2002, \apj, 577, 691 
\bibitem[Henry et al.(2000)]{Henry+00} Henry, R.~B.~C., Edmunds, M.~G., \& K{\"o}ppen, J.\ 2000, \apj, 541, 660 
\bibitem[Hopkins et al.(2017)]{Hopkins+17} Hopkins, P.~F, Wetzel, A., Keres, D., et al.\ 2017, arXiv:1702.06148 
\bibitem[Hummels et al.(2013)]{Hummels+13} Hummels, C.~B., Bryan, G.~L., Smith, B.~D., \& Turk, M.~J.\ 2013, \mnras, 430, 1548 
\bibitem[Issa et al.(1990)]{Issa+90} Issa, M.~R., MacLaren, I., \& Wolfendale, A.~W.\ 1990, \aap, 236, 237 
\bibitem[Johnson et al.(2015)]{Johnson+15} Johnson, S.~D., Chen, H.-W., \& Mulchaey, J.~S.\ 2015, \mnras, 449, 3263 
\bibitem[Johnson et al.(2017)]{Johnson+17} Johnson, S.~D., Chen, H.-W., Mulchaey, J.~S., Schaye, J., \& Straka, L.~A.\ 2017, arXiv:1710.06441 
\bibitem[Kere{\v s} et al.(2005)]{Keres+05} Kere{\v s}, D., Katz, N., Weinberg, D.~H., \& Dav{\'e}, R.\ 2005, \mnras, 363, 2 
\bibitem[Kere{\v s} et al.(2009)]{Keres+09} Kere{\v s}, D., Katz, N., Fardal, M., Dav{\'e}, R., \& Weinberg, D.~H.\ 2009, \mnras, 395, 160 
\bibitem[Kollmeier et al.(2014)]{Kollmeier+14} Kollmeier, J.~A., Weinberg, D.~H., Oppenheimer, B.~D., et al.\ 2014, \apjl, 789, L32 
\bibitem[Lamers \& Cassinelli(1999)]{LamersCassinelli99} Lamers, H.~J.~G.~L.~M., \& Cassinelli, J.~P.\ 1999, Introduction to Stellar Winds, by Henny J.~G.~L.~M.~Lamers and Joseph P.~Cassinelli, pp.~452.~ISBN 0521593980.~Cambridge, UK: Cambridge University Press, June 1999., 452 
\bibitem[Li et al.(2017)]{Li+17} Li, J.-T., Bregman, J.~N., Wang, Q.~D., et al.\ 2017, arXiv:1710.07355 
\bibitem[Liang et al.(2016)]{Liang+16} Liang, C.~J., Kravtsov, A.~V., \& Agertz, O.\ 2016, \mnras, 458, 1164 
\bibitem[Liu et al.(2013)]{Liu+13} Liu, G., Zakamska, N.~L., Greene, J.~E., Nesvadba, N.~P.~H., \& Liu, X.\ 2013, \mnras, 430, 2327 
\bibitem[Madau \& Haardt(2015)]{HaardtMadau15} Madau, P., \& Haardt, F.\ 2015, \apjl, 813, L8 
\bibitem[Maller \& Bullock(2004)]{MallerBullock04} Maller, A.~H., \& Bullock, J.~S.\ 2004, \mnras, 355, 694 
\bibitem[Mandelker et al.(2016)]{Mandelker+16} Mandelker, N., Padnos, D., Dekel, A., et al.\ 2016, \mnras, 463, 3921 
\bibitem[Masaki \& Yoshida(2012)]{MasakiYoshida12} Masaki, S., \& Yoshida, N.\ 2012, \mnras, 423, L117 
\bibitem[Mathews \& Prochaska(2017)]{MathewsProchaska17} Mathews, W.~G., \& Prochaska, J.~X.\ 2017, arXiv:1708.07140 
\bibitem[McKinnon et al.(2016)]{McKinnon+16} McKinnon, R., Torrey, P., \& Vogelsberger, M.\ 2016, \mnras, 457, 3775 
\bibitem[McKinnon et al.(2017)]{McKinnon+17} McKinnon, R., Torrey, P., Vogelsberger, M., Hayward, C.~C., \& Marinacci, F.\ 2017, \mnras, 468, 1505 
\bibitem[McQuinn(2016)]{McQuinn16} McQuinn, M.\ 2016, \araa, 54, 313 
\bibitem[McQuinn \& Werk(2017)]{McQuinnWerk17} McQuinn, M., \& Werk, J.~K.\ 2017, arXiv:1703.03422 (MQW17)
\bibitem[M{\'e}nard \& Chelouche(2009)]{MenardChelouce09} M{\'e}nard, B., \& Chelouche, D.\ 2009, \mnras, 393, 808 
\bibitem[M{\'e}nard et al.(2010)]{Menard+10} M{\'e}nard, B., Scranton, R., Fukugita, M., \& Richards, G.\ 2010, \mnras, 405, 1025 (MSFR)
\bibitem[M{\'e}nard \& Fukugita(2012)]{MenardFukugita12} M{\'e}nard, B., \& Fukugita, M.\ 2012, \apj, 754, 116 
\bibitem[Meyer et al.(1998)]{Meyer+98} Meyer, D.~M., Jura, M., \& Cardelli, J.~A.\ 1998, \apj, 493, 222 
\bibitem[Miller \& Bregman(2013)]{MillerBregman13} Miller, M.~J., \& Bregman, J.~N.\ 2013, \apj, 770, 118 
\bibitem[Miller \& Bregman(2015)]{MillerBregman15} Miller, M.~J., \& Bregman, J.~N.\ 2015, \apj, 800, 14 
\bibitem[Miller et al.(2016)]{Miller+16} Miller, M.~J., Hodges-Kluck, E.~J., \& Bregman, J.~N.\ 2016, \apj, 818, 112 
\bibitem[Moster et al.(2010)]{Moster+10} Moster, B.~P., Somerville, R.~S., Maulbetsch, C., et al.\ 2010, \apj, 710, 903 
\bibitem[Moster et al.(2013)]{Moster+13} Moster, B.~P., Naab, T., \& White, S.~D.~M.\ 2013, \mnras, 428, 3121 
\bibitem[Nelson et al.(2013)]{Nelson+13} Nelson, D., Vogelsberger, M., Genel, S., et al.\ 2013, \mnras, 429, 3353 
\bibitem[Nelson et al.(2016)]{Nelson+16} Nelson, D., Genel, S., Pillepich, A., et al.\ 2016, \mnras, 460, 2881 
\bibitem[Oppenheimer \& Schaye(2013)]{OppenheimerSchaye13} Oppenheimer, B.~D., \& Schaye, J.\ 2013, \mnras, 434, 1063 
\bibitem[Oppenheimer et al.(2016)]{Oppenheimer+16} Oppenheimer, B.~D., Crain, R.~A., Schaye, J., et al.\ 2016, \mnras, 460, 2157 
\bibitem[Oppenheimer et al.(2017)]{Oppenheimer+17} Oppenheimer, B.~D., Segers, M., Schaye, J., Richings, A.~J., \& Crain, R.~A.\ 2017, arXiv:1705.07897 
\bibitem[Peek et al.(2015)]{Peek+15} Peek, J.~E.~G., M{\'e}nard, B., \& Corrales, L.\ 2015, \apj, 813, 7 
\bibitem[Peeples et al.(2014)]{Peeples+14} Peeples, M.~S., Werk, J.~K., Tumlinson, J., et al.\ 2014, \apj, 786, 54 
\bibitem[Planck Collaboration et al.(2016)]{Planck16} Planck Collaboration, Ade, P.~A.~R., Aghanim, N., et al.\ 2016, \aap, 594, A13 
\bibitem[Prochaska et al.(2004)]{Prochaska+04} Prochaska, J.~X., Chen, H.-W., Howk, J.~C., Weiner, B.~J., \& Mulchaey, J.\ 2004, \apj, 617, 718 
\bibitem[Prochaska et al.(2011)]{Prochaska+11} Prochaska, J.~X., Weiner, B., Chen, H.-W., Mulchaey, J., \& Cooksey, K.\ 2011, \apj, 740, 91 
\bibitem[Prochaska et al.(2017)]{Prochaska+17} Prochaska, J.~X., Werk, J.~K., Worseck, G., et al.\ 2017, arXiv:1702.02618 
\bibitem[Rachford et al.(2009)]{Rachford+09} Rachford, B.~L., Snow, T.~P., Destree, J.~D., et al.\ 2009, \apjs, 180, 125 
\bibitem[Rees \& Ostriker(1977)]{ReesOstriker77} Rees, M.~J., \& Ostriker, J.~P.\ 1977, \mnras, 179, 541 
\bibitem[Reines \& Volonteri(2015)]{ReinesVolonteri15} Reines, A.~E., \& Volonteri, M.\ 2015, \apj, 813, 82 
\bibitem[R{\'e}my-Ruyer et al.(2014)]{RemyRuyer+14} R{\'e}my-Ruyer, A., Madden, S.~C., Galliano, F., et al.\ 2014, \aap, 563, A31 
\bibitem[Rosdahl \& Blaizot(2012)]{RosdahlBlaizot12} Rosdahl, J., \& Blaizot, J.\ 2012, \mnras, 423, 344 
\bibitem[Salem et al.(2015)]{Salem+15} Salem, M., Besla, G., Bryan, G., et al.\ 2015, \apj, 815, 77 
\bibitem[Segers et al.(2017)]{Segers+17} Segers, M.~C., Oppenheimer, B.~D., Schaye, J., \& Richings, A.~J.\ 2017, \mnras, 471, 1026 
\bibitem[Shull et al.(2015)]{Shull+15} Shull, J.~M., Moloney, J., Danforth, C.~W., \& Tilton, E.~M.\ 2015, \apj, 811, 3 
\bibitem[Stern et al.(2016a)]{Stern+16} Stern, J., Hennawi, J.~F., Prochaska, J.~X., \& Werk, J.~K.\ 2016a, \apj, 830, 87 (S16)
\bibitem[Stern et al.(2016b)]{Stern+16b} Stern, J., Faucher-Gigu{\`e}re, C.-A., Zakamska, N.~L., \& Hennawi, J.~F.\ 2016b, \apj, 819, 130 
\bibitem[Stinson et al.(2012)]{Stinson+12} Stinson, G.~S., Brook, C., Prochaska, J.~X., et al.\ 2012, \mnras, 425, 1270 
\bibitem[Strateva et al.(2001)]{Strateva+01} Strateva, I., Ivezi{\'c}, {\v Z}., Knapp, G.~R., et al.\ 2001, \aj, 122, 1861 
\bibitem[Suresh et al.(2017)]{Suresh+17} Suresh, J., Rubin, K.~H.~R., Kannan, R., et al.\ 2017, \mnras, 465, 2966 
\bibitem[Thom \& Chen(2008)]{ThomChen08} Thom, C., \& Chen, H.-W.\ 2008, \apj, 683, 22-32 
\bibitem[Tumlinson et al.(2011)]{Tumlinson+11} Tumlinson, J., et al.\ 2011, Science, 334, 948 
\bibitem[Tumlinson et al.(2013)]{Tumlinson+13} Tumlinson, J., Thom, C., Werk, J.~K., et al.\ 2013, \apj, 777, 59 
\bibitem[Upton Sanderbeck et al.(2017)]{UptonSanderbeck+17} Upton Sanderbeck, P.~R., McQuinn, M., D'Aloisio, A., \& Werk, J.~K.\ 2017, arXiv:1710.07295 
\bibitem[Vasiliev et al.(2015)]{Vasiliev+15} Vasiliev, E.~O., Ryabova, M.~V., \& Shchekinov, Y.~A.\ 2015, \mnras, 446, 3078 
\bibitem[Verner \& Yakovlev(1995)]{VernerYakovlev95} Verner, D.~A., \& Yakovlev, D.~G.\ 1995, \aaps, 109, 125 
\bibitem[Vogelsberger et al.(2013)]{Vogelsberger+13} Vogelsberger, M., Genel, S., Sijacki, D., et al.\ 2013, \mnras, 436, 3031 
\bibitem[Wang et al.(2005)]{Wang+05} Wang, Q.~D., Yao, Y., Tripp, T.~M., et al.\ 2005, \apj, 635, 386 
\bibitem[Werk et al.(2012)]{Werk+12} Werk, J.~K., Prochaska, J.~X., Thom, C., et al.\ 2012, \apjs, 198, 3 
\bibitem[Werk et al.(2013)]{Werk+13} Werk, J.~K., Prochaska, J.~X., Thom, C., et al.\ 2013, \apjs, 204, 17 
\bibitem[Werk et al.(2014)]{Werk+14} Werk, J.~K., Prochaska, J.~X., Tumlinson, J., et al.\ 2014, \apj, 792, 8 
\bibitem[Werk et al.(2016)]{Werk+16} Werk, J.~K., Prochaska, J.~X., Cantalupo, S., et al.\ 2016, \apj, 833, 54 
\bibitem[Wetzel \& Nagai(2015)]{WetzelNagai15} Wetzel, A.~R., \& Nagai, D.\ 2015, \apj, 808, 40 
\bibitem[White \& Frenk(1991)]{WhiteFrenk91} White, S.~D.~M., \& Frenk, C.~S.\ 1991, \apj, 379, 52 
\end{thebibliography}

\appendix
\section{A.~Low ions}\label{a:low-ions}

Figure~\ref{f:low ions} plots ion columns versus normalized impact parameter in the COS-Halos+J15 sample. 
The ion columns are taken from \cite{Werk+13} and J15, using only absorption components within $200\kms$ of the central galaxy, as done in Figs.~\ref{f:ovi observations}--\ref{f:hi2ovi} for \ovip\ and \hi.
For comparison, we also mark in the panels the median radius of the \ovip-gas $\Rovi$.
All plotted ions show higher detection rates and columns at lower impact parameters, in contrast with \ovip, for which the column is relatively independent of impact parameter at $R<\Rovi$ (Fig.~\ref{f:ovi observations}). At $\Rimp>\Rovi$ the ions \mgiip, \Siiip, \siiip, \niiip, and \ciip, are never detected, while \Siiiip\ and \niip\ show a single detection, and \ciiip\ shows two detections. The \niip\ detection is at $+183\kms$ from the galaxy J0943+0531\_227\_19, and is likely associated with a \hi\ and \ciiip\ system at $>200\kms$, distinct from the velocities spanned by the \ovip\ absorbers analyzed in this work. 
We conclude that detectable quantities of lower ionization ions typically exist only at $<\Rovi$, with the exception of \ciiip\ and possibly \Siiiip. The low-ions hence reside at a smaller physical distance from the galaxy than most of the \ovip.

\begin{figure*}
 \includegraphics{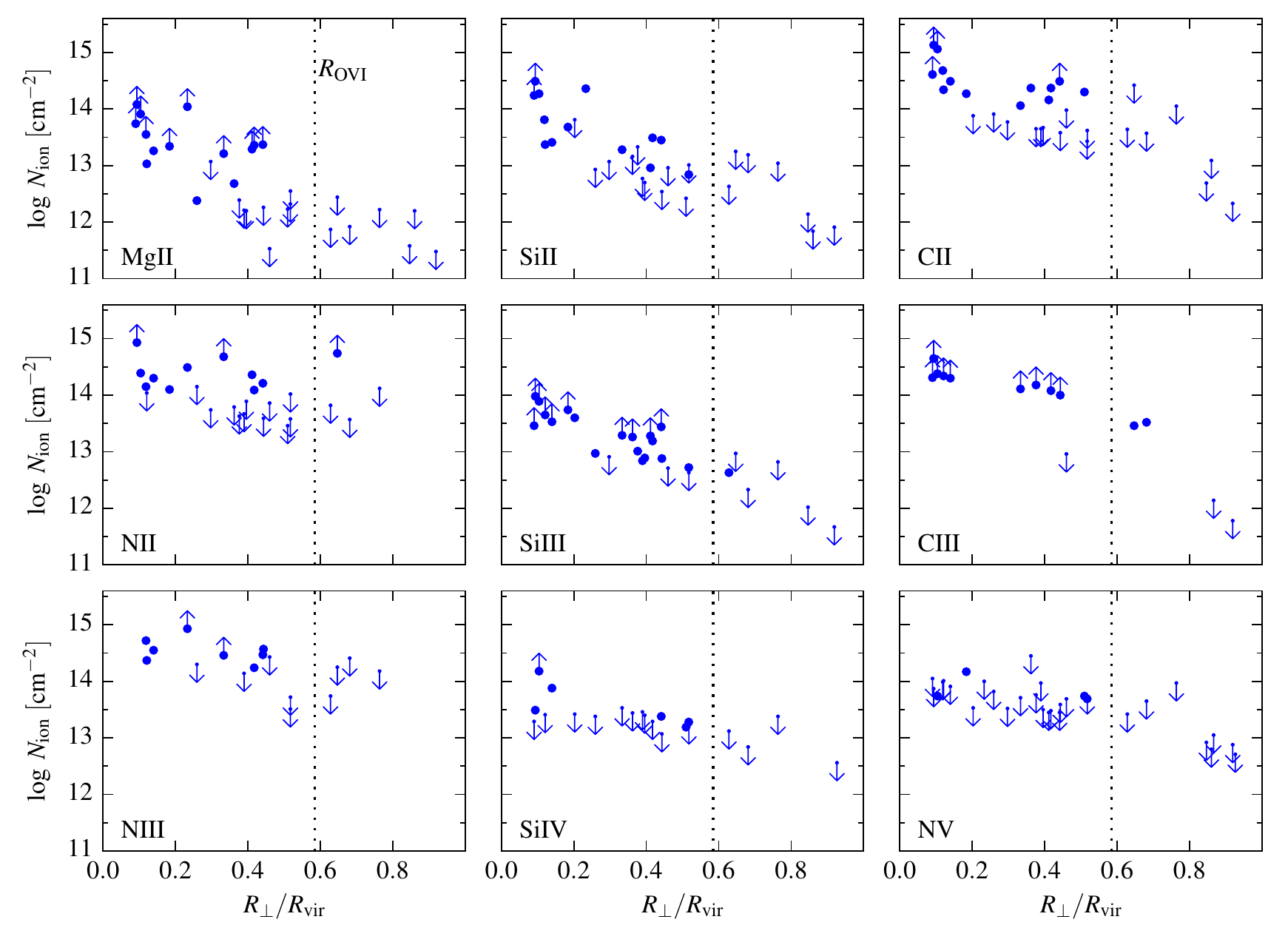}
\caption{Observed ion columns in the CGM of \sLstar\ star-forming galaxies with $0.1<z<0.4$. Data points and upper limits are from the COS-Halos survey and from J15. Non-detections and saturated lines are marked by down- and up-pointing arrows, respectively. Measurement errors are typically $<0.1\dex$. The panels are sorted by the ion ionization energy. The median radius of the \ovip-gas inferred in Fig.~\ref{f:ovi observations}, $\Rovi$, is noted by vertical dotted lines. Note that most ions are observed only at impact parameters smaller than $\Rovi$.
}
\label{f:low ions}
\end{figure*}

\section{B.~Pathlength of the \ovip\ gas assuming photoionization by the UV background}\label{a:NV}

Table \ref{t:pathlengths} lists 28 of the 32 quasar-galaxy pairs in the sample considered in this study in which \nvp~$\lambda1240$ was in the wavelength range probed by the quasar spectrum. We exclude the four objects were both \ovip\ and \nvp\ were not detected. The last three columns in the table give the density, metallicity and pathlength of the \ovip\ gas ($\lovi = \Novi/\novi$), based on the single-density PIE models described in \S\ref{s:other ions} (see also lower-left panel of Fig.~\ref{f:PIE}). The models are constrained using the observed $\Novi$, $\NHI$, and $N_\nvp$. As discussed in \S\ref{s:hi}, in sightlines with $\Rimp<\Rovi$ the \hi\ column is likely dominated by dense gas at smaller scales than \ovip, so we assume $\NHI/\Novi=3$ for the \ovip-gas, which is the median ratio observed at $\Rimp>\Rovi$. For comparison, \cite{Werk+16} performed a similar calculation assuming a constant $\NHI=10^{15}\cm^{-2}$ for all objects. In objects where $N_\nvp$ is an upper limit the derived $Z$ and $\nH$ are upper limits, while the derived $\lovi$ is a lower limit. Table \ref{t:pathlengths} shows that the minimum pathlengths are typically a factor of $2-10$ smaller than $\rvir$, and only in a  single object (J0914+2823\_41\_27) the minimum pathlength is larger than $\rvir$. That is, as long as the actual $N_\nvp$ are not significantly below the measured upper limits, the implied pathlengths are physically possible, and consistent with \ovip\ originating in the ambient medium beyond the shock as suggested by the low pressure scenario discussed in this work. Hence, a photoionization origin for \ovip\ cannot be ruled out based on the existing upper limits on \nvp, in contrast with the conclusion of \cite{Werk+16}. The different conclusion regarding \ovip\ pathlengths is mainly due to the scaling of [N/O] with [O/H] used in this work (see \S3.5) compared to the solar [N/O] assumed by \cite{Werk+16}. 

\begin{table*}
\begin{tabular}{l|c|c|c|c|c|c|c|c|c}
Object & survey & $\rvir$       & $\Rimp/\rvir$ & $\log \Novi$           & $\log \NHI$            & $\log N_{\nvp}$        & $Z^{(a)}$ & $\log \nH^{(a)}$ & $\lovi^{(a)}$ \\
       &        & $[{\rm kpc}]$ &               & $[{\rm cm}^{-2}]$ & $[{\rm cm}^{-2}]$ & $[{\rm cm}^{-2}]$ & $[Z_\odot]$       & $[{\rm cm}^{-3}]$   & $[\kpc]$            \\
\hline
04:07:50.57-12:12:24.0 &    J15 &  275 & 0.9 & 13.6 & 14.0 & $<$12.9 & $<$0.8 & $<$-4.3 &    $>$4  \\
14:37:43.01-01:47:42.6 &    J15 &  243 & 0.9 & 13.8 & 14.8 & $<$12.8 & $<$0.2 & $<$-4.5 &   $>$31  \\
14:37:50.10-01:47:10.2 &    J15 &  149 & 0.9 & 13.9 & 14.1 & $<$12.9 & $<$0.5 & $<$-4.6 &   $>$13  \\
 J1437+5045\_317\_38 & COS-Halos &  187 & 0.8 & 14.4 & 14.5 & $<$14.0 & $<$1.7 & $<$-4.4 &   $>$11  \\
 J1245+3356\_236\_36 & COS-Halos &  165 & 0.7 & 14.3 & 14.7 & $<$13.7 & $<$0.9 & $<$-4.3 &   $>$18  \\
  J0914+2823\_41\_27 & COS-Halos &  162 & 0.6 & 14.7 & 15.5 & $<$13.4 & $<$0.2 & $<$-4.4 &  $>$251  \\
 J1445+3428\_232\_33 & COS-Halos &  219 & 0.5 & 14.4 & 14.8$^{(b)}$ & $<$13.7 & $<$0.9 & $<$-4.4 &   $>$22  \\
 J1241+5721\_208\_27 & COS-Halos &  181 & 0.5 & 14.7 & 15.2$^{(b)}$ & 13.7 & 0.4 & -4.4 &   96  \\
 J1619+3342\_113\_40 & COS-Halos &  190 & 0.5 & 14.2 & 14.7$^{(b)}$ & 13.7 & 1.3 & -4.3 &   10  \\
  J1233+4758\_94\_38 & COS-Halos &  298 & 0.4 & 14.4 & 14.9$^{(b)}$ & $<$13.4 & $<$0.5 & $<$-4.4 &   $>$44  \\
  J0401-0540\_67\_24 & COS-Halos &  189 & 0.4 & 14.5 & 15.0$^{(b)}$ & $<$13.6 & $<$0.5 & $<$-4.4 &   $>$54  \\
  J0910+1014\_34\_46 & COS-Halos &  266 & 0.4 & 14.6 & 15.0$^{(b)}$ & $<$13.4 & $<$0.3 & $<$-4.4 &   $>$88  \\
 J1330+2813\_289\_28 & COS-Halos &  213 & 0.4 & 14.4 & 14.9$^{(b)}$ & $<$13.5 & $<$0.5 & $<$-4.4 &   $>$45  \\
 J1435+3604\_126\_21 & COS-Halos &  212 & 0.4 & 14.6 & 15.1$^{(b)}$ & $<$14.0 & $<$0.9 & $<$-4.4 &   $>$35  \\
  J1233-0031\_168\_7 & COS-Halos &  229 & 0.4 & 14.7 & 15.2$^{(b)}$ & $<$13.8 & $<$0.5 & $<$-4.4 &   $>$86  \\
 J1009+0713\_204\_17 & COS-Halos &  164 & 0.4 & 15.0 & 15.4$^{(b)}$ & $<$14.4 & $<$1.1 & $<$-4.3 &   $>$72  \\
 J1419+4207\_132\_30 & COS-Halos &  256 & 0.3 & 14.4 & 14.9$^{(b)}$ & $<$13.7 & $<$0.7 & $<$-4.4 &   $>$32  \\
 J1133+0327\_164\_21 & COS-Halos &  186 & 0.3 & 14.5 & 15.0$^{(b)}$ & $<$13.5 & $<$0.4 & $<$-4.4 &   $>$57  \\
 J1112+3539\_236\_14 & COS-Halos &  204 & 0.3 & 14.6 & 15.0$^{(b)}$ & $<$13.8 & $<$0.7 & $<$-4.4 &   $>$41  \\
  J1009+0713\_170\_9 & COS-Halos &  189 & 0.2 & 15.0 & 15.4$^{(b)}$ & $<$14.0 & $<$0.5 & $<$-4.4 &  $>$153  \\
14:37:26.68+50:46:07.4 &    J15 &  158 & 0.2 & 14.6 & 15.1$^{(b)}$ & $<$13.5 & $<$0.4 & $<$-4.4 &   $>$84  \\
 J1016+4706\_359\_16 & COS-Halos &  234 & 0.2 & 14.6 & 15.1$^{(b)}$ & 14.2 & 1.2 & -4.3 &   29  \\
  J1555+3628\_88\_11 & COS-Halos &  244 & 0.1 & 14.6 & 15.1$^{(b)}$ & $<$13.9 & $<$0.7 & $<$-4.4 &   $>$48  \\
 J1322+4645\_349\_11 & COS-Halos &  303 & 0.1 & 14.5 & 15.0$^{(b)}$ & $<$14.0 & $<$1.1 & $<$-4.3 &   $>$24  \\
  J1016+4706\_274\_6 & COS-Halos &  193 & 0.1 & 14.9 & 15.3$^{(b)}$ & $<$14.0 & $<$0.5 & $<$-4.4 &  $>$111  \\
  J1241+5721\_199\_6 & COS-Halos &  191 & 0.1 & 14.8 & 15.3$^{(b)}$ & 13.7 & 0.4 & -4.4 &  149  \\
 J1342-0053\_157\_10 & COS-Halos &  366 & 0.1 & 14.6 & 15.1$^{(b)}$ & $<$13.9 & $<$0.7 & $<$-4.4 &   $>$46  \\
  J1435+3604\_68\_12 & COS-Halos &  476 & 0.1 & 14.8 & 15.2$^{(b)}$ & $<$14.1 & $<$0.7 & $<$-4.4 &   $>$66  \\
\end{tabular}
\vspace{0.1cm} 
\caption{}
$(a)$ Implied metallicity, density and pathlength of the \ovip\ gas assuming photoionization by the UV background and the listed ionic columns. In objects where $N_\nvp$ is an upper limit the derived $Z$ and $\nH$ are upper limits, while the derived $\lovi$ is a lower limit. \\
$(b)$ In objects with $\Rimp<\Rovi$ the \hi\ associated with the \ovip\ gas is assumed to satisfiy $\NHI/\Novi=3$ (see Appendix~\ref{a:NV}).
\label{t:pathlengths}
\end{table*}

\end{document}